\documentclass[nofootinbib,aps,prd,10pt,superscriptaddress,showpacs,preprintnumbers,eqsecnum]{revtex4-1}

\usepackage{amsmath}
\usepackage{dsfont}
\usepackage{multirow}
\usepackage{graphicx}
\usepackage{mathrsfs}

\usepackage{amsfonts}
\usepackage{amssymb}
\usepackage[dvips]{color}

\begin{document}
\preprint{IHES/P/10/16}
\newcommand\be{\begin{equation}}
\newcommand\ee{\end{equation}}
\newcommand\bea{\begin{eqnarray}}
\newcommand\eea{\end{eqnarray}}
\newcommand\bseq{\begin{subequations}} 
\newcommand\eseq{\end{subequations}}
\newcommand\bcas{\begin{cases}}
\newcommand\ecas{\end{cases}}
\newcommand{\p}{\partial}
\newcommand{\f}{\frac}

\title{Statistical Properties of Cosmological Billiards}

\author{Thibault Damour}
\email{damour@ihes.fr}
\affiliation{Institut des Hautes Etudes Scientifiques, 35, route de Chartres, 91440 Bures-sur-Yvette, France.}
\affiliation{ICRANet, Pescara, Italy}
\author{Orchidea Maria Lecian}
\email{lecian@ihes.fr}
\affiliation{Institut des Hautes Etudes Scientifiques, 35, route de Chartres, 91440 Bures-sur-Yvette, France.}
\affiliation{ICRANet, Pescara, Italy}
\affiliation{APC, UMR 7164 du CNRS, Universit\'e Paris 7,\\10, rue A. Domon et L. Duquet, 75205 Paris Cedex 13, France.}


\begin{abstract}
Belinski, Khalatnikov and Lifshitz (BKL) pioneered the study of the statistical properties of the never-ending oscillatory behavior (among successive Kasner epochs) of the geometry near a space-like singularity. We show how the use of a ``cosmological billiard'' description allows one to refine and deepen the understanding of these statistical properties. Contrary to previous treatments, we do not quotient the dynamics by its discrete symmetry group (of order $6$), thereby uncovering new phenomena, such as correlations between the successive billiard corners in which the oscillations take place. Starting from the general integral invariants of Hamiltonian systems, we show how to construct invariant measures for various projections of the cosmological-billiard dynamics. In particular, we exhibit, for the first time, a (non-normalizable) invariant measure on the ``Kasner circle'' which parametrizes the exponents of successive Kasner epochs. Finally, we discuss the relation between: (i) the unquotiented dynamics of the Bianchi IX ($a$, $b$, $c$ or mixmaster) model; (ii) its quotienting by the group of permutations of ($a$, $b$, $c$); and (iii) the billiard dynamics that arose in recent studies suggesting the hidden presence of Kac-Moody symmetries in cosmological billiards. 
\end{abstract}

\pacs{04.50.-h Higher-dimensional gravity and other theories of gravity; 05.45.-a Nonlinear dynamics and chaos; 98.80.Jk Mathematical and relativistic aspects of cosmology}

\maketitle
\section{Introduction}
A remarkable achievement of theoretical cosmology has been the construction, by Belinski, Khalatnikov and Lifshitz (BKL), of a general solution to the $4$-dimensional vacuum Einstein equations in the vicinity of a spacelike (``cosmological'') singularity \cite{KB1969a}, \cite{Khalatnikov:1969eg} \cite{BK1970}, \cite{BLK1971}. They found that this solution exhibits a never-ending oscillatory behavior, with strong chaotic properties. They could describe in detail the statistical properties of this never-ending oscillatory behavior by approximating the Einstein field equations (near the singularity) by a system of ODE's for three variables $a$, $b$, $c$ (`anisotropic scale factors'), namely
\begin{subequations}\label{abc}
\begin{align}
&2\frac{d^2\ln a}{d\tau^2}=(b^2-c^2)^2-a^4,\\
&2\frac{d^2\ln b}{d\tau^2}=(c^2-a^2)^2-b^4,\\
&2\frac{d^2\ln c}{d\tau^2}=(a^2-b^2)^2-c^4,
\end{align}
\end{subequations}
where $d\tau=-dt/(abc)$, and by approximately reducing the continuous dynamics of $a$, $b$, $c$ to a sequence of discrete maps. The crucial discrete map introduced by BKL relates the `Kasner exponents' $p_a$, $p_b$, $p_c$ describing the (approximately linear) $\tau$-evolutions of the three scale factors $a$, $b$, $c$ during two successive `epochs' (i.e. two successive segments of the dynamics (\ref{abc}) during which the influence of the right-hand side is negligible). More precisely, BKL, following \cite{Lifshitz:1963ps}, parametrize the three Kasner exponents $p_a$, $p_b$, $p_c$ (constrained to satisfy $1=p_a+p_b+p_c=p_a^2+p_b^2+p_c^2$) by means of one real parameter $u$, and show that the interval $1\le u\le\infty$ is in one to one correspondence with the \textit{unordered} set $\{p_a,p_b,p_c\}$. In terms of this parametrization, BKL showed that the discrete map describing the passage from one epoch to the next is
\begin{subequations}\label{2'ab}
\begin{align}
&{\rm if}\ \ u>2, \ \ u'=u-1\label{2'a}\\
&{\rm while, if}\ \ 1<u<2, \ \ u'=\frac{1}{u-1}.
\end{align}
\end{subequations}
They also defined an `era' as being a set of successive epochs during which $u$ evolves according to the simple law (\ref{2'a}). This led them to realize that the `chaotic' part of the discrete epoch dynamics (\ref{2'ab}) is essentially contained in the ``Gauss iteration map'' \footnote{Here $\{y\}$ denotes the fractional part, $\{y\}=y-[y]$ (where $[y]$ denotes the integer part of $y$) of the (positive) real number $y$. We recall that the Gauss map is at the basis of the expansion of a positive real number into a continued fraction $n_1+1/(n_2+1/(n_3+...))$.}
\be\label{1.1}
x_{n+1}=\{1/x_n\}\equiv 1/x_n-[1/x_n]. 
\ee
Here, $x$ (with $0<x<1$) denotes the fractional part of $u$ during an era. Let us recall that the $u$ parameter of all the epochs belonging to the $n$-th era can be written in two essentially equivalent ways, which depend on the precise way in which one defines an era as a collection of epochs. If one uses the definition of an era such that the corresponding $u$ parameters are always larger than one, the $n^{th}$ era consists of $k_n$ epochs parametrized by values of $u$ of the form $k_n+x_n, k_n-1+x_n, ..., 1+x_n$. The next era then starts by a value of $u$ equal to $k_{n+1}+x_{n+1}=1/x_n$, so that $k_{n+1}=[1/x_n]$, and $x_{n+1}=\{1/x_n\}$. We shall refer to this definition of an era as the ``standard'' one (because it was adopted in the treatise of Landau and Lifshitz \cite{llft}), or as the BKL$_{u>1}$ one. An alternative definition, introduced by BKL in Eq. $(5.4)$ of \cite{BLK1971} (that we shall call the BKL$_{u>0}$ one) leads to considering that the $n$-th era starts with $u=k_n+x_n-1$, and ends with $u=x_n<1$ (with the next era starting with $u=k_{n+1}-1+x_{n+1}$). We shall see below that the latter, (alternative) definition of an era is more natural in the billiard picture, and this is the one we shall actually use in our work. The iteration of the Gauss map (\ref{1.1}) leads to a statistical behavior of the successive values of $x$, with an asymptotic stationary probability distribution over the interval $[0,1]$ \cite{BK1970}, \cite{LLK}, \cite{BLK1971}:
\be\label{2'}
w(x)dx=\frac{1}{\ln2}\frac{dx}{1+x}.
\ee
For general reviews of the works dealing with the BKL singularity , see  \cite{Montani:2007vu} \cite{LLK} \cite{BLK1971}. Later studies have refined the description of the statistical properties of the chaotic BKL oscillations, notably by introducing and studying more complete discrete iteration maps (involving several real variables), and notably by the two-dimensional discrete map \cite{Chernoff:1983zz}, \cite{sinai83}, \cite{sinai85}
\begin{subequations}\label{1.2}
\begin{align}
&x^+_{n+1}=\{1/x^+_n\}\equiv 1/x^+_n-[1/x^+_n],\\
&x^-_{n+1}=1/([1/x^+_n]+x^-_n).
\end{align}
\end{subequations}

The iteration of this two-dimensional map (of the unit square into itself) asymptotically leads to a statistical behavior for $(x^+_n,x^-_n)\in[0,1]\times[0,1]$ with probability distribution \cite{Chernoff:1983zz}, \cite{sinai83}, \cite{sinai85}
\be\label{3'}
w(x^+,x^-)dx^+dx^-=\frac{1}{\ln2}\frac{dx^+dx^-}{(1+x^+x^-)^2}.
\ee
Separately from the work of BKL\footnote{It seems that western physicists, and notably J.A. Wheeler who was in the audience, first heard about the BKL results from a seminar given by Isaak Khalatnikov at the Institut Henri Poincar\'e, Paris, in January 1968; see \cite{dau}.}, and, with a different aim and motivation, Misner realized that generic Bianchi IX homogeneous cosmological models have a ``very complex singularity'' \cite{Misner:1969hg}. He described this complex dynamics by a Hamiltonian formalism, in terms of a ``system point'' $\beta=(\beta _+,\beta_-)$ bouncing against a system of ``potential walls''. A reformulation of this dynamics \cite{Misner:1974qy}, \cite{chi1972} led to the simpler picture of a point moving on a Lobachevsky plane and reflecting upon \textit{fixed} billiard-type cushions. This led Chitre (using earlier mathematical results by Hopf and Hedlung) to remark that the dynamics of the system point is ergodic and mixing, with unique invariant Liouville measure (restricted to a fixed energy shell) \cite{chi1972}, \cite{Misner:1994ge}
\be\label{1.3}
\mu_L=\delta(H(q,p)-E)d^2qd^2p\propto \tfrac{d^2\zeta d\theta}{(1-\mid\zeta \mid ^2)^2}.
\ee 
The description of cosmological singularities in terms of \textit{billiards} in (higher dimensional) Lobachevsky (or Lorentzian) spaces has recently received a new impetus from the discovery that the billiard chambers corresponding to many interesting physical theories can be identified with the ``Weyl chambers'' of certain (infinite-dimensional) Lorentzian Kac-Moody algebras \cite{Damour:2000hv}, \cite{Damour:2001sa}, \cite{Damour:2002fz}. This has raised the conjecture that, hidden below the BKL ``chaos'', there lies a remarkable ``Kac-Moody symmetry'', akin to the duality symmetries of supergravity and string theories \cite{Damour:2002cu}, \cite{Damour:2002et}, \cite{hps2009}.\\
Coming back to the cosmological singularities in ($3+1$)-dimensional General Relativity, the problem of relating the statistical properties of the discrete BKL map, such as Eq. (\ref{2'}) or Eq. (\ref{3'}), to the invariance of the Liouville measure (\ref{1.3}) in the continuous billiard dynamics, \`{a} la Misner-Chitre has been considered in some detail by Kirillov and Montani \cite{Kirillov:1996rd}. These authors have shown, by an explicit calculation, that the Liouville measure $\mu_L$, Eq. (\ref{1.3}), (which is a three-form) could be formally rewritten as the product of the invariant measure of the discrete BKL-type map (\ref{3'}), namely the two-form
\be\label{quatre}
\mu_2=\frac{dx^+\wedge dx^-}{(1+x^+x^-)^2},
\ee
by a one-dimensional measure $d\rho$, measuring the proper (hyperbolic) length along the billiard motion on the Lobachevsky plane. 
One of the aims of the present work is to better understand the link between the two different invariant measures (\ref{1.3}), (\ref{quatre}), and the origin of these measures within the symplectic structure of the (Lorentzian) billiard dynamics. Another aim will be to go beyond the \textit{symmetry quotienting} which has been used in most previous studies of the statistical properties of cosmological billiards. Indeed, there is a basic triality\footnote{Actually, what matters is a six-fold symmetry corresponding to the permutation group of the set $\{a,b,c\}$.} symmetry between the three BKL dynamical variables $a$, $b$, $c$, and the discrete maps (\ref{1.1}) or (\ref{1.2}) arise only if one effectively \textit{quotients} the phase-space dynamics by their symmetry. [An example of this quotienting is the fact that the parameter $u$, taken in the interval $1<u<\infty$, parameterizes the \textit{unordered} set of Kasner exponents $\{p_a, p_b, p_c\}$.] Here, we shall instead consider the richer (continuous and discrete) billiard dynamics in the full, unquotiented phase-space. As we shall see, this full dynamics contains new statistical features that do not appear in the traditionally considered quotiented dynamics. Finally, we shall also compare and contrast the (unquotiented or quotiented) BKL dynamics of the (diagonal type-IX) $a$, $b$, $c$ system with the billiard dynamics that naturally arose in the recent studies that uncovered the hidden presence of Kac-Moody-related structures in cosmological billiards (\cite{Damour:2000hv}, \cite{Damour:2001sa}, \cite{Damour:2002fz}, \cite{Damour:2002cu}, \cite{Damour:2002et}, \cite{hps2009}). Indeed, the billiard dynamics which are most closely connected to such hidden symmetries take place in the Weyl chambers of some Kac-Moody algebras. In the usual case of $4$-dimension vacuum Einstein gravity, this Weyl chamber is what we shall call a `small billiard' , obtained by quotienting the full ($a, b, c$) billiard table by the six-fold permutation group of the three letters $a$, $b$, $c$. As we shall discuss, the billiard dynamics in this quotiented (configuration space) billiard table is \textit{not} equivalent to the quotienting of the billiard dynamics in the full table, though the further quotienting of this small billiard dynamics by modding out the action (in phase space) of the $a-b-c$ permutation \textit{is} equivalent to the (phase-space) quotienting of the full `big billiard' dynamics. Our present paper will focus on the usual case of $4$-dimensional vacuum Einstein gravity. In a sequel paper, we will extend our results to higher-dimensional gravity models, using the generalized ``cosmological billiard'' approach \cite{Damour:2002et}.\\
The paper is organized as follows. In Section II, we introduce cosmological billiards; this leads, in particular, to contrasting the `big billiard' studied by Belinski, Khalatnikov, Lifshitz and Misner, with the `small billiard', connected with the Weyl chamber of a Kac-Moody algebra as shown in Fig \ref{fig2}. In Section III, we discuss two conformal representations of cosmological billiards, i.e. the disk model and the upper-half-plane model; in particular, we outline their representations of epochs and eras. In Section IV, we define integral invariants for general dynamical systems, which allow us to find an invariant measure for the BKL discrete map (full details about integral invariants in Hamiltonian systems are given in the Appendix). In Section V, we analyze the big billiard, describe its dynamics as a ``hopscotch game'' at different levels, and define the corresponding maps. In Section VI, we exploit the symmetries of the big billiard to define a symmetry-quotiented map. In Section VII, we study the main properties of this symmetry-quotiented map. In Section VIII, periodic phenomena in cosmological billiards are considered, and some differences between the complete billiard and the symmetry-quotiented billiard are outlined. In Section IX, the small billiard is introduced: its features are investigated, and its equivalence with the big billiard is discussed. Brief concluding remarks end the paper.  
\section{Reminders and technical preliminaries}
Let us start by defining our notation and recalling some basic facts about ``cosmological billiards''. [We mainly follow the notation of \cite{Damour:2002et}.] In order to describe the evolution of a general inhomogeneous space-time metric near a space-like singularity, it is convenient to use ``pseudo-Gaussian '' coordinates, with vanishing ``shift'' $N^i=0$, but with some convenient choice of the ``lapse'' $N$:
\be\label{2.1}
ds^2=-\left( N(x^0,x^i)dx^0\right)^2+g_{ij}(x^0,x^k)dx^idx^j.
\ee 
Here $x^0$ denotes the coordinate time associated to any particular way of choosing the value of the lapse function $N$. The indices $i,j=1, ..., d$ denote the various spatial dimensions. In the present work we shall consider the case $d=3$, but many of the general technical results recalled in this section are valid for any value of the space dimension $d$. There are two useful choices of the lapse $N$ for exhibiting the ``billiard nature'' of the dynamics of $g_{ij}$ near the singularity. The choice 
\be\label{2.2}  
N=\sqrt{g},
\ee
where $g$ denotes the determinant of the spatial metric $g_{ij}$, corresponds to using as coordinate time $x^0$ the parameter $\tau$ introduced by BKL, i.e. the quantity
\be\label{2.3}
d\tau=-\tfrac{dt}{\sqrt{g}},
\ee
where $dt=Ndx^0$ denotes the (local) proper time. [A minus sign is introduced in (\ref{2.3}) so that the cosmological singularity conventionally located at $t\rightarrow 0^+$ (``big bang'') is approached as $\tau$$\rightarrow$$+\infty$ with respect to the $\tau$-coordinate time.]\\
The choice of (\ref{2.2}), (\ref{2.3}) leads to an asymptotic description of the gravitational dynamics in terms of a ``Lorentzian billiard in $\beta$-space''. More precisely, one first performs an \textit{Iwasawa decomposition} of the spatial metric, i.e. one (locally) replaces the $d(d+1)/2$ functions $g_{ij}(x^0,x^k)$ by the $d$ functions $\beta^a(x^0, x^k)$, together with the $d(d-1)/2$ functions $\mathcal{N}^a_{\ \ i}(x^0,x^k)$ parametrizing an upper triangular matrix with $1$'s on the diagonal, according to
\be\label{2.4}
g_{ij}=\sum_{a=1}^d e^{-2\beta^a}\mathcal{N}^a_{\ \ i}\mathcal{N}^a_{\ \ j}.
\ee
In the near-singularity (or BKL) limit ($t\rightarrow0^+$ or $\tau\rightarrow+\infty$, or $\sum_a\beta^a\rightarrow+\infty$), one finds that the upper triangular matrix $\mathcal{N}^a_i$ has a limit \cite{Damour:2002et} and that the only parts of the metric which have a ``chaotic behavior'' are the ``diagonal degrees of freedom'' parametrized by the $d$ functions $\beta^a(x^0,x^k)$. Then one finds that, at each point of space, the $\beta^a$'s asymptotically follow a Lorentzian billiard dynamics: namely, the $\beta^a(\tau)$'s undergo a succession of constant-velocity straight-line flights interrupted by collisions (and reflections) on some hyperplanes in $\beta$-space.\\
The free-flight dynamics of the $\beta$ particle between wall collisions is described
 by the free action
 \be\label{2.5}
 \int \frac{1}{2}d\tau\sum_{a,b=1}^{d}G_{ab}\tfrac{d\beta^a}{d\tau}\tfrac{d\beta^b}{d\tau},
 \ee
 submitted to the constraint
\be\label{2.6}
\sum_{a,b=1}^{d}G_{ab}\tfrac{d\beta^a}{d\tau}\tfrac{d\beta^b}{d\tau}=0.
 \ee
Here, the $\beta$-space metric $G_{ab}$ is defined by
\be\label{2.7}
d\sigma^2=\sum_{a,b=1}^{d}G_{ab}d\beta^a d\beta^b=\sum_a\left(d\beta^a\right)^2-\left(\sum_ad\beta^a\right)^2.
\ee 
The metric $G_{ab}$ endows the d-dimensional space of the $\beta$'s with a \textit{Lorentzian} structure (signature $-++...+$). [In the case of $3+1$ dimensional General Relativity, the 3-dimensional $\beta$ space metric has signature $-++$. Note, however, that the coordinates $\beta^1, \beta^2, \beta^3$ are not of the canonical Lorentzian form. Indeed one has $d\sigma^2=-2\left(d\beta^1d\beta^2+d\beta^2d\beta^3+d\beta^3d\beta^1\right)$.] Note that the constraint (\ref{2.6}) means that, between collisions, the $\beta$ particle goes ``with the velocity of light'' (in the sense of the Lorentzian structure $G_{ab}$ in $\beta$ space.) In other words, the free-flight dynamics deduced from the action (\ref{2.5}), namely 
\be\label{2.7a}
\tfrac{d^2\beta^a}{d\tau^2}=0 \Rightarrow \beta^a(\tau)=\beta^a_0+v^a\tau,
\ee
is restricted by the quadratic constraint
\be\label{2.8}
0=\sum_{a,b=1}^d G_{ab}v^av^b=\sum_a(v^a)^2-\left(\sum_a v^a \right) ^2.
\ee
The free-flight dynamics (\ref{2.7a}) in $\beta$ space (and in the $\tau$ parametrization) corresponds, in the BKL language, to a ``Kasner epoch'' (between two successive wall collisions). The usually considered Kasner exponents $p_a$, $a=1,..,d$, corresponding to a Gaussian (or synchronous) gauge (i.e. $N=1$), are related to the $d$-dimensional velocity vector $v^a$ in $\beta$ space (and pseudo-Gaussian $\tau$ gauge, $N=\sqrt{g}$) via
\be\label{2.9}
p_a=\tfrac{v^a}{\sum_bv^b}.
\ee
Note that while $v^a$ satisfies the unique quadratic constraint (\ref{2.8}) (proportional to the combination $\sum_a\left(p_a\right)^2-\left(\sum_ap_a\right)^2$), the Kasner parameters $p_a$ satisfy the two well-known constraints
\be\label{2.10}
\sum_ap_a=1=\sum_ap_a^2.
\ee
The free flight dynamics (\ref{2.7}) is only valid if the `point' $\beta$ is sufficiently far from certain (Lorentzian) `wall hyperplanes' in $\beta$-space. The equations of these wall hyperplanes depend on the field content of the theory that one considers, (e.g. Einstein-Maxwell versus Einstein, etc...). They are of the general form
\be\label{2.11}
w^A(\beta)\equiv\sum_{a=1}^d w^A_a\beta^a=0.
\ee 
More precisely, the dynamics of the $\beta$ particle is given by a Hamiltonian $H$ of the form $H=H_0+V$, where $H_0$ is a free kinetic term describing (in Hamiltonian form) the free-flight part of the dynamics, namely 
\be
H_0=\frac{1}{2}\sum_{a,b=1}^d G^{ab}\pi_a\pi_b,
\ee 
where $G^{ab}$ is the  inverse of the covariant metric tensor $G_{ab}$ (in $\beta$ space), introduced in (\ref{2.7}). Its components in the $\beta^a$ (Iwasawa-related) coordinates are explicitly given by
\be\label{2.15}
\sum_{a,b}G^{ab}\pi_a\pi_b=\sum_a(\pi_a)^2-\tfrac{1}{d-1}\left(\sum_a\pi_a\right)^2.
\ee
As for the potential $V(\beta)$ in the Hamiltonian $H=H_0+V$, it is a sum of ``Toda-like'', i.e. exponential, terms: $V(\beta)\sim\sum_Ac_A\exp(-2 w^A(\beta))$. As recalled in the Appendix, in the near-singularity limit the potential $V(\beta)$ can be replaced by its sharp-wall limit $V_\infty(\beta)=\sum_A\Theta_\infty(-2w^A(\beta))$ where $\Theta(x):=0$ if $x<0$ and $\Theta(x):=+\infty$ if $x>0$.\\
For a generic inhomogeneous metric, the set of linear wall forms $w^A(\beta)$ always includes \textit{curvature} (or \textit{gravitational}) walls ($ w^g_{(abc)}(\beta)$) and \textit{symmetry} (or \textit{centrifugal}) walls ($w^S_{ab}$). They are explicitly defined by
\be\label{2.12}
 w^g_{(abc)}(\beta)\equiv\beta^a-\beta^b-\beta^c+\sum_e\beta^e \ \ (b\neq c),
\ee
\be\label{2.13}
w^S_{(ab)}\equiv\beta^b-\beta^a \ \ (a<b).
\ee
Beware of the fact that the indices with parentheses appearing on the left-hand sides (l.h.s.) of these definitions should be considered as \textit{labels} (like the label $A$ in Eq. (\ref{2.11})), and not as $\beta$-space tensor indices. E.g. in the linear form of the $\beta$'s $  w^g_{(abc)}(\beta)=\sum_e w^g_{(abc)e}\beta^e$ only the summed-over index $e$ must be considered as a tensor index. In addition, note that the index $e$ on $ w^g_{(abc)e}$ is \textit{covariant}, while the index $e$ on $\beta^e$ is \textit{contravariant}. This means that, when computing the Lorentzian scalar product between two wall forms, $w^A(\beta)=\sum_aw^A_a\beta^a$ and $w^B(\beta)=\sum_aw^B_a\beta^a$, one should use the \textit{contravariant} $\beta$-space metric $G^{ab}$:
\be\label{2.14}
w^A\cdot w^B\equiv\sum_{a,b}G^{ab}w^A_aw^B_b.
\ee
Among all possible walls entering the Hamiltonian, only the subset of ``leading'' walls (those not ``hidden behind'' another wall) should be retained to define the $\beta$-space billiard defining the asymptotic BKL-like dynamics. Indeed, the billiard chamber is defined as the intersection of the \textit{positive sides} of the set of wall hyperplanes, i.e. the domain where all the linear forms $w^A(\beta)$ are \textit{positive}. For instance, in the case of three spatial dimensions, there are $6$ gravitational walls, and $3$ symmetry ones. However, among these, some walls are ``subleading'' in that they are always behind some other walls. E.g. the symmetry wall $w^S_{(13)}=\beta^3-\beta^1$ can be identically expressed as $w^S_{(13)}=\beta^3-\beta^2+\beta^2-\beta^1\equiv w^S_{(12)}(\beta)+w^S_{(23)}(\beta)$. Therefore the inequality $w^S_{(13)}(\beta)>0$ is a consequence of the two inequalities $w^S_{(12)}(\beta)>0$ and $w^S_{(23)}(\beta)>0$, meaning that the wall $w^S_{(13)}(\beta)$ is behind the two walls $w^S_{(12)}(\beta)$ and $w^S_{(23)}(\beta)$ and is therefore subleading. Similarly, one finds that among the gravitational walls $ w^g_{(abc)}$ the ones where the first label is equal either to $ b$ or $c$ ($b\neq c$), i.e. $\mu_c(\beta)=-\beta^c+\sum_e\beta^e$ are always subleading. In $d=3$, this leaves only $3$ a priori leading gravitational walls
\be\label{2.16}
 w^g_{(123)}(\beta)=2\beta^1,\ \  w^g_{(231)}(\beta)=2\beta^2,\ \   w^g_{(312)}(\beta)=2\beta^3. 
\ee 
Moreover, the same reasoning which allows one to conclude that the symmetry wall $w^S_{(13)}(\beta)$ is `behind' the two other symmetry walls $w^S_{(12)}(\beta)$ and $w^S_{(23)}(\beta)$ shows that the gravitational walls $ w^g_{(231)}(\beta)$ and $ w^g_{(312)}(\beta)$ are both `behind' the combination of the walls $\{w^S_{(12)}, w^S_{(23)},  w^g_{(123)}\}$. Therefore, for a \textit{generic inhomogeneous} metric the asymptotic billiard chamber in $\beta$ space is defined by the following three independent inequalities
\begin{subequations}\label{2.17}
\begin{align}
&w^S_{(12)}(\beta)\equiv \beta^2-\beta^1\ge0,\\
&w^S_{(23)}(\beta)\equiv \beta^3-\beta^2\ge0,\\
& w^g_{(123)}(\beta)\equiv 2\beta^1\ge0.
\end{align}
\end{subequations}
Note that the boundary of this billiard chamber is made of two (portions of) symmetry walls, and one (portion of) gravitational wall. The occurrence of the symmetry walls here comes from terms in the Hamiltonian associated with the kinetic energy of the off-diagonal components, $\mathcal{N}^a_i$, in the Iwasawa decomposition (\ref{2.4}) of the metric. Note that an alternative way of seeing the `constraining' effect of the off-diagonal components of the metric consists of using, instead of an Iwasawa decomposition, a \textit{Gauss decomposition} of the spatial metric: $g_{ij}=\sum_a e^{-2\beta^a}R^a_{\ \ i}R^a_{\ \ j}$, with $R^a_{\ \ i}$ being a rotation matrix, parametrized by three Euler angles. Such a Gauss decomposition was introduced by Belinski, Khalatnikhov and Ryan in \cite{Ryan1972}. As shown there it entails the presence of \textit{centrifugal walls} which are simply related to the (Iwasawa) exponential symmetry walls $\exp(-2w^S_{(ab)})$ via $V_{(ab)}^{\rm centrif}\propto [\sinh (w^S_{(ab)})]^{-2}$.\\
In the special case of a \textit{homogeneous vacuum} model of Bianchi-type IX, it is possible to restrict oneself (without loss of generality) to considering a metric $g_{ij}\theta^i\theta^j$ which is \textit{diagonal} in a co-frame $\theta^i=e^i_{\ \ m}dx^m$ of left-invariant one-forms. In that case, the kinetic energy terms associated to the off-diagonal components of $g_{ij}$ vanish, so that the symmetry walls do not appear. As a consequence, the billiard chamber for the special \textit{diagonal} Bianchi-IX case is defined by the three leading gravitational walls (\ref{2.16}), i.e. by the three inequalities
\begin{subequations}\label{2.18}
\begin{align} 
& w^g_{(123)}(\beta)=2\beta^1\ge0,\\
& w^g_{(231)}(\beta)=2\beta^2\ge0,\\
& w^g_{(312)}(\beta)=2\beta^3\ge0.
\end{align}
\end{subequations}
Note that these three billiard walls correspond to the leading terms that appear on the right-hand side (r.h.s.) of the BKL $a$, $b$, $c$ system (\ref{abc}) when using the exponential parametrization $a=e^{-\alpha}$, $b=e^{-\beta}$, $c=e^{-\gamma}$. Indeed, in terms of these variables, the $a$, $b$, $c$ system (\ref{abc}) reads
\begin{subequations}\label{eeq}
\begin{align}
&\frac{d^2\alpha}{d\tau^2}=\frac{1}{2}\left[ e^{-4\alpha}-e^{-4\beta}-e^{-4\gamma}+2e^{-2(\beta+\gamma)}\right],\\
&\frac{d^2\beta}{d\tau^2}=\frac{1}{2}\left[ e^{-4\beta}-e^{-4\gamma}-e^{-4\alpha}+2e^{-2(\alpha+\gamma)}\right],\\
&\frac{d^2\gamma}{d\tau^2}=\frac{1}{2}\left[ e^{-4\gamma}-e^{-4\alpha}-e^{-4\beta}+2e^{-2(\beta+\alpha)}\right].
\end{align}
\end{subequations} 
the terms $\propto e^{-4\alpha}, e^{-4\beta}, e^{-4\gamma}$ exactly correspond to the three wall forms (\ref{2.18}), i.e. $ e^{-4\beta^1}, e^{-4\beta^2}, e^{-4\beta^3}$. They appear even more clearly in the  Hamiltonian constraint of the $a$, $b$, $c$ system which has the form $H\equiv\tfrac{1}{2}G_{ab}\dot{\beta}^a\dot{\beta}^b+V(\beta)=0$, i.e., explicitly, 
\be\label{hcon}
H\equiv-\frac{d\alpha}{d\tau}\frac{d\beta}{d\tau}-\frac{d\alpha}{d\tau}\frac{d\gamma}{d\tau}-\frac{d\beta}{d\tau}\frac{d\gamma}{d\tau}+\tfrac{1}{4}\left[e^{-4\alpha}+e^{-4\beta}+e^{-4\gamma} -2e^{-2\alpha}e^{-2\beta}-2e^{-2\alpha}e^{-2\gamma}-2e^{-2\beta}e^{-2\gamma}\right]=0.
\ee
\vspace{0.1cm} 

It is easily seen that the three symmetry walls $w^S_{(12)}(\beta)$, $w^S_{(23)}(\beta)$, $w^S_{(31)}(\beta)$ partition the $\beta$ space chamber (\ref{2.18}) into six sub-chambers which are all congruent (with respect to the Lorentzian geometry of $\beta$-space) to the billiard chamber (\ref{2.17}) corresponding to a generic inhomogeneous (and generically non-diagonal) metric. In view of this, and for brevity, we shall refer, in the following, to the diagonal Bianchi-IX chamber (\ref{2.18}) as being the \textit{big billiard} by contrast to the \textit{small billiard} (\ref{2.17}) associated to a generic inhomogeneous metric. Note that, in the introduction, we referred to the big billiard either as `the full $a$, $b$, $c$ billiard' or as the `unquotiented $a$, $b$, $c$ billiard'. In addition, note that the (six times smaller) billiard table of the small billiard is obtained by quotienting the big billiard table by the permutation group of $\{\beta^1, \beta^2, \beta^3\}$.
\subsection{Hyperbolic billiards}\label{hyperbolicbilliards}
So far we have recalled how the use of the time gauge (\ref{2.2}), (\ref{2.3}) leads to a description of the asymptotic dynamics of the metric, near a space-like singularity, in terms of a billiard motion in an auxiliary $d$-dimensional Lorentzian space parametrized by the `logarithmic scale factors' $\beta^a$. [Note that, in the diagonal Bianchi-IX case the BKL scale factors $a$, $b$, $c$ are related to the $\beta$'s via $a=\exp (-\beta^1)$, $b=\exp (-\beta^2)$, $c=\exp (-\beta^3)$]. A convenient reformulation of this Lorentzian billiard consists of decomposing the motion in $\beta$-space into \textit{radial}, $\rho$, and \textit{angular}, $\gamma^a$, parts. Here, the terms ``radial'' and ``angular'' refer to Lorentzian analogs of the usual Euclidean decomposition of a position vector $\mathbf{x}$ as $\mathbf{x}=r\mathbf{n}$, with $r\equiv(\mathbf{x}^2)^\frac{1}{2}$ and $\mathbf{n}$ being a unit vector. Namely, one decomposes the (time-like) `position vector' $\beta^a$ in Lorentzian space as
\be\label{2.19}
\beta^a\equiv\rho\gamma^a;\ \ \rho\equiv\left( -G_{ab}\beta^a\beta^b\right)^{1/2}.
\ee 
When using such a decomposition, it is convenient to redefine the time gauge, and to replace the condition (\ref{2.2}) by \cite{Misner:1969hg}, \cite{Damour:2002et}
\be\label{2.20}
N=\rho^2\sqrt{g}.
\ee
One then finds that the radial motion asymptotically decouples from the angular one and leads to a uniform motion of the logarithmic variable $\lambda\equiv\ln\rho$ w.r.t. the coordinate time, say $T$, associated to the new gauge (\ref{2.20}), i.e. 
\be\label{2.21}
dT=-\frac{d\tau}{\rho^2}=-\frac{dt}{\rho^2\sqrt{g}}.
\ee
\begin{figure*}[htbp]
\begin{center}
\includegraphics[width=0.4\textwidth]{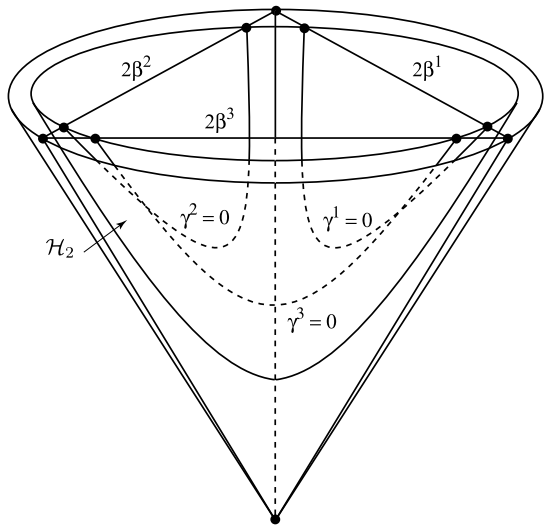}
\caption{\label{fig1} The hyperbolic billiard on the unit hyperboloid $\mathcal{H}_2$. The `big billiard' chamber defined by the three gravitational walls is sketched.}
\end{center}
\end{figure*}
As for the ``angular motion'' $\gamma^a(T)$ on the (future) unit $(d-1)$-dimensional hyperboloid, say
\be\label{2.22}
\mathcal{H}_{d-1}:\ \ G_{ab}\gamma^a\gamma^b=-1,
\ee
it is found to be asymptotically described by a \textit{hyperbolic billiard}, i.e. by a succession of constant-velocity (in $T$ time, Eq. (\ref{2.21})) \textit{geodesic} flights on the unit hyperboloid $\mathcal{H}_{d-1}$ (\ref{2.22}), interrupted by collisions (and reflections) on hyperbolic walls located on some geodetic hyperplane in $\gamma$-space. The $\gamma$-space chamber within which this angular billiard dynamics takes place is simply the projection (seen from the origin) of the corresponding $\beta$-space billiard chamber onto the unit hyperboloid $\mathcal{H}_{d-1}$, Eq. (\ref{2.22}). For instance, in the case of $D=4$ vacuum Einstein gravity that we shall focus on in this paper, we end up with a non-compact, but finite volume billiard chamber on a $2$-dimensional hyperboloid $\mathcal{H}_2$ bounded by three geodetic lines. In the case of a generic, inhomogeneous metric, the billiard chamber is defined by the $\mathcal{H}_2$ projection of the inequalities (\ref{2.17}), i.e.
\be\label{2.23}
\rm{`small\ \  billiard'}:\ \  \gamma^2-\gamma^1\ge0;\ \ \gamma^3-\gamma^2\ge0;\ \ 2\gamma^1\ge0. 
\ee
On the other hand, in the special diagonal Bianchi-IX case, the billiard chamber is the $\mathcal{H}_2$ projection of (\ref{2.18}), i.e.
\be\label{2.24}
\rm{`big\ \  billiard'}:\ \ 2\gamma^1\ge0;\ \ 2\gamma^2\ge0;\ \ 2\gamma^3\ge0.
\ee
The link between the projected `big billiard' chamber on $\mathcal{H}_2$ and the corresponding big Lorentzian billiard in $\beta$ space is sketched in Fig. \ref{fig1}.\\
We have defined here the billiards on $\mathcal{H}_2$ in terms of the three components of the unit Lorentzian vector $\gamma^a$ (satisfying $G_{ab}\gamma^a\gamma^b=-2\left(\gamma^1\gamma^2+\gamma^2\gamma^3+\gamma^3\gamma^1\right)=-1$, see Eq. (\ref{2.7}) with $d=3$). This is a hyperbolic analog of defining a billiard chamber on the unit sphere $S_2$ by writing three linear inequalities $w_A(\mathbf{n})\ge0$ in the three components $n^1$, $n^2$, $n^3$ of a unit Euclidean vector (satisfying $(n^1)^2+(n^2)^2+(n^3)^2=1$). For many purposes, the knowledge of the linear forms $w^S_A(\gamma)$ defining the billiard walls (e.g. $w^S_{(12)}=\gamma^2-\gamma^1$ for the first wall of the ``small billiard'' (\ref{2.23})) is all that is needed to compute most quantities of physical interest. For instance, the (hyperbolic-geometry) angle $\theta_{AB}$ between the walls $w_A(\gamma)=w_{Aa}\gamma^a$ and $w_B(\gamma)=w_{Ba}\gamma^a$ is given by
\be\label{2.25}
\cos\theta_{AB}=\frac{w_A\cdot w_B}{\sqrt{w_A\cdot w_A}\sqrt{w_B\cdot w_B}},
\ee
where $w_A\cdot w_B\equiv G^{ab}w_{Aa}w_{Bb}$, Eq. (\ref{2.14}), $G^{ab}$ denoting the contravariant metric, Eq. (\ref{2.15}). E.g. one easily checks that the three angles on the small billiard (\ref{2.23}) are $0$, $\frac{\pi}{3}$ and $\frac{\pi}{2}$, while the three angles between the three sides of the big billiard (\ref{2.24}) are $0$, $0$, $0$. In addition, the law of reflection of the $\gamma$-space $T$-time velocity vector, say $V^a=d\gamma^a/dT$, on a certain wall $w_A(\gamma)$ is simply given (at the location of the collision, i.e. when $w_A(\gamma)=0$) by 
\be\label{2.26}
V^{'a}=V^a-2\frac{w_A(V)w^a_A}{w_A\cdot w_A},
\ee 
where $w_A(V)\equiv w_{Aa}V^a$, and where $w^a_A\equiv G^{ab}w_{Ab}$ is the \textit{contravariant} vector associated to the covariant components $w_{Aa}$ entering the wall form $w_A(\gamma)=w_{Aa}\gamma^a$. Eq. (\ref{2.26}) relates $\beta$-space vectors that are (at the location of the collision) all tangent to $\mathcal{H}_{d-1}$. It is obtained by projecting the corresponding $\beta$-space, $\tau$-time collision law \cite{Damour:2000hv}
\be\label{2.27}
v^{'a}=v^a-2\frac{w_A(v)w_A^a}{w_A\cdot w_A},
\ee
which, contrary to (\ref{2.26}), involves time-independent vectors.\\
\vspace{0.1cm}

However, for some purposes, it is convenient to use an explicit parametrization of the billiard dynamics on $\mathcal{H}_{d-1}\equiv\mathcal{H}_{n}$ by means of $n\equiv d-1$ intrinsic coordinates. This can be done in several ways. Let us first emphasize that the unit hyperboloid $\mathcal{H}_{n}$ is a model of the $n$-dimensional Lobachevsky space (i.e. it is diffeomorphic to $\mathbb{R}^n$ and has a constant sectional curvature $-1$). As shown long ago by Beltrami (see, e.g., the textbook \cite{yellow} and the review paper \cite{Balazs:1986uj}), $\mathcal{H}_n$ admits several useful representations in an $n$-dimensional Euclidean space. Among these, the \textit{conformal} representations are useful because, as they preserve angles, they allow one to express the reflection law (\ref{2.26}) as a usual (locally Euclidean) reflection law of the (local) velocity vector on the wall $w_A$. The two main conformal representations are:\\
(i) the `ball model', say $\mathcal{B}_n$, which represents $\mathcal{H}_n$ as a unit ball $\mathbf{x}^2\le1$ in $n$-dimensional Euclidean space $\mathbf{x}\in\mathbb{R}^n$, with metric $ds^2=4d\mathbf{x}^2/(1-\mathbf{x}^2)^2$; and\\
(ii) the `upper half-space model', or `Poincar\'e model', say $\mathcal{P}_n$, which represents $\mathcal{H}_n$ as the half-space $v\ge0$, $\mathbf{u}\equiv (u^1, ..., u^{n-1})\in\mathbb{R}^{n-1}$, with metric
\be\label{2.28}
ds^2=\frac{d\mathbf{u}^2+dv^2}{v^2}.
\ee 
The `ball' conformal representation can be geometrically realized within the $(n+1)$-dimensional Lorentzian $\beta$-space by stereographically projecting (from the `South Pole' $\gamma_S$, i.e. a center of projection located on the \textit{past} unit hyperboloid) the \textit{future} unit hyperboloid (\ref{2.22}) onto a ($n$-dimensional) hyperplane passing through the origin in $\beta$-space.\\
The Poincar\'e model can be obtained by a suitable geometric inversion of the ball model. In both models, the geodesics of $\mathcal{H}_n$ become Euclidean circles orthogonal to the boundary (the boundary being a unit ($n-1$)-dimensional sphere for $\mathcal{B}_n$ and the plane $v=0$ for $\mathcal{P}_n$), while walls, i.e. geodetic hyperplanes, become $(n-1)$-dimensional spheres orthogonal to the boundary. Note that the `boundary' corresponds to the `absolute' of $\mathcal{H}_n$, i.e. its domain at infinity (which corresponds to the future null cone, $G_{ab}\beta^a\beta^b=0$ when replacing the $\gamma^a$'s with projective $\beta^a$ coordinates). In addition to these Euclidean space representations (which naturally define $n$ coordinates, ($x^1, ..., x^n$), or ($u^1, ..., u^{n-1}, v$), on $\mathcal{H}_n$), it might also be useful to coordinatize $\mathcal{H}_n$ by means of the hyperbolic analog of the polar coordinates on a sphere. E.g. in the case $n=2$, one can represent the metric on $\mathcal{H}_2$ as $ds^2=d\theta^2+\sinh^2\theta d\phi^2$.
\section{Conformal representation of the D=4 cosmological billiards}\label{conformalrepresentation}
\begin{figure*}[htbp]
\begin{center}
\includegraphics[width=0.4\textwidth]{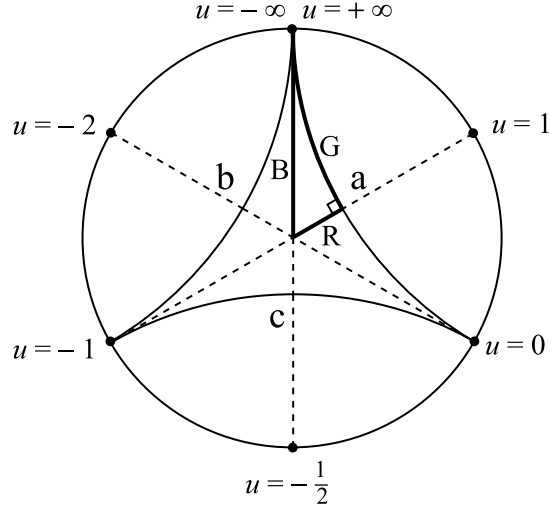}
\caption{\label{fig2} The disk model $\mathcal{B}_2$ of the hyperbolic billiard. Both the big billiard (with walls $a$, $b$, $c$) and the small billiard (with walls $G$, $B$, $R$) are sketched. The $6$ fundamental Kasner intervals are indicated on the boundary of the disk, which is identified with the Kasner circle.}
\end{center}
\end{figure*}
\begin{figure*}[htbp]
\begin{center}
\includegraphics[width=0.4\textwidth]{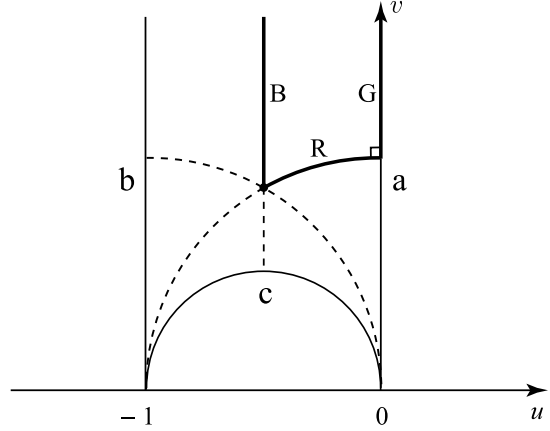}
\caption{\label{fig3} The Poincar\'e model $\mathcal{P}_2$ of the hyperbolic billiard. Both the big billiard and the small billiards are sketched.}
\end{center}
\end{figure*}

In space-time dimensions $D=d+1=4$, the walls reduce to (geodesic) lines on  the $2$-dimensional Lobachevsky plane $\mathcal{H}_2$.
As recalled in the previous section, one can consider two different pure gravity billiards, in $D=d+1=4$:
(i) the big billiard (\ref{2.24}) (defined by an `ideal triangle` on $\mathcal{H}_2$, i.e. a triangle whose three sides meet at infinity with pairwise vanishing angles) or (ii) the small billiard (\ref{2.23}) (which has angles $0$, $\frac{\pi}{3}$ and $\frac{\pi}{2}$ and only one vertex at infinity). The most symmetric representation of the big billiard (which manifestly respects the symmetry group, of order $3!=6$, of the inequalities (\ref{2.24})) is obtained by using a disk model centered at the point $\gamma^1=\gamma^2=\gamma^3$. See Fig. \ref{fig2} which also exhibits the small billiard (\ref{2.23}). By the Gauss-Bonnet theorem, ($A+B+C-\pi=\int KdS$) the (hyperbolic) area of the billiard is equal to $\pi$, while that of the small billiard is $\pi/6$ (consistently with the fact that there are six congruent copies of the small billiard within the big one). By using a Euclidean geometric inversion with respect to the `cusp' (i.e. the vertex on the absolute) of the small billiard, one obtains a Poincar\'e model of the billiard in which that cusp is represented by the point at infinity of the upper half plane ($v=\infty$). In this representation (see Fig. \ref{fig3}) the geodesics $\gamma^1=0$, $\gamma^2=0$ and $\gamma^2-\gamma^1=0$ are all represented as vertical straight lines.\\

As the explicit form of the transformations relating the original gravitational variables $\beta^a$ successively to $\gamma^a$, and to its images in the ball and Poincar\'e models, tend to be unwieldy and not very illuminating, let us sketch how the form of the final results can be obtained essentially without calculations, by using various geometric considerations. In any Poincar\'e representation (say of coordinates $u, v$) a general wall $0=w_A(\gamma)\equiv w_{A1}\gamma^1+w_{A2}\gamma^2+w_{A3}\gamma^3$ must be (projectively) equivalent to the equation of a circle. Therefore each $\gamma^a$ must be of the form $\gamma^a(u,v)=\lambda(u,v)\delta^a(u,v)$, with
\be\label{3.1}
\delta^a(u,v)=A^a(u^2+v^2)+B^au+C^av+D^a,
\ee 
for some constants $A^a$, $B^a$, $C^a$, $D^a$. Moreover, if we choose to put the cusp of the small billiard at infinity in the Poincar\'e plane, $\delta^1(u,v)=0$ and $\delta^2(u,v)=0$ must be the equations of two vertical lines (see Fig. \ref{fig3}). Therefore, $\delta^1$ and $\delta^2$ must simply be of the form $\delta^1=B^1u+D^1$, $\delta^2=B^2u+D^2$. By contrast, $\delta^3(u,v)=0$ must be the equation of a circle centered on the $v=0$ axis, i.e. $\delta^3(u,v)=A^3(u^2+v^2)+B^3u+D^3$, and cutting the $v$ axis at the two points $\delta^1(u)=0$ and $\delta^2(u)=0$ (see Fig. \ref{fig3}). In addition, the infinity of $\mathcal{H}_2$ projectively corresponds to the light cone in $\beta$-space so that the infinity of the Poincar\'e model (i.e. $v=0$) must correspond to $G_{ab}\delta^a\delta^b=0$. All these conditions fix the expressions of $\delta^a(u,v)$ modulo an overall factor and modulo the parabolic subgroup of the symmetry group ($SL(2,\mathbb{R})$) of $\mathcal{H}_2$ leaving fixed the cusp: i.e. $u'=au+b$, $v'=au$. It was shown by Kirillov and Montani \cite{Kirillov:1996rd} that a particular choice of $a$ and $b$ leads to expressions for $\delta^a(u,v)$ which are nicely compatible with the $u$-parametrization of Kasner parameters which has been used by BKL \cite{BLK1971}, \cite{Lifshitz:1963ps}. This choice leads to the expressions
\begin{subequations}\label{3.2}
\begin{align}
&\delta^1(u,v)=-u,\\
&\delta^2(u,v)=u+1,\\
&\delta^3(u,v)=u(u+1)+v^2,
\end{align}
\end{subequations}
which entail [in view of the quadratic constraint (\ref{2.22})] $\gamma^a(u,v)=\delta^a(u,v)/(\sqrt{2}\:v)$, i.e. explicitly 
\begin{subequations}\label{3.3}
\begin{align}
&\gamma^1(u,v)=-\frac{u}{\sqrt{2}\:v},\\
&\gamma^2(u,v)=\frac{u+1}{\sqrt{2}\:v},\\
&\gamma^3(u,v)=\frac{u(u+1)+v^2}{\sqrt{2}\:v}.
\end{align}
\end{subequations}
As shown in Fig. \ref{fig3}, in this normalization the gravitational wall $\gamma^1=0$ (i.e. the $a$ wall) which is common to the small and the big billiard is located at $u=0$; the symmetry wall $\gamma^1=\gamma^2$ of the small billiard ($B$ wall) is located at $u=-1/2$; and the other vertical gravitational wall of the big billiard, $\gamma^2=0$ ($b$ wall), is at $u=-1$. On the other hand, the remaining walls (either $\gamma^3=0$ or $\gamma^3-\gamma^2=0$) are circles orthogonal to the $v$ axis.\\
\vspace{0.1cm}

An essential role will be played in the following by the images in the Poincar\'e model of the `Kasner epochs' of the cosmological billiard, i.e. the free flights between two successive wall collisions. In $\beta$-space, these free-flight segments are described by uniform motion (in $\tau$-time) and in straight line, see Eq. (\ref{2.7}). The ($\beta$-space) `velocity' of these free flights is described by the Lorentzian vector $v^a$, submitted to the constraint of being null, Eq. (\ref{2.8}). The `Kasner parameters' $p_a$ of each Kasner epoch are (projectively) related to the components of the $\beta$-space velocity $v^a$ by the relation (\ref{2.9}).\\
\begin{figure*}[htbp]
\begin{center}
\includegraphics[width=0.4\textwidth]{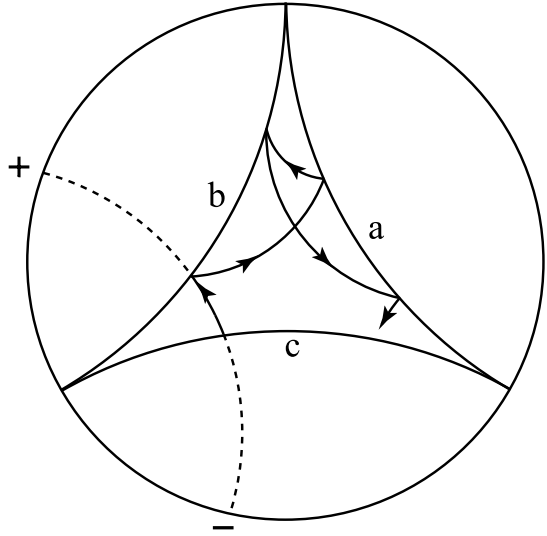}
\caption{\label{fig4} Kasner epochs of the big billiard in the disk model $\mathcal{B}_2$ of the hyperbolic billiard.}
\end{center}
\end{figure*}
\begin{figure*}[htbp]
\begin{center}
\includegraphics[width=0.4\textwidth]{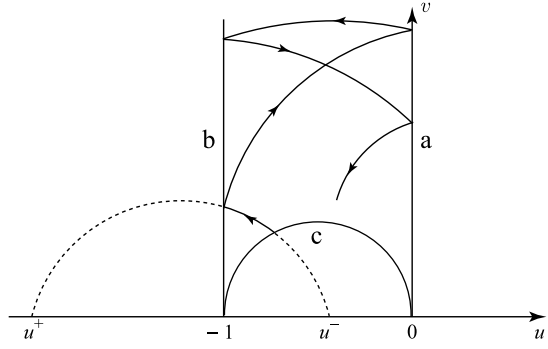}
\caption{\label{fig5} Kasner epochs of the big billiard in the Poincar\'e model $\mathcal{P}_2$ of the hyperbolic billiard.}
\end{center}
\end{figure*}
The disk-model ($\mathcal{B}_2$) projection of consecutive $\beta$-space free flights is made of geodesic segments in $\mathcal{B}_2$, i.e. arcs of circles orthogonal to the boundary circle of $\mathcal{B}_2$. See Fig. \ref{fig4}, which represents the `reflection' of these geodesics on the three gravitational walls of the big billiard.\\
\vspace{0.1cm}

When represented in the Poincar\'e model, instead of the disk model, each $\beta$-space straight-line segment $\beta(\tau)=\beta_0+v^a\tau$ (with $\tau_1<\tau<\tau_2$) gets mapped into a geodesic segment of the Poincar\'e plane $\mathcal{P}_2$, i.e. a segment of a circle orthogonal to the horizontal axis $v=0$, say 
\begin{subequations}\label{3.4}
\begin{align}
&u=\frac{1}{2}(u^++u^-)-\frac{1}{2}(u^+-u^-)\cos\theta,\\
&v=\frac{1}{2}\mid u^+-u^-\mid \sin\theta,
\end{align}
\end{subequations}
with $0<\theta_1\le\theta\le\theta_2<\pi$ Here $\theta_1$ (which corresponds to $\tau_1$ in the $\beta$-space ``upstairs'') corresponds to the last collision, and $\theta_2$ ($\leftrightarrow\tau_2$) to the next one. See Fig \ref{fig5}, which represents the big billiard in the Poincar\'e model $\mathcal{P}_2$. Here, we are considering \textit{oriented} circles whose formal extension to the full interval $0<\theta<\pi$ would start, when $\theta=0$, at the location $u=u^-$ on the $v$ axis, and end, when $\theta=\pi$, at the location $u=u^+$ on the $v$ axis. [Note that the radius of the circle is $\frac{1}{2}\mid u^+-u^-\mid $, while its center is located at $u_c=\frac{1}{2}(u^++u^-)$, $v_c=0$.]\\
\vspace{0.1cm}

For a given billiard table, the oriented pair of real parameters ($u^+, u^-$) (exemplified in Fig. \ref{fig5}) uniquely determines the (oriented) geodesic segment corresponding to some Kasner epoch. More precisely, it is easy to see geometrically that the `end' parameter $u^+$ uniquely parameterizes the family of $\beta$-space straight-line segments $\beta^a=\beta^a_0+v^a\tau$ (without considering their $\tau$ parametrization) that share a common (formal) asymptotic direction $\propto v^a$. In other words, $u^+$ \textit{uniquely parameterizes the three Kasner exponents of the considered Kasner epoch}. The precise technical link between $u^+\in\mathbb{R}$ and the three $p_a$'s is obtained by relating successively $p_a$: (i) to $v^a$ (see Eq. (\ref{2.9})); (ii) to $\beta^a(\tau)$ in the formal $\tau\rightarrow+\infty$ limit (via Eq. (\ref{2.7})); (iii) to $\gamma^a(\tau)=\beta^a/\rho$ in the same limit, and thereby (iv) to $\gamma^a(u,v)$ in the corresponding (formal) limit $u\rightarrow u^+$, $v\rightarrow0$ corresponding to the future endpoint of the circle in the Poincar\'e model representing the formal extension of the Kasner epoch. In other words,
\be\label{3.5}
p_a=\frac{v^a}{\sum_bv^b}=\lim_{\tau\rightarrow+\infty}\frac{\beta_0^a+v^a\tau}{\sum_b(\beta_0^b+v^b\tau)}=\lim_{\tau\rightarrow+\infty}\frac{\gamma^a(\tau)}{\sum_b\gamma^b(\tau)}=\lim_{v\rightarrow0, u\rightarrow u^+}\frac{\gamma^a(u,v)}{\sum_b\gamma^b(u,v)}.
\ee
Using the Poincar\'e model expressions (\ref{3.3}), this leads to the result
\be\label{3.6}
p_a=p^{BKL}_a(u^+),
\ee
where the three functions $p^{BKL}_a(u)$ are the well-known $u$-parametrization of Kasner exponents introduced by Belinski, Khalatnikhov and Lifshitz, namely
\begin{subequations}\label{3.7}
\begin{align}
&p_1^{BKL}(u)\equiv-\frac{u}{u^2+u+1},\\
&p_2^{BKL}(u)\equiv\frac{u+1}{u^2+u+1},\\
&p_3^{BKL}(u)\equiv\frac{u(u+1)}{u^2+u+1}.
\end{align}
\end{subequations}
\vspace{0.1cm}

In other words, as was found in Ref. \cite{Kirillov:1996rd}, the BKL $u$-parameter can be interpreted as the location on the real axis of the future end point of the circle representing the considered Kasner flight in a suitably defined Poincar\'e model of $\mathcal{H}_2$.\\
\vspace{0.1cm}

On the other hand, if we now consider the formal extension of the considered Kasner flight towards the `past' (in $\tau$-time), there occurs a subtlety: the past end point $u^-$ of the corresponding Poincar\'e circle \textit{does not} correspond (as one might naively expect) to the formal $\tau\rightarrow-\infty$ limit of $\beta^a(\tau)=\beta^a_0+v^a\tau$, but only to the finite past limit $\tau\rightarrow\tau_0$ such that $\beta^a_{\rm Kasner}(\tau)=\beta^a_0+v^a\tau$ intersects the $\beta$-space (future) light cone: $G_{ab}\beta^a_{\rm Kasner}(\tau_0)\beta^b_{\rm Kasner}(\tau_0)=0$, i.e. $\rho(\tau_0)=0$. As this limit again corresponds to a point at infinity for $\gamma^a_{\rm Kasner}(\tau)\equiv\beta^a_{\rm Kasner}(\tau)/\rho(\tau)\in\mathcal{H}_2$, one finds, by changing the limits in Eq. (\ref{3.5}) that the past end point $u=u^-$, $v=0$ on the (oriented) Poincar\'e circle parametrizes $\beta^a_{\rm Kasner}(\tau_0)$ in a projective manner:
\be\label{3.8}
\frac{\beta^a_{\rm Kasner}(\tau_0)}{\sum_b\beta^b_{\rm Kasner}(\tau_0)}=p_a^{BKL}(u^-).
\ee 
As we shall need them below, let us note at this stage some features of the BKL $u$-parametrization of Kasner exponents (\ref{3.7}). First, let us emphasize that the manifold of Kasner parameters, $p_a$, restricted by the two constraints (\ref{2.10}), is topologically a ($d-2$)-dimensional sphere [indeed Eqs. (\ref{2.10})) represent the intersection of a ($d-1$)-dimensional sphere by an hyperplane]. In the case considered here where $d=3$, this means that the three Kasner parameters $p_1$, $p_2$, $p_3$ run over a topological circle. In fact, this `Kasner circle' can be identified, in the disk representation of $\mathcal{H}_2$, with the boundary of the unit disk, i.e. with the absolute of $\mathcal{H}_2$. See Fig. \ref{fig2}. In particular, the $u$-parameter in Eqs. (\ref{3.7}) should be considered as running on the extended real line $\bar{\mathbb{R}}=\mathbb{R}\cup\{\infty\}$. The extended line $u\in\bar{\mathbb{R}}$ is then naturally divided into the six permutations of the three letters ($p_1$, $p_2$, $p_3$) which constitute the symmetry group of the two Kasner constraints (\ref{2.10}). As the latter symmetry group is generated by reflections in the symmetry walls, it is natural to divide the $u$ line, i.e. the boundary of $\mathcal{H}_2$, by means of these symmetry walls: $\gamma^2-\gamma^1=0$, $\gamma^3-\gamma^2=0$, $\gamma^3-\gamma^1=0$, that is, by considering the solutions of the equations $p_1^{BKL}(u)=p_2^{BKL}(u)$, or $p_2^{BKL}(u)=p_3^{BKL}(u)$, or $p_1^{BKL}(u)=p_3^{BKL}(u)$. This leads to dividing the $u$-line into the intervals:
\be\label{3.9}
(-\infty, -2);\ \ (-2,-1);\ \ (-1, -\frac{1}{2}); \ \ (-\frac{1}{2}, 0);\ \ (0, 1);\ \ (1, +\infty). 
\ee
If, following BKL, we consider the interval $1<u<+\infty$ as a `fundamental interval', over which the Kasner exponents are ordered as $p_1^{BKL}(u)<p_2^{BKL}(u)<p_3^{BKL}(u)$, the other five possible orderings of $p_1$, $p_2$, $p_3$ will be obtained by applying to the variable $u$ a transformation implementing a composition of geometric reflections accross some symmetry walls. It is well known that a geometric reflection acts on the complex variable $z\equiv u+iv$ of the Poincar\'e plane according to
\be\label{3.10}
z'=-\frac{a\bar{z}+b}{c\bar{z}+d},
\ee
with $a, b, c, d \in \mathbb{R}$ and $ad-bc=+1$. When acting on the boundary of the Poincar\'e model ($v=0$), and composing several reflections, this leads to transformations of the form
\be\label{3.11}
u'=\pm\frac{au+b}{cu+d};\ \ ad-bc=1.
\ee
The explicit expression of the five transformations $u'=f_i(u)$, $i=1, ..., 5$ of the form (\ref{3.10}) that map the first five intervals (\ref{3.9}) into the last ('fundamental') one are given in Table \ref{table1}. In the following, we shall refer to these as ``Kasner transformations''. Note that each boundary Kasner transformation $u'=f_i(u)$ uniquely determines the way the corresponding combination of reflections acts on the interior of the Poincar\'e model: if the sign in Eq. (\ref{3.11}) is $+$ (even isometry) it is $z'=(az+b)/(cz+d)$, while if the sign in Eq. (\ref{3.11}) is $-$ (odd isometry), it is given by Eq. (\ref{3.10}). If one adds the identity transformation, say $k_0$ ($k_0(u)\equiv u$), the six transformations $\{k_0, k_1, ..., k_5\}$ constitute (under their composition) a realization of the permutation group of three objects (say, the three walls $a$, $b$, $c$).

\begin{table}
\begin{center}
    \begin{tabular}{ | l | l | l | l | l | l | }
    \hline
    $k_1$ & $0<u<1$ & $u'=1/u$ & $p_1(u')=p_1(u)$ & $p_2(u')=p_3(u)$ & $p_3(u')=p_2(u)$\\ \hline
    $k_2$ & $-1/2<u<0$ & $u'=-(1+u)/u$ & $p_1(u')=p_3(u)$ & $p_2(u')=p_1(u)$ & $p_3(u')=p_2(u)$ \\ \hline
    $k_3$ & $-1<u<-1/2$ & $u'=-u/(u+1)$ & $p_1(u')=p_3(u)$ & $p_2(u')=p_2(u)$ & $p_3(u')=p_1(u)$ \\ \hline
    $k_4$ & $-2<u<-1$ & $u'=-1/(u+1)$ & $p_1(u')=p_2(u)$ & $p_2(u')=p_3(u)$ & $p_3(u')=p_1(u)$\\ \hline
    $k_5$ & $-\infty<u<-2$ & $u'=-u-1$ & $p_1(u')=p_2(u)$ & $p_2(u')=p_1(u)$ & $p_3(u')=p_3(u)$\\ \hline
    \end{tabular}
\end{center}
\caption{\label{table1} Kasner transformation, $k_1, ..., k_5$ mapping the indicated intervals of the $u$ line (``Kasner circle'') onto the fundamental interval $1<u<+\infty$.}
\end{table}
\section{\label{reducedforms} Reduced symplectic form for billiards}
Let us briefly recall various ways in which one can define integral invariants for Hamiltonian systems. (For more details, see, e.g., \cite{ycb1968}, \cite{arnold} and the Appendix below) For a general (possibly time dependent) Hamiltonian dynamics, with (Hamiltonian) action ($i=1, ..., n$)
\be\label{4.1new}
S=\int dt\left[p_i\dot{q}^i-H(\mathbf{q}, \mathbf{p},t)\right],
\ee
one can use the Poincar\'e-Cartan two-form
\be\label{4.2new}
\omega^{(2)}_{PC}=\sum_idq^i\wedge dp_i-dt\wedge dH(q, p, t)
\ee
to define a whole hierarchy of integral invariants of the (unparametrized or parametrized) Hamiltonian flow. Leaving a more general discussion to the Appendix, we shall consider here the case of a time-independent Hamiltonian $H(q,p)$, and focus on invariants of the flow on a given ($2n-1$)-dimensional energy hypersurface, say $\mathcal{E}^{(2n-1)}_E$, satisfying $H(q,p)=E$. For this case, the invariance of the \textit{energy-shell reduced} Liouville measure $\Omega^{(2n-1)}_{L,E}\propto\delta(H(p,q)-E)dq^1\wedge dq^2\wedge...\wedge dq^n\wedge dp_1\wedge dp_2\wedge...\wedge dp_n$ is well-known. Less well-known is the construction of integral invariants based on the existence of a \textit{reduced} symplectic $2$-form
\be\label{4.3new}
\omega^{(2)}_{\rm red}:=\left[\sum_i dq^i\wedge dp_i\right]_{Q_E^{(2n-2)}},
\ee
defined on the \textit{quotient} $Q_E^{(2n-2)}\equiv\mathcal{E}_E^{(2n-1)}/\mathcal{F}_H$ of the energy hypersurface $H(q,p)=E$ by the unparametrized Hamiltonian flow. In other words, $Q_E^{(2n-2)}\equiv\mathcal{E}_E^{(2n-1)}/\mathcal{F}_H$ is the ($2n-2$)-dimensional space of unparametrized Hamiltonian motions on $\mathcal{E}_E^{(2n-1)}$. [The reduced symplectic form (\ref{4.3new}) is linked to the general theory of reduction of phase spaces with symmetry; it was used in Ref. \cite{Gibbons:1986xk} as a way to define a measure in cosmology.] A concrete representation of the quotient space $Q_E^{(2n-2)}$ can be obtained by considering any ``transverse section'' $\mathcal{F}_H$ of $\mathcal{E}_E^{(2n-1)}$, i.e any ``initial conditions'' for the (unparametrized) Hamiltonian flow. The fact that $\omega^{(2)}_{PC}$ ``descends'' to the quotient means that its restriction to any transverse section of $\mathcal{F}_H$ is independent of the choice of section.\\
As a concrete example of the reduced (symplectic) form (\ref{4.3new}) on the space of motions (or initial conditions), one can have in mind the symplectic form on the manifold of (unparametrized) straight lines in a Euclidean plane. A straight line $L$, i.e. $\mathbf{x}(s)$ (where $s$ measures the length along the line), can be parametrized by two vectors, $\mathbf{b}$, $\mathbf{n}$, submitted to the two constraints: $\mathbf{n}^2=1$, $\mathbf{b}\cdot \mathbf{n}=0$, namely $\mathbf{x}(s)=\mathbf{b}+\mathbf{n}s$. In any Cartesian coordinate system, one can explicitly parametrize ($\mathbf{b},\mathbf{n}$) by two real numbers: $\mathbf{b}=(-b\sin\alpha,b\cos\alpha)$, $\mathbf{n}=(\cos\alpha,\sin\alpha)$, where $b$ is the impact parameter between the origin and $L$, and $\alpha$ is the angle between the $x$ axis and $L$. The reduced symplectic form on the $2$-dimensional manifold of (unparametrized) straight lines $L$ can be obtained by starting either from the unparametrized action $S_1=\int\sqrt{dx^2+dy^2}$, or from the parametrized one $S_2=\int dt \frac{1}{2}(\dot{x}^2+\dot{y}^2)$ submitted to a fixed energy constraint. For instance, $S_2$ corresponds to a $4$-dimensional phase space ($x,y,p_x,p_y$) with Hamiltonian $H=\frac{1}{2}\left(p_x^2+p_y^2\right)$. Let us consider the $2$-form (\ref{4.3new}) reduced both by the energy constraint $H=1/2$ (i.e. $p_x^2+p_y^2=1$, or $p_x=\cos\alpha$, $p_y=\sin\alpha$), and by restricting it, e.g., to the section $0=y(s_0)=b\cos\alpha+s_0\sin\alpha$, i.e. $s_0=-b\cot\alpha$. These two constraints reduce the phase space to a $2$-dimensional one parametrized either by $x_0,p_x$, where $x_0=[x]_{y=0}=-b\sin\alpha+s_0\cos\alpha=-b/\sin\alpha$ and $p_x=\cos\alpha$, or by ($b,\alpha$). The symplectic form reduced to this section yields
\be\label{4.12a}
\omega^{(2)}_{\rm red}=dx_0\wedge dp_x=d\left(-\frac{b}{\sin\alpha}\right)\wedge d(\cos\alpha)=db\wedge d\alpha.
\ee  
We see that the impact parameter $b$ and the angle $\alpha$ constitute canonical coordinates on the $2$-dimensional manifold of straight lines. The reduced symplectic form (\ref{4.12a}) can also be written in a form that is manifestly invariant under the group of Euclidean symmetries, namely
\be\label{4.13a}
\omega^{(2)}_{\rm red}=d\mathbf{b}\wedge \mathbf{n}\equiv db_x\wedge dn_x+db_y\wedge dn_y
\ee
with the algebraic constraints $\mathbf{n}^2=1$, $\mathbf{b}\cdot\mathbf{n}=0$.  It is also easily checked that any other section (e.g. $x=0$ instead of $y=0$) yields the same reduced form. Note that, in the present example, the reduced phase space of Euclidean straight lines is $2$-dimensional so that the reduced symplectic form $\omega^{(2)}_{\rm red}$ furnishes directly a measure on the space of straight lines.\\
\vspace{0.1cm}

We spent some time on this simple example because we shall be interested, in this paper, with the hyperbolic-plane generalization of this structure. More precisely, we shall find useful to consider the reduced symplectic form (\ref{4.3new}) in the case where the Hamiltonian dynamics is that of a (radially-projected) billiard motion on $\gamma$-space, i.e. on the hyperbolic space $\mathcal{H}_n$. More precisely, we consider the case where $n=2$, and use the Poincar\'e half-plane model. The corresponding phase-space is $4$-dimensional, say ($u,v,p_u,p_v$) in the Poincar\'e model, and the corresponding Hamiltonian reads
\be\label{4.24}
H_{\gamma}(u,v,p_u,p_v)=\frac{1}{2}v^2\left(p_u^2+p_v^2\right)+V_\infty(\gamma(u,v)).
\ee  
The energy surface $\mathcal{E}_{E_\gamma}=\{H_\gamma(u,v,p_u,p_v)=E_\gamma\}$ is $3$-dimensional. Finally, the quotient space $Q_\gamma=\mathcal{E}_{E_\gamma}/\mathcal{F}_{H_\gamma}$ is simply $2$-dimensional. Therefore, in that case, the general fact that the ambient symplectic form $-d\sigma^{(1)}_{PC}$ can ``descend'' onto the reduced symplectic form $\omega^{(2)}_{\gamma \rm red}$ on the abstract quotient space $Q_{\gamma}$ (and, therefore, on any section of the Hamiltonian flow) means that $\omega^{(2)}_{\gamma \rm red}$ directly provides an integral invariant of arbitrary ``snapshots'' of the billiard motion. In particular, if we take the snapshots corresponding to collisions on the wall, we conclude that $\omega^{(2)}_{\gamma \rm red}$ is invariant under the discrete billiard map $\mathcal{T}$. Moreover, as we are considering here a case where the quotient space $Q_\gamma$ is two-dimensional, $\omega^{(2)}_{\gamma \rm red}$ directly defines an invariant \textit{measure} of the discrete billiard map $\mathcal{T}$. [See the Appendix for a general discussion of the link between the measure on $Q_{E}^{(2n-2)}$ associated to $\omega^{(2)}_{\gamma \rm red}$ and the energy-shell Liouville measure.]\\
\vspace{0.1cm}

There are many ways to compute the reduced symplectic form  $\omega^{(2)}_{\gamma {\rm red}}$. Let us first note that it is the generalization of the measure discussed above, on the $2$-dimensional manifold of straight lines in a Euclidean plane. Here, indeed,  $\omega^{(2)}_{\gamma {\rm red}}$ is a measure on the $2$-dimensional manifold of geodesic lines in a Lobachevsky plane. As in the Euclidean calculation above, Eq. (\ref{4.12a}),  $\omega^{(2)}_{\gamma {\rm red}}$ can be obtained by reducing the ambient symplectic form $du\wedge dp_u+ dv\wedge dp_v$ by two conditions: the energy-shell condition $v^2(p_u^2+p_v^2)=2E_\gamma$, and a cross-section condition locally transverse to the Hamiltonian flow. It is easily seen that, if we consider for simplicity the energy shell $E=\frac{1}{2}$ for a geodesic (i. e. that the geodesic motion on $\mathcal{H}_2$ proceeds with unit speed), the general geodesic line in $\mathcal{E}_{E_\gamma=\frac{1}{2}}$ (i.e. a circle orthogonal to the measure boundary) can be parametrized as
\be\label{4.25}
u=U-V\cos\theta,\ \ v=\mid V\mid \sin\theta,
\ee
\be\label{4.26}
p_u={\rm sign}(V)\frac{\sin\theta}{v}=\frac{1}{V},\ \ p_v=\frac{\cos\theta}{v}=\frac{\cot\theta}{| V|},
\ee
where $\theta$ grows, according to $d\theta/dT=\sin\theta$, from $0$ to $\pi$ as $T$ formally varies from $-\infty$ to $+\infty$. The two constants of integration ($U,V$) parametrize the two-dimensional manifold of geodesic lines on the Poincar\'e half plane. They are related to the parameters $u^+,u^-$ used in Eq. (\ref{3.4}) above via
\be\label{4.27}
U=\frac{1}{2}\left(u^++u^-\right),\ \ V=\frac{1}{2}\left(u^+-u^-\right).
\ee
If, for instance, we use as cross-section to restrict $\omega^{(2)}_\gamma=du\wedge dp_u+dv\wedge dp_v$ any $v=const$ slice (i.e. $0=dv=\sin\theta d|V|+|V|\cos\theta d\theta$) we get
\be\label{4.28}
\omega^{(2)}_{\rm red}=du\wedge dp_u=d(U-V\cos\theta)\wedge d\left(\frac{1}{V}\right)=-\frac{dU\wedge dV}{V^2}.
\ee
Rewritten in terms of ($u^+,u^-$), Eq. (\ref{4.27}), this reads
\be\label{4.29}
\omega^{(2)}_{\rm red}=2\frac{du^+\wedge du^-}{(u^+-u^-)^2}.
\ee
The result (\ref{4.28}) or (\ref{4.29}) is similar to the Euclidean measure (\ref{4.12a}) or (\ref{4.13a}) on the manifold of Euclidean straight lines. Analogously to the fact that the measure (\ref{4.13a}) was invariant under the group of Euclidean symmetries (translations, rotations and reflections), the measure (\ref{4.28}), (\ref{4.29}) is invariant under the group of symmetries of the hyperbolic plane. This group is $SL_2(\mathbb{R})\times\mathbb{Z}_2$, and it acts on the boundary of the Poincar\'e model (i.e. on the parameters $u^+$ and $u^-$) by transformations of the form (\ref{3.11}). This group of transformations is generated by $u'_\pm=au_\pm+b$, $u'_\pm=-1/u_\pm$ and $u'_\pm=-u_\pm$. It is then easily seen that (\ref{4.29}) is indeed invariant under each one of these generating transformations. Note that this also gives a direct proof that the reduced symplectic form (\ref{4.29}) is invariant under the Hamiltonian flow of the billiard. Indeed, this flow is made of two types of evolutions: (i) a free-flight evolution during which $u^+$ and $u^-$ do not vary, and (ii) collisions on the walls, during which the geodesic undergoes a hyperbolic reflection, i.e. a transformation of the type
(\ref{3.10}) leading to Eq. (\ref{3.11}) with a minus sign.

\section{\label{bigbilliard}Hopscotch dynamics of the Big Billiard}
We shall start our investigation of the various possible cosmological billiards for pure gravity in $d=3$ spatial dimensions by considering the `big billiard table' delimited by the three gravitational walls appearing in diagonal homogeneous Bianchi IX cosmological models, i.e. the walls $\omega^g_{(123)}(\beta)$, $\omega^g_{(231)}(\beta)$ and $\omega^g_{(312)}(\beta)$, Eq.(\ref{2.16}), which respectively correspond to the terms $a^4\equiv e^{-4\alpha}$, $b^4\equiv e^{-4\beta}$ and $c^4\equiv e^{-4\gamma}$ in the usual BKL representation [see Eqs. (\ref{eeq}), (\ref{hcon})]. Using Eqs. (\ref{2.16}), which express $\gamma^a\propto\beta^a$ in terms of the coordinates ($u,v$) of the Poincar\'e model, we see that, in this model, the `$a$ wall' (or $\beta^1$ wall) is located along the vertical line $u=0$; the `$b$ wall' ($\beta^2$) along the vertical line $u=-1$; while the `$c$ wall' ($\beta^3$) is located along the circle $u(u+1)+v^2=0$. The reflection laws of a point $z=u+iv\in\mathcal{P}_2$ through these geodesics must have the form (\ref{3.10}). One can determine the values of the coefficients $a$, $b$, $c$, $d$ entering the transformation (\ref{3.10}) by requiring that the transformation  leaves point-wise fixed the circle through which one is `reflecting' (in a hyperbolic-geometry sense) the point $z=u+iv$. Indeed, the condition $z'=z$, i.e. $z=-(a\bar{z}+b)/(c\bar{z}+d)$ yields as a locus of fixed points the circle
\be\label{5.1}
0=cz\bar{z}+dz+a\bar{z}+b=c(u^2+v^2)+(d+a)u+(d-a)v+b
\ee
which degenerates to a straight line when $c=0$.\\
\vspace{0.1cm}

By successively identifying the point-wise fixed circle (\ref{5.1}) to the three diagonal Bianchi IX walls, one finds the following reflection laws (acting on the end points $u^\pm$, according to Eq. (\ref{3.11}) with a minus sign). For the `$a$ wall' ($u=0$), $u^\pm\rightarrow A(u^\pm)$ with
\be\label{5.2}
A(u^\pm)=-u^\pm;
\ee
for the `$b$ wall' ($u=-1$), $u^\pm\rightarrow B(u^\pm)$ with
\be\label{5.3}
B(u^\pm)=-u^\pm-2;
\ee 
and for the `$c$ wall' ($u(u+1)+v^2=0$), $u^\pm\rightarrow C(u^\pm)$ with
\be\label{5.4}
C(u^\pm)=-\frac{u^\pm}{2u^\pm+1}.
\ee

Note that all these reflection laws act \textit{diagonally} (i.e. separately) on $u^+$ and $u^-$. 
A billiard motion in the presently considered big billiard is a succession of geodesic flights (or `Kasner epochs') connecting two different walls. For instance (as illustrated in Fig. \ref{fig5})
\be\label{5.5}
...c\rightarrow b\rightarrow a\rightarrow b\rightarrow a \rightarrow c\rightarrow...
\ee
In this work, we shall define a `Kasner era' as a set of Kasner epochs joining the same two walls, with the condition that the epochs preceding and following the considered Kasner era involve the third wall. The \textit{length} of an era is defined as the number of epochs (i.e. geodesic flights) it contains. For instance, in the sequence (\ref{5.5}) there is an era of length $3$ between the walls $a$ and $b$, namely
\be\label{5.6}
E_3(b,a):\ \ b\rightarrow a \rightarrow b\rightarrow a.
\ee

Here, we have introduced the notation $E_k(x,y)$ for an era of length $k=1,2,3,...$ whose \textit{first} free flight is from the wall $x$ to the wall $y$ (were $x, y\in\{a,b,c\}, x\neq y$). Note that ($x,y$) is an \textit{ordered} pair as one should distinguish an era which starts on $x$ and then goes to $y$, from an era which starts on $y$ and then goes to $x$. Note also that the beginning of an era is defined by checking (as sketched in (\ref{5.5})) that the previous connecting flight started from a wall $z\neq x$ and $y$ (similarly for the end of an era). Note that the above definition of a Kasner era corresponds to one of the two different definitions of an era considered by BKL. More precisely, it is the definition they consider in Eq. (5.4) of the review \cite{BLK1971}. In this definition (called BKL$_{u>0}$ in the Introduction of this paper), the BKL $u$ parameter varies (using the notation of the Introduction) from $k-1+x$ to $x$ (with $0<x<1$ and $k\in\mathbb{R}$). [This contrasts with the other (more standard) definition of an era used by BKL, the BKL$_{u>1}$ in which the $u$ parameter varies from $k+x$ to $1+x$ (so that $u$ stays in the interval $[1,+\infty]$)]. As noted in the second footnote on p. 753 of \cite{BLK1971}, the former (less standard) definition of an era (that we shall use here) is more natural when considering the dynamics of the variables $\ln a$, $\ln b$, $\ln c$. In terms of the billiard picture, this more natural character does correspond to the definition we gave above of collecting all the epochs joining the same two walls. By contrast, in the other BKL definition ($k+x$ to $1+x$), the era corresponding, for instance, to the sequence (\ref{5.5}) would consist of the three epochs $c \rightarrow b\rightarrow a\rightarrow b$, and what is in our definition the last epoch of the era (of the type $b\rightarrow a$) would be considered as the first epoch of the \textit{next} era. Note that in both definitions the Kasner era has the same length: in our example a length $3$; the last epoch $c\rightarrow b$ of the preceding era having been added as a first epoch, in replacement of the last $b\rightarrow a$ in Eq. (\ref{5.6}).\\
\vspace{0.1cm}

The shortest possible length of an era is $k=1$, i.e. an era corresponding to only one epoch. E.g., as we shall see later, the simplest periodic big-billiard orbit (involving the golden ratio) proceeds along the equilateral (hyperbolic) triangle geodesically connecting the `middles' of the three $a$, $b$, $c$ walls, and is made of only one-epoch eras, say
\be\label{5.7}
...c\rightarrow b\rightarrow a\rightarrow c\rightarrow b\rightarrow a\rightarrow c...,
\ee
or the reverse.

\subsection{Hopscotch dynamics}
As just recalled, the dynamics of the big billiard is described as a sequence of eras $E_k(x,y)$, where each era $E_k(x,y)$ is made of $k$ epochs, i.e. $k$ `arrows', $x\rightarrow y\rightarrow x\rightarrow...$, connecting the walls $x$ and $y$ ($x\neq y,x,y\in\{a,b,c\}$). Note that the last wall involved in the era $E_k(x,y)$ will be $y$ if $k$ is odd, and $x$ if $k$ is even. To describe mathematically the discrete big-billiard dynamics induced by the effect of the successive collisions, i.e. of the corresponding discrete `billiard maps' $\mathcal{T}_{xy}$transforming the phase-space variables on a $x$ wall (just after the $x$-collision) to the phase-space variables on the following $y$ wall, it is useful to use the Poincar\'e plane variables. [Note, however, that though the Poincar\'e-plane variables are \textit{algebraically} more convenient, it is generally more enlightening to \textit{geometrically} visualize the billiard dynamics on the disk model. See Fig. \ref{fig4}.] As recalled above, in the Poincar\'e plane each epoch trajectory (i.e. each geodesic segment connecting two successive sides) is uniquely parametrized by an ordered pair ($u^-,u^+$), where $u^-$ (respectively $u^+$) is the end point (resp. starting point) on the $v=0$ axis (or `absolute') of the corresponding, extended geodesic. In terms of these variables, the billiard dynamics induces a discrete map $\mathcal{T}$ transforming a point in the $(u^-,u^+)$ plane (describing some epoch) into another point $(u^{'-},u^{'+})$ (describing the next epoch): $(u^{'-},u^{'+})=\mathcal{T}(u^-,u^+)$. Looking at Fig. \ref{fig5}, it is clear that the knowledge of $(u^-,u^+)$, i.e. the knowledge of the initial epoch, \textit{uniquely determines} the wall on which it will next collide, and therefore uniquely determines the explicit form of the transformation $\mathcal{T}$, among the three possible explicit forms $A$, $B$, $C$, listed in Eqs. (\ref{5.2}), (\ref{5.3}) and (\ref{5.4}). It is also clear that the transformation $\mathcal{T}$ is \textit{one-to-one} because its inverse $\mathcal{T}^{-1}$ is defined by ``reversing the time evolution'', i.e. exchanging the roles of $u^+$ and $u^-$. As for the iteration of $\mathcal{T}$, $\mathcal{T}\circ\mathcal{T}\circ\mathcal{T}\circ...$, it corresponds to composing a sequence of $(u^-,u^+)$ transformations [among Eqs. (\ref{5.2}), (\ref{5.3}) and (\ref{5.4})] corresponding to a sequence of wall collisions. E.g. the sequence (\ref{5.5}) will correspond to successively composing the actions of
\be\label{5.8}
...C\rightarrow B\rightarrow A\rightarrow B\rightarrow A\rightarrow C\rightarrow...
\ee
on the ($u^-,u^+$) plane, i.e. the combined map (in reverse order)
\be\label{5.9}
...C\circ A\circ B\circ A\circ B\circ C...
\ee
Note also that the composition of maps (\ref{5.9}) can also be expressed as the corresponding matrix product of the matrices 
\[ \left( \begin{array}{cc}
-a & -b \\
c & d \end{array} \right)\]
(with $ad-bc=1$) corresponding [via $u'_{\pm}=-(au_\pm+b)/(cu_\pm+d)=(-au_\pm-b)/(cu_\pm+d)$] to the fractional linear transformations (\ref{5.2}), i.e.
\be\label{5.10}
...C.A.B.A.B.C...
\ee
where
\be\label{5.11}
A= \left( \begin{array}{cc}
-1 & 0 \\
0 & 1 \end{array} \right),\ \ 
B= \left( \begin{array}{cc}
-1 & -2 \\
0 & 1 \end{array} \right),\ \
C= \left( \begin{array}{cc}
-1 & 0 \\
2 & 1 \end{array} \right),
\ee
and where the dots in Eq. (\ref{5.10}) denote the ordinary matrix product.\\
\vspace{0.1cm}

Summarizing so far: the representation of the big-billiard dynamics in the $(u^-,u^+)$ plane is the following: (i) during each epoch (i.e. free flight) the reduced phase-space point $(u^-,u^+)$ stays fixed; (ii) the effect of each collision on the wall $a$, $b$ or $c$ consists in transforming the phase-space point ($u^-, u^+$) into a new point $(u'^-, u'^+)=\mathcal{T}(u^-, u^+)$, where the explicit expression of $(u'^-, u'^+)$ is uniquely defined by the initial phase-space point $(u^-, u^+)$\footnote{As we shall discuss below, the initial value of the $u^+$ alone suffices to determine the explicit form of $\mathcal{T}$ among $A$, $B$, $C$.}, and is either of the form $(A(u^-), A(u^+))$, $(B(u^-), B(u^+))$ or $(C(u^-), C(u^+))$ [with $A(u^\pm)$, $B(u^\pm)$, $C(u^\pm)$ given by Eqs. (\ref{5.2}), (\ref{5.3}) and (\ref{5.4}), respectively], where the choice between $A$, $B$ or $C$ is determined by the $a$, $b$, $c$ wall that is next crossed by the oriented geodesic defined by $(u^-, u^+)$. In other words, we can think of the $(u^-,u^+)$ plane as a big `hopscotch court'\footnote{The child game called `hopscotch' in English is called `marelle' in French and `campana' in Italian.} on which the representative phase-space point $(u^-,u^+)$ jumps around, in a deterministic manner, $(u^-,u^+)\rightarrow\mathcal{T}(u^-,u^+)\rightarrow\mathcal{T}\circ\mathcal{T}(u^-,u^+)\rightarrow...$ according to a sequence of `jumps' whose concrete form is of the type (\ref{5.8}). These jumps act diagonally, i.e. in the same way on $u^+$ and $u^-$.\\
\vspace{0.1cm}

As we shall discuss below, though each ``collision transformation'' $A$, $B$ or $C$ acts on $u^+$ (respectively, on $u^-$) independently of $u^-$ (respectively, of $u^+$), one needs to keep track of the successive values of the pairs $(u^-,u^+)$ to determine the entire (two-sided) sequence of maps, such as Eq. (\ref{5.9}), corresponding to the billiard dynamics. The big-billiard hopscotch dynamics $\mathcal{T}$ just defined differs from the usually discussed BKL dynamics in several respects. Indeed, in order to simplify their discussion, and go to the essence of the Bianchi IX dynamics, Belinski, Khalatnikhov and Lifshitz did not keep track of the order of the Kasner exponents during an era (i.e., in their notation, whether ($p_l,p_m$) is ($p_1,p_2$) or ($p_2,p_1$) in an era of oscillations between the $a$ and $b$ walls). Moreover, BKL further simplified their discussion by using the $6$-fold permutation symmetry among ($a,b,c$), so that they also did not keep track of which unordered pair $\{a,b\}$, $\{b,c\}$ or $\{c,a\}$ an era referred to, nor of the ordering of the first pair in a given era. By contrast, our description above explicitly keeps track of both ordering and labeling issues. Below, we shall discuss how one can `quotient', in a precise manner, the more complete big-billiard dynamics down to the usual BKL discrete dynamics.\\
\vspace{0.1cm}

As an example of fuller description in the big-billiard representation, note that the particular length-$3$ era $E_3(b,a)$, Eq. (\ref{5.6}), corresponds to applying, successively, to the values ($u^+_{Fba},u^-_{Fba}$) parametrizing the \textit{first} epoch of $E_3(b,a)$ (i.e. the first arrow in (\ref{5.6})) the transformations $A$,$B$, and then $A$. This yields successively
\begin{subequations}\label{5.12}
\begin{align}
&u'_\pm=A(u^\pm_{F_{ba}})=-u^\pm_{Fba},\\
&u''_\pm=B(u'_\pm)=u^\pm_{Fba}-2,\\
&u^\pm_{Fac}\equiv u'''_\pm=A(u''_\pm)=-u^\pm_{Fba}+2,
\end{align}
\end{subequations}
where, as indicated by the notation, $u^\pm_{F_{ac}}=A\circ B\circ A(u^\pm_{F_{ba}})$ are the phase-space parameters of the \textit{first} epoch of the following era (which oscillates between the $a$ and $c$ walls, starting on $a$). Note that the transformations appearing in the era composition $A\circ B\circ A$ do not include the effect of the first wall $b$ in (\ref{5.6}). Indeed, the collision on $b$ would be (conventionally) included in the composition of transformations appearing in the previous era (which oscillated between $c$ and $b$, ending on $b$). By contrast, we conventionally include the effect of the collision on the last wall of an era in the composition of transformations associated to this era (e.g. the last transformation in (\ref{5.12}) represents the last collision, on the wall $a$ of the era (\ref{5.6})). The number of transformations which are composed during an era $E_k(x,y)$ is equal to the length $k$ of the era, i.e. to the number of arrows (or of epochs) in the diagram (\ref{5.6}) of the era.\\
\vspace{0.1cm}

The successive transformations,
\be\label{5.13}
u\rightarrow-u\rightarrow u-2\rightarrow-u+2\rightarrow u-4\rightarrow...
\ee
that appear in a long era of big-billiard oscillations between the $a$ and $b$ walls differ from the standard BKL result for oscillations between $a$ and $b$, namely:
\be\label{5.14}
u\rightarrow u-1\rightarrow u-2\rightarrow u-3\rightarrow u-4\rightarrow...
\ee
However, this difference is only due to the fact that the big-billiard description is keeping track of an information that BKL did not wish to keep track of: namely, the precise order between the Kasner exponents $p_l,p_m$ (in the notation of BKL) associated to the `oscillating' diagonal metric components $a^2\sim t^{2p_l}$, $b^2\sim t^{2p_m}$. As an unordered set $\{p_l,p_m\}=\{p_1,p_2\}$ (with $p_1\le p_2\le p_3$ as in section \ref{conformalrepresentation} above). Indeed, we see in the list of `Kasner transformations', Table \ref{table1}, labeling the possible permutations of the Kasner exponents, that the transformation
\be\label{5.15}
u'=k_5(u)\equiv-u-1,
\ee
which corresponds to the permutation between $p_1$ and $p_2$, maps each one of the apparently discrepant values of $u$ (namely $-u$, $-u+2$, etc) in the big-billiard sequence (\ref{5.13}) into the corresponding usual BKL one (\ref{5.14}). To wit
\be\label{5.16}
k_5(-u)=u-1;\ \ k_5(-u+2)=u-3;\ \ \rm{etc}. 
\ee

\begin{figure*}[tb]
\begin{center}
\includegraphics[width=0.8\textwidth]{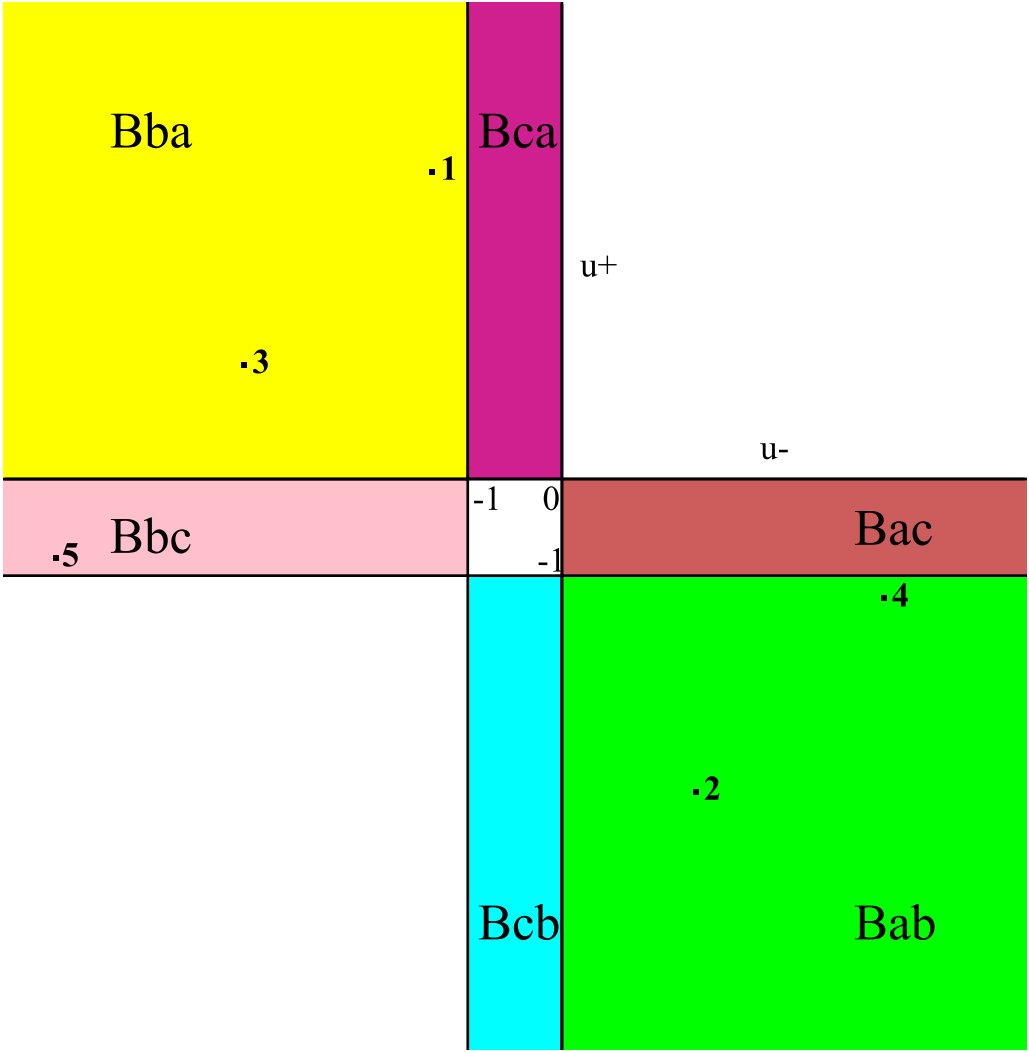}
\caption{\label{biliardino11} Billiard phase space in the $u^-u^+$ parametrization: the epoch hopscotch court. The $B_{xy}$ regions are sketched, and filled with different colors (shades of gray).}
\end{center}
\end{figure*}

Let us now clarify what is the shape of the `hopscotch court', i.e. the part of the $(u^-,u^+)$ plane which parametrizes the dynamics of the big billiard. This full hopscotch court is naturally divided into six separate `boxes':(1) a box, say $B_{ab}$, parametrizing the epochs going from $a$ to $b$, (2) a box, say $B_{ba}$, parametrizing the epochs going form $b$ to $a$, etc, when considering the other ordered pairs ($x,y$) with ($x,y)\in \{a,b,c\}$. The precise boundaries of the box $B_{xy}$ are easily obtained by requiring that, in the Poincar\'e half plane, there exists an oriented half circle (orthogonal to the boundary) crossing the walls $x$ and $y$ in that order. For instance, it is easily seen that the box $B_{ab}$ is defined by the inequalities
\be\label{5.17}
B_{ab}:\ \ 0<u^-<+\infty,\ \ -\infty<u^+<-1.
\ee
The inequalities defining all the boxes $B_{xy}$ are gathered in Table \ref{table2}, and the corresponding regions in the $(u^-,u^+)$ plane are represented in Fig \ref{biliardino11}.
\begin{table}
\begin{center}
    \begin{tabular}{ | l | l | l| }
    \hline
    $B_{ab}$ & $-\infty<u^+<-1$ & $0<u^-<\infty$ \\ \hline
    $B_{ba}$ & $0<u^+<\infty$ & $-\infty<u^-<-1$ \\ \hline
    $B_{ac}$ & $-1<u^+<0$ & $0<u^-<\infty$ \\ \hline
    $B_{bc}$ & $-1<u^+<0$ & $-\infty<u^-<-1$ \\ \hline
    $B_{ca}$ & $0<u^+<\infty$ & $-1<u^-<0$ \\ \hline
    $B_{cb}$ & $-\infty<u^+<-1$ & $-1<u^-<0$ \\ \hline
    \end{tabular}
\end{center}
\caption{\label{table2} Hopscotch Court}
\end{table}
Two important remarks concerning these boxes are: (i) all the boxes have a rectangular shape, and (ii) the union of all the boxes (together with their boundaries) does not cover the full $(u^-,u^+)$ plane. More precisely, the domain of the $(u^+,u^-)$ plane which does not parametrize any epoch is the union of the following three `vacuum boxes'
\begin{subequations}\label{5.18}
\begin{align}
&V_a:\ \ 0<u^-<+\infty,\ \ 0<u^+<+\infty,\\
&V_b:\ \ -\infty<u^-<-1,\ \ -\infty<u^+<-1,\\
&V_c:\ \ -1<u^-<0,\ \ -1<u^+<0.
\end{align}
\end{subequations}
For instance, we illustrate in Fig. \ref{biliardino11} a long era of epochs oscillating between $b$ and $a$ (see the points marked $1,2,3,4,5$) which starts (point $1$) in the upper right part of the box $B_{ba}$, and then jumps successively from $B_{ba}$ toward $B_{ab}$ and back until it `exits' by terminating in $B_{bc}$ (point $5$). Then the $(u^-,u^+)$ point will jump from $B_{bc}$ to $B_{cb}$, as part of a next era of the $E_k(bc)$ type.\\
\vspace{0.1cm}

We have seen above that the (bijective) applications $A$, $B$, $C$ (Eqs. (\ref{5.2}), (\ref{5.3}), (\ref{5.4})) corresponding to the collisions on the walls $a$, $b$, $c$, respectively, leave invariant the two-form $\omega^{(2)}_{\rm red}$, Eq. (\ref{4.29}), that we shall simply denote in the following as
\be\label{5.19}
\omega=2\frac{du^+\wedge du^-}{(u^+-u^-)^2}.
\ee
Therefore the $2$-form $\omega$ defines an oriented measure on the full hopscotch court that is invariant under the hopscotch discrete map $\mathcal{T}$ defined above. We would seem to be able to straightforwardly apply the tools and results of ergodic theory to our full hopscotch game. In particular, we know that this hopscotch game must be \textit{ergodic} in the sense that it cannot leave invariant a subdomain, having a non-zero measure with respect to the form $\omega$, Eq. (\ref{5.19}), of the full hopscotch court. Indeed, our hopscotch game is a projection of the billiard motion within an ideal triangle on the hyperbolic plane $\mathcal{H}_2$. If the projected billiard motion could leave `unvisited', for an infinite `time', a continuous subdomain of the full hopscotch court, this would be inconsistent with the fact that the billiard dynamics on an ideal triangle on $\mathcal{H}_2$ is known (since the classic work of Hedlung and Hopf) to be ergodic, and therefore to visit the full $3$-dimensional phase-space $(u^-,u^+,s)$ which lies `above' our hopscotch court. \\
However, there is a catch in that, contrary to what is usually assumed in most investigations of ergodic theory, the full invariant measure of our (projected) phase-space, i.e. the integral of the two-form $\omega$, Eq. (\ref{5.19}), on the full hopscotch court defined in Table \ref{table2} is \textit{infinite}! Indeed, the form (\ref{5.19}) is singular along the line $u^+=u^-$, as well as at infinity $|u^+|\sim|u^-|\sim\infty$, where the integral of $\omega$ is logarithmically divergent. We see on Fig \ref{biliardino11} that the singular line $u^+=u^-$ lies mostly in the excluded ('vacuum') part of the hopscotch court. However, this line touches the boundaries of the court around the points $(u^-,u^+)=(0,0)$ and $(u^-,u^+)=(-1,-1)$. In addition, the boxes $B_{ab}$ and $B_{ba}$ extend at infinity, where $\int\omega$ diverges logarithmically. It is also easily seen that the integral of $\omega$ over $B_{ca}$ and $B_{ac}$ diverges logarithmically near $(u^+,u^-)=(0,0)$, and that the same is true for $\int\omega$ near $(u^+,u^-)=(-1,-1)$. The three points $(0,0), (-1,-1), (\infty,\infty)$ correspond to the three `cusps' of the ideal triangle on $\mathcal{H}_2$. Indeed, another way to understand why the integral of $\omega$ is infinite is to remember that $\omega^{(2)}_{\rm red}=\omega$ can also be written (in `Birkhoff coordinates') as $dl\wedge d(\sin\alpha)=\cos\alpha dl\wedge d\alpha$, where $l$ measures the length of the boundary of the billiard, and where $\alpha$, $-\pi/2<\alpha<\pi/2$, is the angle between the normal to the boundary and the velocity vector. The integral of $\sin\alpha$ yields a factor $2$, while the integral over $dl$ yields the total length of the boundary of the billiard. In the case of the ideal triangle, the length diverges logarithmically at each corner.\\
\vspace{0.1cm}

Before discussing other issues concerning the invariant measure $\omega$, Eq. (\ref{5.19}), in the $2$-plane ($u^-,u^+$), let us note that, by \textit{marginalizing} the variable $u^-$, we can deduce from $\omega$ an invariant measure for the dynamics of $u^+$ alone. We already noticed that the hopscotch map $\mathcal{T}$ acts \textit{diagonally} on ($u^-,u^+$) (i.e. separately, and actually in the same way, on $u^-$ and $u^+$). We warned the reader above that, in spite of this diagonal action, one needs to keep track of the action of $\mathcal{T}$ on the two variables $(u^-, u^+)$ in order to determine the full, \textit{two-sided} sequence of collisions corresponding to the billiard dynamics taking place within the chamber of the big billiard. However, if one ignores the variable $u^-$, and only considers the action of $\mathcal{T}$ on $u^+$, it is easily seen that the sole knowledge of the initial value of $u^+$ suffices to determine  the explicit expression of $\mathcal{T}$ (among $A$, $B$ or $C$) and therefore all the \textit{future} values of $u^+$. Indeed, a look at  Fig. \ref{fig5} shows that there are three, and only three, possible cases: (i) if $u^+$ belongs to the interval $[-\infty, -1]$, the next collision will be on the $b$ wall so that the action of $\mathcal{T}$ is $u'^+=\mathcal{T}(u^+)=B(u^+)$; (ii) if $u^+\in[-1,0]$, the next collision is on the  $c$ wall, so that $\mathcal{T}(u^+)=C(u^+)$; and (iii) if $u^+\in[0,+\infty]$, one has $\mathcal{T}(u^+)=A(u^+)$. Therefore, once we know $u^+$, we can uniquely determine all its $\mathcal{T}$ iterates. [Reciprocally, it is easy to see that the knowledge of the initial value of $u^-$ suffices to determine all the \textit{past} values of $u^-$, i.e. all its $\mathcal{T}^{-1}$ iterates.] In other words, if we simply ignore the variable $u^-$, the map $\mathcal{T}$ defines a dynamics for $u^+$ alone, which is an unquotiented version of the usual BKL dynamics on the single variable $u$ recalled in the Introduction. [Remember that the BKL variable $u$ actually coincides with our variable $u^+$.] These remarks show that the unquotiented generalization of the BKL $u$-map defined by $u'_+=\mathcal{T}(u_+)$ will admit \textit{an invariant one-dimensional measure} $w(u^+)du^+$ obtained by marginalizing (i.e. integrating upon) the variable $u^-$ in the two-dimensional measure (\ref{5.19}). Explicitly, we can then define $w(u^+)du^+$ as
\be\label{1dimmeasure}
w(u^+)du^+\equiv\frac{1}{2}\int_{u^-}\omega du^+=\int\frac{du^-}{(u^+-u^-)^2},
\ee  
so that
\be
w(u^+)=\sum_b\epsilon_b\frac{1}{u^+-u^-_b(u^+)}.
\ee
Here, $u^-_b(u^+)$ denote the various boundaries of the integration domain on the $u^-$ axis, and $\epsilon_b$ the associated signs. As one sees on Fig. \ref{biliardino11}, these boundaries are piecewise-constant functions of $u^+$. For instance, when $u^+>0$, there are two boundaries: $u^-_{\rm min}=-\infty$ (with $\epsilon_{\rm min}=-1$), and $u^+_{\rm max}=0$ (with $\epsilon_{\rm max}=+1$). This leads to $w(u^+)=1/u^+$. When $-1<u^+<0$, one has four different boundaries (on each side of the vacuum domain in the middle of Fig. \ref{biliardino11}). Finally, we can conclude that the unquotiented, big billiard dynamics of the variable $u^+$ considered separately leaves invariant the measure $w(u^+)du^+$ where
\begin{subequations}\label{119}
\begin{align}
&{\rm if}\ \ 0<u^+<+\infty, \ \ w(u^+)=\frac{1}{u^+},\\
&{\rm if}\ \ -1<u^+<0, \ \ w(u^+)=\frac{1}{u^++1}-\frac{1}{u^+}=-\frac{1}{u^+(u^++1)},\\
&{\rm if}\ \ -\infty<u^+<-1, \ \ w(u^+)=-\frac{1}{u^++1}.
\end{align}
\end{subequations}
The existence of this invariant one-dimensional measure for the (unquotiented) BKL dynamics is not well known, and the explicit expression of $w(u^+)du^+$ has, as far as we know, never been written down before.\\
It was indicated in Fig. \ref{fig2} that $u^+\in\bar{\mathbb{R}}$ should really be considered as a coordinate on the Kasner circle, i.e. the manifold of solutions $p_1, p_2, p_3$ of the two Kasner constraints. Note that one could also parametrize the Kasner circle by an angle $\theta$ (with the usual $2\pi$ period). In addition, one can require that $\theta$ be equal, say, to $0$ at the $ab$ corner ($u^+=+\infty$), to $2\pi/3$ at the $bc$ one ($u^+=-1$) and to $4\pi/3$ at the $ca$ one ($u^+=0$). With these requirements, and remembering that the disk model of Fig. \ref{fig2} is related with the half-plane model of Fig. \ref{fig3} (in which $u^+$ appears as a natural coordinate along the \textit{absolute}) by a fractional linear transformation between the complex coordinates $z\equiv u+iv$ of Fig. \ref{fig3}, and $\zeta=x+iy$ of Fig. \ref{fig2}, of the form 
\be
\zeta=\tfrac{2z+1-i\sqrt{3}}{2z+1+i\sqrt{3}},
\ee
the transformation between $u^+$ and $\theta$ (as defined above) is given by (with $\zeta=ie^{i\theta}$ on the Kasner circle, i.e. the absolute)
\be\label{120}
e^{i\theta}=\frac{u^++\tfrac{1}{2}-i\tfrac{\sqrt{3}}{2}}{u^++\tfrac{1}{2}+i\tfrac{\sqrt{3}}{2}}
\ee
The expression of the invariant measure $w(u^+)du^+$, Eqs. (\ref{119}), in terms of the angle $\theta$ on the Kasner circle (in the disk representation) is then easily obtained by using Eq. (\ref{120}). It is found to only depend on the angular distances between $\theta$ and the two big-billiard ``corners'' surrounding it. Let us denote the angular location of the ``corner $ab$'', between the walls $a$ and $b$, see Fig. \ref{fig2}, by $\theta_{ab}$, and similarly for the angular location $\theta_{bc}$ for the corner $bc$, and $\theta_{ca}$ for the corner $ca$. These three angles correspond to $u^+=+\infty$, $-1$ and $0$, respectively, and with our chosen normalization, take the values $\theta_{ab}=0$, $\theta_{bc}=2\pi/3$ and $\theta_{ca}=4\pi/3$ or $-2\pi/3$. With this notation, the invariant measure $w(\theta)d\theta$ on the Kasner circle is given, when $\theta_{ab}<\theta<\theta_{bc}$ (i.e. when $\theta$ is on the ``negative'', or ``shadow'', side of the $b$ wall) by
\be\label{121}
w(\theta)d\theta=\frac{\sqrt{3}}{4}\frac{d\theta}{\sin\tfrac{\theta-\theta_{ab}}{2}\sin\tfrac{\theta_{bc}-\theta}{2}}\ \ \ \ (\theta_{ab}<\theta<\theta_{bc}).
\ee
Its expression in the two other intervals $\theta_{bc}<\theta<\theta_{ca}$ and $\theta_{ca}<\theta<\theta_{ab}$ is obtained by cyclic permutations $abc\rightarrow bca\rightarrow cab$. Note that this measure is invariant around the middle points, e.g. $(\theta_{ab}+\theta_{bc})/2$ for the interval $[\theta_{ab}, \theta_{bc}]$ of Eq. (\ref{121}), and is logarithmically divergent as $\theta$ tends to the extremities. For instance, as $\theta\rightarrow\theta_{ab}=0$, $\int_\theta w(\theta')d\theta'\simeq\int_\theta d\theta'/\theta'\simeq\ln1/\theta$ in keeping with the logarithmic divergence $\simeq\ln(-u^+)$ of the corresponding $u^+$ interval (\ref{119}c) as $u^+\rightarrow-\infty$, given the link $\theta\simeq-\sqrt{3}/u^+$ deduced from Eq. (\ref{120}) as $u^+\rightarrow\pm\infty$.

Let us note finally that it is straightforward to check directly the invariance of the measure $w(u^+)du^+$ defined above under the explicit transformation laws $A, B, C$ defined in Eqs. (\ref{5.2}), (\ref{5.3}), (\ref{5.4}). However, to do that one must note that, contrary to the two-dimensional map $\mathcal{T}$ acting on the ($u^-, u^+$) plane (which is one-to-one), the one-dimensional action of $\mathcal{T}$ on the real line of the variable $u^+$ is no longer one-to-one, but rather two-to-one. [Indeed, the preimage of a certain $u^+$ located within one of the three intervals $[-\infty,-1]$, $[-1,0]$, $[0,+\infty]$, can lie in either one of the two other intervals.] In such a case, one must remember \cite{corn1982} that the definition of the invariance of a measure $\mu$ under $\mathcal{T}$ is that, for any measurable set $U$, $\mu(U)=\mu(\mathcal{T}^{-1}U)$, where $\mathcal{T}^{-1}$ is the \textit{pre-image} (rather then the image) of $U$. When applied to an infinitesimal interval $I=u^+\pm\tfrac{1}{2}du^+$, one must then take into account that $\mathcal{T}^{-1}I$ consists of \textit{two} 
separate infinitesimal intervals.\\
\vspace{0.1cm}

The invariant measure $w(u^+)du^+$ (or $w(\theta)d\theta$) on the Kasner circle is not normalizable. Indeed, we pointed out that its integral diverges logarithmically near each one of the three corners of the big billiard (i.e. near $u^+=\infty, -1$ and $0$, or $\theta=0, 2\pi/3$ or $4\pi/3$). Therefore, the invariant measure of \textit{both} the two-dimensional map $\mathcal{T}$ and its one-dimensional restriction are not normalizable. It can seem strange to have an invariant measure of a projected phase-space which is infinitely large, while the invariant measure of the original, unprojected, phase-space was finite. Indeed, the invariant Liouville measure of the billiard on an ideal triangle, namely
\be\label{5.20}
\Omega^{(3)}_L=\frac{du\wedge dv\wedge d\beta}{v^2}
\ee
where the angle $\beta$ ($0\leq\beta<2\pi$) parametrizes the angular direction of the unit velocity vector, integrates to the product of the finite area of the idea triangle $\int\int du dv/v^2=2\pi$ and of $\int d\beta=2\pi$. As proven in full generality (for any time-independent Hamiltonian system) in the Appendix, see Eq. (\ref{4.17a}), the energy-shell-reduced Liouville measure is simply equal to the reduced product of the symplectic measure $\Omega^{(2n-1)}_{\rm red}\propto (\omega^{(2)}_{\rm red})^{\wedge(n-1)}$ by $ds$, where $s$ denotes a phase-space coordinate which is canonically conjugate to the Hamiltonian (so that $ds/dt=1$ along the Hamiltonian flow). In the present case, where $n=d-1=2$, this yields
\be\label{5.21}
\Omega^{(3)}_L=\omega(u^-,u^+)\wedge ds
\ee
where $s$ measures (when considering a unit velocity billiard) the hyperbolic length along the billiard orbit. This agrees (modulo an unimportant factor) with the result ($25$) of \cite{Kirillov:1996rd}. In terms of the phase-space coordinates ($U,V,\theta$), or equivalently ($u^+,u^-,\theta$)\footnote{The unit (Euclidean) velocity vector $(vp_u,vp_u)\equiv(\cos\beta,\sin\beta)=({\rm sign}(V)\sin\theta,\cos\theta)$ so that we have a link of the form $\beta+\pi/2=-{\rm sign}(V)\theta$ which ensures that the inequality $0\le\theta\le\pi$, together with the fact that $V$ can be either positive or negative, corresponds to an angular direction $\beta$ of the velocity vector varying over a $2\pi$ range.} of Eqs. (\ref{4.25})-(\ref{4.27}), we have $ds=d\theta/\sin\theta$, so that
\be\label{5.22}
s(u^-,u^+,\theta)=s_0(u^-,u^+)+\ln \tan \frac{\theta}{2}=s_0(u^-,u^+)+\frac{1}{2}\ln\frac{1-\cos\theta}{1+\cos\theta}.
\ee
Here, we can choose the $u^\pm$-dependent integration constant as we wish. For instance, we can choose it so that, for any given ($u^+,u^-$), $s(u^-,u^+,\theta)$ varies between $0$ and some maximum value, say $\sigma(u^-,u^+)$ as $\theta$ varies between the starting wall and the ending wall along the oriented geodesic defined by ($u^-,u^+$). With this choice, we see that the full three-dimensional big-billiard phase-space has the shape, in ($u^-,u^+,s$) coordinates, of a `slab', above the ($u^-,u^+$) hopscotch court, of varying thickness $0\le s\le\sigma(u^-,u^+)$. Using Eqs. (\ref{5.22}) and (\ref{4.25}), one can express the thickness $\sigma(u^-,u^+)$ of this slab in terms of the values of the $u$ coordinates (in the Poincar\`e model) of the starting and ending walls along the geodesic (say $u_{\rm start}(u^-,u^+)$, $u_{\rm end}(u^-,u^+)$), namely
\be\label{5.23}
\sigma(u^-,u^+)=\frac{1}{2}\left[\ln\frac{u_{\rm end}-u^-}{u^+-u_{\rm end}}-\ln\frac{u_{\rm start}-u^-}{u^+-u_{\rm start}}\right]=\frac{1}{2}\ln\frac{(u_{\rm end}-u^-)(u^+-u_{\rm start})}{(u^+-u_{\rm end})(u_{\rm start}-u^-)}.
\ee 
It is then easily checked that $\sigma(u^-,u^+)$  tends to zero near each corner of the billiard, thereby ensuring the convergence of the Liouville measure (\ref{5.21}), i.e. the convergence of $\int\int\omega(u^-,u^+)\sigma(u^-,u^+)$. [For instance, near the corner $u^-\rightarrow\infty$, $u^+\rightarrow\infty$] (with $u_{\rm start}=0$, $u_{\rm end}=-1$ or the reverse), the result (\ref{5.23}) yields a thickness $\sigma(u^-,u^+)\simeq \frac{1}{2}(u_{\rm end}-u_{\rm start})(u_+^{-1}-u_-^{-1})\rightarrow0$.\\
\vspace{0.1cm}

The ergodic theory of transformations preserving a measure on an infinite-measure space (or \textit{infinite ergodic theory}) is an active field of current mathematical research \footnote{See, e.g., \cite{gs2010} for an entry into the literature on infinite ergodic theory.} in which, however, there are many less concrete general results than for the case of finite measure. In order to be able to avail ourselves of the usual ergodic theorems (such as the equality between the `time average' and the `measure average'), it is useful to transform the problem onto another one exhibiting a finite measure. Several different strategies are possible for doing so. \\
\vspace{0.1cm}

As a first strategy, we could lift the full big billiard hopscotch game back to the hyperbolic billiard it came from. This would mean considering the ergodic properties of functions on the three-dimensional phase-space ($u^+,u^-,s$), with the finite Liouville measure (\ref{5.21}). In this case, we would be considering a continuous Hamiltonian flow, so that the relevant ergodic theorem would assert that, for almost every phase-space point $x$,
\be\label{5.24}
\lim_{T\rightarrow\infty}\frac{1}{T}\int^{T}_{0}dT'f\left(\mathcal{F}_{T'}(x)\right)=\frac{\int\mu(x')f(x')}{\int\mu(x')}.
\ee
Here $f(x)$ is a (measurable) function on phase-space, $\mathcal{F}_{T}(x)$ denotes the Hamiltonian flow over a time\footnote{As recalled in Section \ref{hyperbolicbilliards}, the appropriate time variable for the hyperbolic billiard is the $T$-time of Eq. (\ref{2.21}).} $T$, and $\mu$ denotes the relevant, finite measure, i.e. $\mu=\Omega^{(3)}_L$, Eq. (\ref{5.21}). Then, if we were interested in the ergodic properties of phase-space functions $f(x)=f(u^-,u^+,s)$ that do not depend on $s$, i.e. on functions $f(u^-,u^+)$ that live on the ($u^+,u^-$) hopscotch court, we can conclude from (\ref{5.24}) (using the fact that $ds/dT=1$ along each geodesic segment) that
\be\label{5.25}
\lim_{N\rightarrow\infty}\frac{\sum_{n=0}^{N-1}\sigma\left(\mathcal{T}^n(u^-,u^+)\right)f\left(\mathcal{T}^n(u^-,u^+)\right)}{\sum_{n=0}^{N-1}\sigma\left(\mathcal{T}^n(u^-,u^+)\right)}=\frac{\int\omega(u^-,u^+)\sigma(u^-,u^+)f(u^-,u^+)}{\int\omega(u^-,u^+)\sigma(u^-,u^+)}
\ee
where $\sigma(u^-,u^+)$ is the thickness (in the $s$ direction) of the phase-space slab `above' the point $(u^-,u^+)$, and where $\mathcal{T}$ denotes the billiard map, i.e. the discrete hopscotch map transforming any $(u^-,u^+)$  parametrizing one Kasner epoch, into the values $(u'_-,u'_+)$ parametrizing the next Kasner epoch. As explained above, $\mathcal{T}$ is equal to $A$, $B$ or $C$ depending on the wall on which the considered geodesic segment will collide. The notation $\mathcal{T}^n$ denotes the $n$-th iteration $\mathcal{T}\circ\mathcal{T}\circ...\circ\mathcal{T}$, i.e. a composed transformation of the type of Eq. (\ref{5.9}). Note how the continuous time average of Eq. (\ref{5.24}) has reduced itself (for functions depending only on $u^-$ and $u^+$) to a discrete-time average, i.e. to a discrete sum (\ref{5.25}) involving the successive iterates of the hopscotch map. However, the continuous-time origin of (\ref{5.25}) is recalled through the occurrence of the `weights' $\sigma\left(\mathcal{T}^n(u^-,u^+)\right)$ involving the successive thicknesses of the phase-space slabs, encountered along the billiard trajectory in $(u^-,u^+,s)$ space. [In $(u^-,u^+,s)$ space the billiard motion becomes a so-called \textit{special flow} \cite{corn1982}, i.e. a combination of uniform motion in the `vertical' $s$ direction, with $0\le s<\sigma(u^-,u^+)$, with discrete jumps in $(u^-,u^+)$ and in $s$ (back to zero when it reaches $\sigma(u^-,u^+)$).]\\
\vspace{0.1cm}

\begin{table}
\begin{center}
    \begin{tabular}{ | l | l | l| }
    \hline
    $F_{ab}$ & $-\infty<u^+<-1$ & $0<u^-<1$ \\ \hline
    $F_{ba}$ & $0<u^+<\infty$ & $-2<u^-<-1$ \\ \hline
    $F_{ac}$ & $-1<u^+<0$ & $1<u^-<\infty$ \\ \hline
    $F_{bc}$ & $-1<u^+<0$ & $-\infty<u^-<-2$ \\ \hline
    $F_{ca}$ & $0<u^+<\infty$ & $-1<u^-<1/2$ \\ \hline
    $F_{cb}$ & $-\infty<u^+<-1$ & $-1/2<u^-<0$ \\ \hline\hline 
    $L_{ab}$ & $-2<u^+<-1$ & $0<u^-<\infty$ \\ \hline
    $L_{ba}$ & $0<u^+<1$ & $-\infty<u^-<-1$ \\ \hline
    $L_{ac}$ & $-1<u^+<-1/2$ & $0<u^-<\infty$ \\ \hline
    $L_{bc}$ & $-1/2<u^+<0$ & $-\infty<u^-<-1$ \\ \hline
    $L_{ca}$ & $1<u^+<\infty$ & $-1<u^-<0$ \\ \hline
    $L_{cb}$ & $-\infty<u^+<-2$ & $-1<u^-<0$ \\ \hline 
    \end{tabular}
\end{center}
\caption{\label{table3} Starting and ending subregions}
\end{table}
A second strategy for reducing the problem to a discrete map having a finite measure is to follow BKL in lumping together the epochs into eras, and to focus on the statistical properties of \textit{eras} rather than \textit{epochs}. In order to do this, we need to know on which subregions of the full hopscotch court, Fig. \ref{biliardino11}, each type of era $E_*(x,y)$ must start. [here, as above, ($x,y$) denotes an oriented pair of walls, and $E_*$ denotes the union of all $E_k$'s, i.e. an era of arbitrary length $k=1,2,3,...$ ] This is straightforwardly obtained by using the transformation rules $A$, $B$, $C$ discussed above. For instance, the sub-region say $F_{ab}$ (where $F$ stands for `First') of the hopscotch court corresponding to the start of an era of the $E_*(a,b)$ type must come from a $c$ wall, and include the process $c\rightarrow a\rightarrow b$. Using either some simple geometric reasoning, or working with the algebraic relations defining the transformations $A$, $B$ and $C$, Eq. (\ref{5.12}), one finds that the $F_{ab}$ subregion is the rectangular subdomain of the $B_{ab}$ box defined by the inequalities
\be\label{5.26}
F_{ab}:\ \ 0<u^-<1,\ \ -\infty<u^+<-1. 
\ee
The full set of inequalities defining the six possible starting subregions $F_{xy}$, with $x,y\in\{a,b,c\}$ are listed in Table \ref{table3}. For compactness, we also indicate the six possible subregions on which an era of the type $E_*(x,y)$ can \textit{end}. They are denoted by $L_{xy}$ (where $L$ stands for `Last'). If we consider the overlap domain $F_{xy}\bigcap L_{xy}$ between the start and the end of some ($x,y$)-type era, it must correspond to an era $E_1(x,y)$ of length $k=1$, i.e. containing only one epoch. For instance, we see on Table \ref{table3} that the intersection $F_{ba}\bigcap L_{ba}$ corresponds to the small box
\be\label{5.27}
F_{ba}^1:\ \ -2<u^-<-1,\ \ 0<u^+<1.
\ee 
The box $F^1_{ba}$. Eq. (\ref{5.27}), describes the starting domain, in the ($u^+,u^-$) plane, of all the one-epoch eras of the $ba$-type. More generally, it is not difficult to write down the inequalities defining the starting domains, say $F^k_{xy}$, of all the $k$-epoch eras (with $k=1,2,3,...$) of the (starting) $xy$-type. They are given by intersecting the full $F_{xy}$ with the condition
\be\label{5.28}
n^{xy}(u^+)=k
\ee
where $n^{xy}(u^+)$ is the integer-valued \footnote{\label{fn}The notation $[x]$ for $x\in\mathbb{R}$ denotes the usual integer part of $x$ when $x\ge0$ (e.g. $[\pi]=3$), and $-[-x]\le0$ when $x\le0$ (so that $[-\pi]=-3$). We did not find useful to introduce other definitions of the integer part (e.g. the `floor', `ceiling', or `Hurwitz' ones).} function listed in the second column of Table \ref{table4} which yields the length of the era starting at some given point ($u^-,u^+$) in phase-space. This leads to the `era hopscotch court' of Fig. \ref{biliardone} which represents the six era-starting domains, and their division in $k$-epoch subregions $F^k_{xy}$. Note that the function $n^{xy}(u^+)$ giving the length of each era depends \textit{only} on $u^+$, and not on $u^-$. This corresponds to the fact that on Fig. \ref{biliardone} all the boundaries between the $F^k_{xy}$ boxes are horizontal. We have also indicated in Fig. \ref{biliardone} the special points $(u^-,u^+)=(-\phi-1,\phi)$, $(u^-,u^+)=(-\phi, \phi+1)$, $(u^-,u^+)=(-1/(2+\phi), -1/(1-\phi)$ (where $\phi=(\sqrt{5}-1)/2\simeq0.618$ denotes the `small' golden ratio) corresponding to the simplest periodic hopscotch orbit corresponding to the infinite succession of one-epoch eras (\ref{5.7}). [There exists also the `time-reverse' version of (\ref{5.7}), namely $a\rightarrow b\rightarrow c\rightarrow a...$ which jumps between $F^1_{ab}\rightarrow F^1_{bc}\rightarrow F^1_{ca}\rightarrow...$.]\\
\vspace{0.1cm}

For completeness, we have also indicated in Table \ref{table4} the discrete sequence of values of $u^\pm_m$, where $1\le m\le n^{xy}$ describing the successive epochs ``contained'' within an era that starts from some $u^\pm\in F_{xy}^{n^{xy}}$ (one example of such sequence of epochs was drawn in Fig. \ref{biliardino11}). In Table \ref{table4}, $u_{m_{xy}}$ denotes either $u_{m_{xy}}^+$ or $u_{m_{xy}}^-$ [we indeed recall that the discrete hopscotch map $\mathcal{T}$ acts on a \textit{diagonal} manner on $u^+$ and $u^-$: $u'_+=\mathcal{T}(u^+)$ and $u'_-=\mathcal{T}(u^-)$]. Moreover, within some era $E_{n^{xy}}(x,y)$ starting with an epoch of the $x\rightarrow y$ type, roughly half of the epochs contained in $E_{n^{xy}}(x,y)$ are of the $x\rightarrow y$ type (namely those corresponding to $m=1,3,5,...$) while the other half are of the $y\rightarrow x$ type (those corresponding to $m=2,4,...$).
\begin{table}
\begin{center}
    \begin{tabular}{ | l | l | l | l |}
    \hline
    $F_{ab}$ & $n^{ab}=[-u^+_{F_{ab}}]$  & $u_{m_{ab}}=u_{F_{ab}}+m-1$ & $u_{m_{ba}}=-u_{F_{ab}}-m$   \\ \hline
    $F_{ba}$ &  $n^{ba}= [u^+_{F_{ba}}]+1$  & $u_{m_{ba}}=u_{F_{ba}}-n+1$ & $u_{m_{ab}}=-u_{F_{ba}}+m-2$   \\ \hline
    $F_{ac}$ & $n^{ac}=\left[-\tfrac{1}{u^+_{F_{ac}}}\right]$ & $u_{m_{ac}}=\tfrac{1}{m-1+1/u_{F_{ac}}}$ & $u_{m_{ca}}=-\tfrac{1}{m+1/u_{F_{ac}}}$  \\ \hline
    $F_{ca}$ & $n^{ca}=\left[\tfrac{1}{u^+_{F_{ca}}}\right]+1$ & $u_{m_{ca}}=-\tfrac{1}{m-1-1/u_{F_{ca}}}$ & $u_{m_{ac}}=\tfrac{1}{m-2-1/u_{F_{ca}}}$ \\ \hline
    $F_{bc}$ & $n^{bc}=\left[\frac{1}{u^+_{F_{bc}}+1}\right]$ & $u_{m_{bc}}=-1-\tfrac{1}{m-1-\tfrac{1}{1+u_{F_{bc}}}}$ & $u_{m_{cb}}=-1+\tfrac{1}{m-\tfrac{1}{1+u_{F_{bc}}}}$ \\ \hline
    $F_{cb}$ & $n^{cb}=\left[\frac{1}{1+1/u^+_{F_{cb}}}\right]$ & $u_{m_{cb}}=-1+\tfrac{1}{m-1-\tfrac{1}{1+u_{F_{cb}}}}$ & $u_{m_{bc}}=-1-\tfrac{1}{m-2+\tfrac{1}{1+u_{F_{cb}}}}$ \\ \hline 
    \end{tabular}
\end{center}
\caption{\label{table4} Epoch Hopscotch}
\end{table}

\subsection{Era hopscotch dynamics}
\begin{figure*}[tb]
\includegraphics[width=15cm]{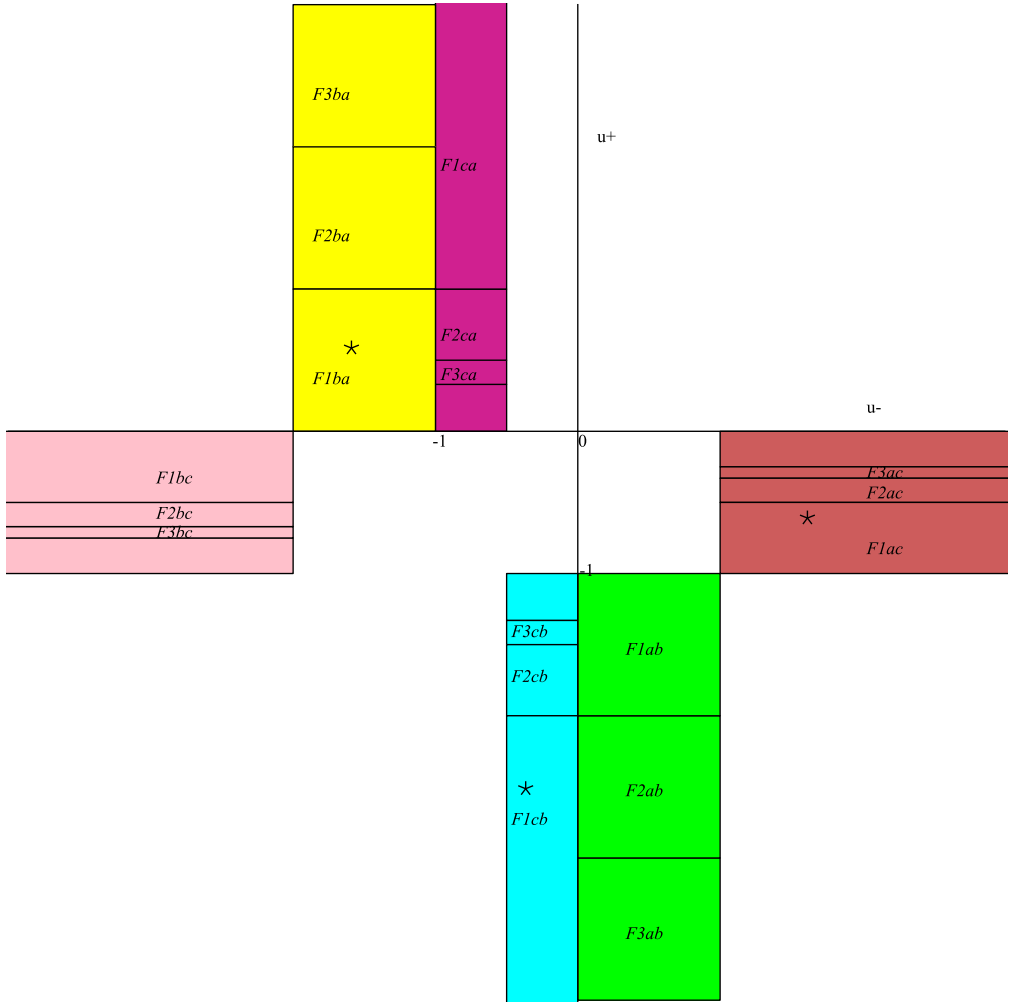}
\caption{\label{biliardone} The era hopscotch court in the ($u^-,u^+$) parametrization. The six starting boxes $F_{xy}$ are sketched, and filled with different colors (shades of gray), according to Fig. \ref{biliardino11}, as explained in Table \ref{table3}. For each starting box $F_{xy}$, the starting sub-boxes corresponding to $k$-epoch eras (as described in Table \ref{table4}) are indicated for $k=1,2,3$; namely $F^1_{xy}$, $F^2_{xy}$ and $F^3_{xy}$. The three ($u^-,u^+$) points of the simplest periodic orbit are denoted as asterisks. Please note that, for typographical reasons, the $F^k_{xy}$'s are in this figure indicated as $Fkxy$.}
\end{figure*}

If, following the spirit of Belinski, Khalatnikhov and Lifshitz, we focus on the discrete dynamics of successive \textit{eras}, we can consider a hopscotch game based on the `era hopscotch court' represented in Fig. \ref{biliardone}, i.e. the six era-starting domains $F_{xy}$ (further divided in sub-boxes labelling the length of the era). The resulting discrete era-transition maps $\mathcal{T}_{\rm era}$ (mapping the ($u^-,u^+$) point of the first epoch of an era to that of the first epoch of the next era) will be obtained by composing the individual epoch-transition maps contained in the considered era. E.g. an era $E_3(a,b)$, i.e. $a\rightarrow b\rightarrow a\rightarrow b$, would correspond to $\mathcal{T}_{era}=B\circ A\circ B$. Depending on the parity of the number $n^{xy}$ of epochs contained in the considered era $E_{n^{xy}}(x,y)$, the era-transition map $\mathcal{T}_{\rm era}$ will map the initial starting rectangle $F_{xy}$ to a \textit{next} starting rectangle, say $F'_{x'y'}$ where the labels $x'$ and $y'$ are fully determined by the knowledge of ($x,y$) and of the parity of $n^{xy}$ (i.e. whether it is even or odd). The explicit rules giving $F'_{x'y'}$ for each $F_{xy}$ are given in the first columns of Table \ref{table5}. In addition, the explicit form of the corresponding era-transition map,
\be\label{5.29}
u^\pm_{F'_{x'y'}}=\mathcal{T}_{n^{xy}}\left(u^\pm_{F_{xy}}\right),
\ee
transforming the phase-space point $u^\pm_{F_{xy}}$of the first epoch in some era $E_{n^{xy}}(x,y)$ into the phase-space point $u^\pm_{F'_{x'y'}}$ of the first epoch in the \textit{next} era $E_{n^{x'y'}}(x',y')$ are explicitly given in the last column of Table \ref{table5}. For instance, if we consider $\mathbf{u}_{F_{ba}}=(u^-_F,\phi)$, $\phi\simeq0.618$ denoting the small golden ratio as above, we shall have (from Table \ref{table4}) $n^{ba}=[\phi]+1=1$, which is odd, so that (from Table \ref{table5}) the next era will be $F'_{ac}$, and the new starting phase-space point in $F'_{ac}$ will have as coordinates (from the last column of Table \ref{table4})
\be\label{5.30}
u^-_{F'_{ac}}=-u^-_F,\ \ u^+_{F'_{ac}}=-\phi.
\ee
\begin{table}
\begin{center}
\begin{tabular}{|l|l|l|l|}

  \hline
  
   $F_{ab}$  & $n^{ab}$ odd   & $F'_{bc}$ & $u_{F_{bc}}=-u_{F_{ab}}-n^{ab}-1$  \\ 
      & $n^{ab}$ even & $F'_{ac}$ & $u_{F_{ac}}=u_{F_{ab}}+n^{ab}$ \\
  \hline
  $F_{ba}$  & $n^{ba}$ odd  & $F'_{ac}$ & $u_{F_{ac}}=-u_{F_{ba}}+n^{ba}-1$  \\
      & $n^{ba}$ even & $F'_{bc}$ & $u_{F_{bc}}=u_{F_{ba}}-n^{ba}$ \\
      \hline
  $F_{ac}$  & $n^{ac}$ odd  & $F'_{cb}$ & $u_{F_{cb}}=-\frac{1}{n^{ac}+1+\tfrac{1}{u_{F_{ac}}}}$  \\
      & $n^{ac}$ even & $F'_{ab}$ & $u_{F_{ab}}=\frac{1}{n^{ac}+\tfrac{1}{u_{F_{ac}}}}$ \\
      \hline
  $F_{ca}$  & $n^{ca}$ odd  & $F'_{ab}$ & $u_{F_{ab}}=\frac{1}{n^{ca}-1-\tfrac{1}{u_{F_{ca}}}}$  \\
      & $n^{ca}$ even & $F'_{cb}$ & $u_{F_{cb}}=-\frac{1}{n^{ca}-\tfrac{1}{u_{F_{ca}}}}$ \\
      \hline
  $F_{bc}$  & $n^{bc}$ odd  & $F'_{ca}$ & $u_{F_{ca}}=-1+\frac{1}{n^{bc}+1-\tfrac{1}{1+u_{F_{bc}}}}$  \\
      & $n^{bc}$ even & $F'_{ba}$ & $u_{F_{ba}}=-1-\frac{1}{n^{bc}-\tfrac{1}{1+u_{F_{bc}}}}$ \\
      \hline
   $F_{cb}$  & $n^{cb}$ odd  & $F'_{ba}$ & $u_{F_{ba}}=-1-\frac{1}{n^{cb}-1+\tfrac{1}{1+u_{F_{cb}}}}$  \\
      & $n^{cb}$ even & $F'_{ca}$ & $u_{F_{ca}}=-1+\frac{1}{n^{cb}+\tfrac{1}{1+u_{F_{cb}}}}$ \\
      \hline 
        
\end{tabular}
\end{center}
\caption{\label{table5} Era Hopscotch}
\end{table}
We have proven above that the $2$-form $\omega$, (\ref{5.19}) was left invariant by each individual wall collision transformation $A$, $B$ or $C$ (and more generally by any symmetry transformation of $\mathcal{H}_2$). Therefore, $\omega$ will be invariant under the era-transition maps $\mathcal{T}_{n^{xy}}$, which are certain compositions of $n^{xy}$ wall-collision transformations, e.g. $\mathcal{T}_{E_3(a,b)}=B\circ A\circ B$.\\
\vspace{0.1cm}

A crucial property of the era-Hopscotch court, Fig. \ref{biliardone}, is that the integral of $\omega$ over the era court is \textit{finite}. Indeed, the points ($0,0$), ($-1,-1$) and ($\infty,\infty$) leading to the logarithmic divergence of the $\omega$-measure of the full epoch hopscotch court of Fig. \ref{biliardino11} are well separated from the six $F_{xy}$ era rectangles of the era hopscotch court. For instance, the region at infinity of the $F_{ba}$ rectangle is $-2<u^-<-1$, $u^+\rightarrow+\infty$, which leads to convergence for $\int\omega(u^-,u^+)$.\\
\vspace{0.1cm}

We are therefore in the usual conditions for applying the results of ergodic theory on a finite-measure space. In other words, after many iterations of the era map $\mathcal{T}_{\rm era}$ we can consider that the phase-space point ($u^-,u^+$) behaves in a stochastic manner, described by a `probability measure'  equal to $\omega/I$, where $I$ is the integral of $\omega$ over the \textit{era} hopscotch court of Fig. \ref{biliardone}. Note that, in the present context of the iteration of a discrete map $\mathcal{T}$, which is ergodic and admits an invariant measure $\omega$, the `probability' measure is $\omega$ itself, and its meaning is that the ratio $\int_{\mathcal{A}}\omega/\int_{\mathcal{D}}\omega$ (where $\mathcal{A}\subset\mathcal{D}$ is a subregion of the full domain $\mathcal{D}$ of the era hopscotch court) yields the $n\rightarrow\infty$ limit of the fraction $n_{\mathcal{A}}/n$ of the number of eras $n_{\mathcal{A}}$ spent in $\mathcal{A}$ among the $n$ first iterates of an arbitrary initial phase-space point. In other words, the word `probability' refers here to a limiting era-frequency. We shall explicitly compute some probabilities in the era-hopscotch dynamics below.
\section{Symmetry-quotienting the big billiard}
As already mentioned, the basic $a$, $b$, $c$ system, Eq. (\ref{abc}), underlying the big billiard dynamics is formally invariant under the six-fold group of permutations of the $3$ letters $a$, $b$, $c$, say $S_3$. This group $S_3$ is the symmetry group of the (ideal) triangle, in the Lobachevsky plane, represented on Fig \ref{fig1}. It comprises (when seen in the disk model) two rotations by $\pm2\pi/3$ (that exchange the corners among themselves), and three reflections with respect to the lines bisecting the corners (that permute two sides among themselves). Note also that the action of the six elements of $S_3$ on the boundary of the disk correspond to the $5$ Kasner transformations given in Table (\ref{table1}), together with identity transformation, say $k_0$ (with $u'=k_0(u)\equiv u$).\\
\vspace{0.1cm}

We can use the symmetry group $S_3$ to \textit{quotient} the dynamics of the big billiard. There are (at least) two ways of thinking about this quotienting. One way would be to consider a \textit{kaleidoscopic} version of the big billiard dynamics in which the single ``moving ball'' of the billiard, shown e.g. in Fig. \ref{fig4}, is augmented by its $5$ (generically distinct) images under $S_3$. This leads to a billiard game where $6$ (symmetry related) balls simultaneously move within the same billiard table, and (simultaneously) bounce on its bounding walls. The phase-space points of this kaleidoscopic billiard is a symmetry orbit of the original (single ball) phase-space point, i.e. an unordered set of (at most) six (two-dimensional) $q^i$'s and six (two-dimensional) $p_i$'s (restricted by the condition $g^{ij}p_ip_j=1$). [Some of these variables are allowed to coincide when the original ball crosses one, or several, of the fixed sets of $S_3$ (bisecting lines).].\\
A second way to look at the quotiented dynamics is to replace the latter kaleidoscopic phase-space point\\ $\{q^i_{(1)},q^i_{(2)}, ..., q^i_{(6)}; p_i^{(1)}, p_i^{(2)}, ..., p_i^{(6)}\}$ by its unique representative, say $q^i_{\rm rep}$, within a fundamental domain of $S_3$, together with its corresponding momenta $p^i_{\rm rep}$. For instance, one can use as fundamental domain the \textit{small billiard}, with sides $G, B, R$ in Fig. \ref{fig2}. [We shall use this relation when considering the small billiard dynamics below.] When passing from the continuous billiard dynamics to the discrete billiard map from an epoch to the next epoch, the quotienting of the big billiard leads to a quotiented version of the epoch hopscotch game of Fig. \ref{biliardino11}. For instance, the kaleidoscopic version of Fig. \ref{biliardino11} would replace each ($u^-,u^+$) point indicated there by six symmetry-related points, i.e. $u^\pm$ and its five transforms under the Kasner transformations of Table \ref{table1}, namely $k_1(u^\pm)$, $k_2(u^\pm)$, $k_3(u^\pm)$, $k_4(u^\pm)$, $k_5(u^\pm)$. These six points would then simultaneously ``jump'', after a (six-fold) collision on  a gravitational wall, to their next six-fold positions in the ($u^-, u^+$) plane. For instance, the $S_3$ orbit $\{1^I, 1^{II}, 1^{III}, 1^{IV}, 1^V, 1^{VI}\}$ of the point $1$ in $B_{ba}$ (Fig. \ref{biliardino11}) would jump onto the new $S_3$ orbit $\{2^I, 2^{II}, 2^{III}, 2^{IV}, 2^V, 2^{VI}\}$. In the alternative, fundamental-domain, version of the quotiented dynamics  we could replace each $S_3$ orbit in the ($u^-,u^+$) plane by its unique representative located within, say, the box $B_{ba}$. [Indeed, the six boxes $B_{xy}$ of Fig. \ref{biliardino11} are exchanged under $S_3$.] In that view, the discrete quotiented big billiard would become a map from the box $B_{ba}$ onto itself. For instance, in the example shown in Fig. \ref{biliardino11}, the initial point $1$ would first jump to the point $2'$ (midway between the points $1$ and $3$), then to a point $4'$ (midway between $2$ and $5$, and belonging to $B_{ba}$). The next epoch would be the image of the point $5\in B_{bc}$ which belongs to $B_{ba}$. As suggested by our description of the example of Fig. \ref{biliardino11}, one finds that each era gets quotiented into a succession of ($u^-, u^+$) representative points within $B_{ba}$ which lie on a straight (Euclidean) line of slope $+1$ (as the line passing through the points $1$ and $3$ in Fig. \ref{biliardino11}). More precisely, if we denote by ($u^-_F,u^+_F$)$\in B_{ba}$ the first epoch of a quotiented era, the $B_{ba}$-representative of the version of the considered era is made of the points 
\be\label{6.1new}
(u^-_F,u^+_F);\ \ (u^-_F-1,u^+_F-1);\ \ (u^-_F-2,u^+_F-2);\ \ ...;\ \ (u^-_F-[u^+_F],u^+_F-[u^+_F])
\ee
obtained by successively subtracting $1$ both from $u^-_F$ and $u^+_F$. As indicated, the length of the era is simply given by $k=[u^+_F]+1$, so that the last point of an era is reached when its $u^+$ coordinate is between$0$ and $1$: $u^+_L=u^+_F-[u^+_F]\equiv \{u^+_F\}$. Note, however, that the $u^-$ coordinate of the last epoch is given by $u^-_L=u^-_F-k+1=u^-_F-[u^+_F]$, so that it depends both on $u^-_F$ and on $u^+_F$ (while the sequence of the $u^+$ values depends only on the starting value of $u^+_F$ of $u^+$). Then, it is easily seen that the next epoch (i.e. the first point of the next era) will be (when mapped back to $B_{ba}$ by $S_3$)
\be\label{6.2new}
\left( \frac{1}{u^-_F-[u^+_F]}-1,\ \ \frac{1}{u^+_F-[u^+_F]}-1 \right).
\ee 
If we ignore the $u^-$ coordinate, we see that the law giving the successive values of the $u^+$ coordinate coincides with the law found long ago by BKL (when using the BKL$_{u>0}$ definition of an era, as discussed above) namely: $u^+_F=x+k-1$, $u^+_F-1=x+k-2$, down to $u^+_L=x$, with $k=[u^+_F]+1$ denoting the length of the era. This shows that the BKL discrete dynamics of the variable $u$ is obtained by: (i) quotienting our more complete hopscotch dynamics by the permutation group $S_3$, and (ii) ignoring the $u^-$ coordinate and identifying the BKL variable $u$ with $u^+$. Note again that this link between the hopscotch dynamics and the BKL dynamics is particularly simple if one uses the BKL$_{u>0}$ definition of an era, rather than the BKL$_{u>1}$.\\
The $S_3$-quotiented hopscotch dynamics, i.e. the discrete dynamics mapping $B_{ba}$ onto itself, defined by Eqs. (\ref{6.1new}) and (\ref{6.2new}), constitutes a two-variable generalization of the BKL map, say
\be\label{6.3new}
(u^-, u^+)\in B_{ba}\rightarrow (u^{- '}, u^{+'})=\mathcal{T}_{ba} (u^-, u^+)\in B_{ba}.
\ee
Like the full (unquotiented) epoch hopscotch dynamics, the quotiented discrete map, $\mathcal{T}_{ba}$, (\ref{6.3new}) leaves invariant the measure $\omega$ However, like in the unquotiented case, the integral of $\omega$ on the domain (and image) $B_{ba}$ of the map
$\mathcal{T}_{ba}$ is infinite.\\
\vspace{0.1cm}

Before discussing the obtention of a finite-measure discrete map associated with $\mathcal{T}_{ba}$, let us note that, as we did for the unquotiented hopscotch dynamics, we can also consider the action of the quotiented billiard map $\mathcal{T}_{ba}$ on the single (BKL-like) variable $u^+$. It is defined as
\be
u^+_F\rightarrow u^+_F-1\rightarrow ...\rightarrow u^+_F-[u^+_F]\rightarrow\frac{1}{u^+_F-[u^+_F]}-1\rightarrow ...
\ee
In other words, it is just the usual BKL map on $u$ recalled in the Introduction (in its BKL$_{u>0}$ version). Note the appearance of $-1$ in the definition of the new ``first $u^+$'' of the next era. This takes into account that the unordered triplet of Kasner exponents of $\{u^+_F\}<1$ is identical to that of $\{u^+_F\}^{-1}$ and should therefore not appear twice in the sequence of transforms of $u^+$.\\
As in the unordered case discussed above, we can obtain an invariant measure for the one-dimensional version, say $\mathcal{T}_{ba}^{(1)}$ of the quotiented billiard map by \textit{marginalizing} over $u^-$ the two-dimensional invariant measure $\omega_{ba}$ (i.e. the restriction of $\omega$ to the region $B_{ba}$). We use the same definition as above, Eq. (\ref{1dimmeasure}). The difference is that, now, $u^+$ is restricted to be in the interval $0<u^+<+\infty$, and the boundaries of integration over $u^-$ are $u^-_{\rm min}=-\infty$ and $u^-_{\rm max}=-1$ (where $u^-_{\rm max}$ differs from its previous value). This leads to the invariant one-dimensional measure
\be\label{omegaba}
w_{ba}(u^+)du^+=\frac{du^+}{u^++1}.
\ee
It is again easy to check directly that $w_{ba}(u^+)du^+$ is invariant (in the sense recalled in Section \ref{reducedforms} and in the Appendix) under the discrete map $\mathcal{T}_{ba}^{(1)}$ [one needs to take into account that $\mathcal{T}_{ba}^{(1)}$ is two-to-one so that the preimage of an infinitesimal interval $u^+\pm\tfrac{1}{2}du^+$ consists of \textit{two} infinitesimal intervals.] Note that if, instead of considering the definition above of $\mathcal{T}_{ba}^{(1)}$, one considers the ``standard'' BKL$_{u>1}$ map (where $u$ remains always $>1$ and decreases by units of $1$ until the value $1+\{u\}$, after which it jumps to $1/\{u\}$), the expression of the invariant measure reads
\be\label{omegabkl}
w_{BKL_{u>1}}(u)du=\frac{du}{u}
\ee
As far as we know, the results in Eqs. (\ref{omegaba}) and (\ref{omegabkl}) have not been explicitly discussed before in the literature.\\
It should be noted that the one-dimensional measure $w_{ba}(u^+)du^+$ differs from the restriction to $B_{ba}$ of its unquotiented analog discussed in  Section \ref{bigbilliard} above. [Indeed, we had before $w(u^+)du^=(du^+/u^+$ when $u^+>0$.] This is an effect of the quotienting which means that one must fold back onto $B_{ba}$ the symmetric images that were elsewhere (and notably in $B_{ca}$). Finally, we note that, as before, the invariant measure $w(u^+)du^+$ is not normalizable. Indeed, it diverges logarithmically when $u^+\rightarrow+\infty$. Note, however, that it converges at the lower boundary $u^+\rightarrow0$. [The same is true of the invariant measure $w_{{\rm BKL}_{u>1}}(u)du=1/u$ of the standard BKL $u$-map, with $1<u<+\infty$].\\
\vspace{0.1cm}

Let us now come back to discussing the \textit{two-dimensional} quotiented map $\mathcal{T}_{ba}$, acting on $B_{ba}$. To get a discrete map preserving a finite measure, we need to consider the quotiented analog of the \textit{era} hopscotch dynamics. In fact, we obtain an \textit{era} hopscotch dynamics simply by ignoring the intermediate epochs and focusing on the map transforming the (quotiented) first epoch of an era $(u^-_F, u^+_F)$ into the first epoch $(u^{- '}_F, u^{+'}_F)$ of the next era. We shall denote this quotiented era map as $\mbox{\large{$\mathsf{T}$}}$. When using, as we did above, a representative of the $S_3$ orbit within the $B_{ba}$ box, the quotiented era map $\mbox{\large{$\mathsf{T}$}}$ is a map of $F_{ba}$ onto itself. We recall that $F_{ba}$ is the domain of the first points of eras starting as $b\rightarrow a$. It is the rectangular domain $-2<u^-_F<-1$, $0<u^+_F<+\infty$ (see Fig. \ref{biliardone}). The explicit expression of the map $\mbox{\large{$\mathsf{T}$}}$ is given by Eq. (\ref{6.2new}), i.e.
\begin{table}
\begin{center}
    \begin{tabular}{ | l | l | }
    \hline
    $k_5$ & $u_K^{ba}=-u_{ab}-1$ \\ \hline
    $k_2$ & $u_K^{ba}=-(u_{ac}+1)/u_{ac}$ \\ \hline
    $k_3$ & $u_K^{ba}=-u_{bc}/(u_{bc}+1)$ \\ \hline
    $k_1$ & $u_K^{ba}=1/u_{ca}$ \\ \hline
    $k_4$ & $u_K^{ba}=-1/(u_{cb}+1)$ \\ \hline  
    \end{tabular}
\end{center}
\caption{\label{table6} The Kasner maps for each starting box of the big billiard table}
\end{table}
\be\label{cblksks}
\mbox{\large{$\mathsf{T}$}}u_F^\pm=+\frac{1}{u_F^\pm-[u_F^+]}-1.
\ee
This map leaves invariant the \textit{restriction} of the two-form $\omega$ to the domain $F_{ba}$, i.e.
\be\label{omegaf}
\omega_F=2\frac{du^+_F\wedge du^-_F}{(u^+_F-u^-_F)^2}.
\ee
By contrast with the original measure $\omega$ on the full hopscotch court, this restricted measure has now a \textit{finite integral}, namely
\be\label{intomegaf}
\int_{\omega_{F_{ba}}}\omega_F=2\ln2
\ee 
We shall also refer to the map $\mbox{\large{$\mathsf{T}$}}$ as being the \textit{Chernoff- Barrow- Lifshitz- Khalatnikov- Sinai- Khanin- Shchur map}  for the big billiard, or CB-LKSKS map in short. This one-to-one map between two variables was introduced in \cite{Chernoff:1983zz}\footnote{The definition of the map in \cite{Chernoff:1983zz} and our definition match when one relates the variable denoted as $x$ in \cite{Chernoff:1983zz}, which we shall call $X_{CB}$, with our $u^-_F$ by $X_{CB}=-1/(u^-_F+1)$. For the sake of completeness, let us also note that the names of the Kasner exponents $p_1$ and $p_2$ are exchanged in \cite{Chernoff:1983zz}.}, while Ref. \cite{sinai83} and \cite{sinai85} showed how such a two-variable map appears as a completion of the original BKL analysis, when keeping not only the original BKL variable $u^+$ (parametrizing $p_1$, $p_2$ and $p_3$), but also a variable $\delta$ related to the amplitude of oscillations of the $a$, $b$, $c$ metric variables during an era (the precise relation between our variables and those used in \cite{sinai83} and \cite{sinai85} will be given below). The CB- LKSKS map is one-to-one over its domain $F_{ba}$. Note, however, that the original map $\mathcal{T}$ on the era hopscotch court, whose quotienting leads to $\mbox{\large{$\mathsf{T}$}}$, is such that both the image, $\mathcal{T}F_{ba}$, and the pre-image $\mathcal{T}^{-1}F_{ba}$, of $F_{ba}$ are the union of an infinite number of rectangular domains belonging to two different $F_{xy}$ boxes. For instance, the image $\mathcal{T}F_{ba}$ is of the form $P_{bc}\cup P_{ac}$, where $P_{bc}$ is the union of an infinite number of disconnected rectangles contained within $F_{bc}$, while $P_{ac}$ is the union of an infinite numbers of disconnected rectangles contained within $F_{ac}$.\\
\vspace{0.1cm}

Note that $u^+_F$ and $u^-_F$ play asymmetric roles in the $\mbox{\large{$\mathsf{T}$}}$ map. Indeed, for all the starting boxes $F_{xy}$, the boundaries of the subdomains corresponding to a given era-length are \textit{horizontal} (see Fig. \ref{biliardone}). Therefore, whether a given rectangle in $F_{ba}$ is $\mbox{\large{$\mathsf{T}$}}$-mapped into one or several rectangles only depends on the range of $u^+_F$ \textit{independently of the range of} $u^-_F$. As a consequence, the `mixing' character of the $\mbox{\large{$\mathsf{T}$}}$ map is essentially contained in the $u^+$ direction. Actually, we see by differentiating Eq. (\ref{cblksks}) that
\be\label{139}
d\left(\mbox{\large{$\mathsf{T}$}}u^-_F\right)=-\frac{du^-_F}{(u^-_F-[u^+_F])^2}.
\ee
As $-2\le u^-_F\le-1$ and $[u^+_F]\in\mathbb{N}$, the denominator $(u^F-[u^+_F])^2$ is always strictly larger than one. Therefore, $\left| d\mbox{\large{$\mathsf{T}$}}u^-_F\right|/du^-_F<1$, i.e. the $\mbox{\large{$\mathsf{T}$}}$ map is \textit{contracting} in the $u^-_F$ direction.\\
\vspace{0.1cm}

Finally, let us clarify the link between the CB-LKSKS map as defined above and the unit-square map given in \cite{sinai83} and \cite{sinai85}. The range of the variables $u^-, u^+$ in the above-defined CB-LKSKS map is the infinite vertical rectangle $F_{ba}$, i.e. $-2<u^-<-1$, $0<u^+<+\infty$. By contrast, the statistical analysis developed in \cite{sinai83} and \cite{sinai85} is based on two variables $x_+, x_-$, defined in the unit square, i.e. $0<x_+<1$, $0<x_+<1$.\\
Let us consider the following transformation mapping $F_{ba}$ into the unit square $0<x_+<1$ and $0<x_-<1$:
\begin{subequations}\label{ust}
\begin{align}
&x_+=\frac{1}{u^+_F+1}\\
&x_-=-u^-_F-1.
\end{align}
\end{subequations}
In terms of the unit-square variables ($x_-,x_+$), Eqs. (\ref{ust}), the CB-LKSKS map reads 
\begin{subequations}\label{usr1}
\begin{align}
&\mbox{\large{$\mathsf{T}$}}x_+=\left\{\tfrac{1}{x_+}\right\},\\
&\mbox{\large{$\mathsf{T}$}}x_-=\frac{1}{x_-+\left[\tfrac{1}{x_+}\right]},
\end{align}
\end{subequations}
consistently with the results of \cite{sinai85}. As the unit-square transformation in (\ref{ust}) does not consist of applying the \textit{same} fractional linear transformation to \textit{both} variables $u^+$ and $u^-$, the two-form $\omega$ is not invariant under (\ref{ust}), but becomes
\be\label{usr2}
\omega_F=2\frac{dx_+\wedge dx_-}{(1+x_+x_-)^2}.
\ee
The choice of $B_{ba}$ (and $F_{ba}$) as representative boxes for the quotiented big billiard dynamics does not affect the unit-square results (\ref{usr1}) and (\ref{usr2}). Had we considered another representative box $B_{xy}$, and defined a correspondingly modified version of the unit-square transformation (\ref{ust}), we would have ended up with the same results (\ref{usr1}) and (\ref{usr2}).  
\section{Some properties of the symmetry-quotiented dynamics}
In this section, we shall start from the $F_{ba}$-box version of the CB-LKSKS map (or, simply, $\mbox{\large{$\mathsf{T}$}}$ map), Eq. (\ref{cblksks}), and study some of its properties, recalling, when needed, some results obtained by previous authors and focusing on new results.
\subsection{Probabilities}
Let us indicate how, in our set-up, one can compute the probability for an era to contain a given number of epochs.\\
The probability $P_{n_1}$ for an era to contain a number $n_1$ of epochs, $1<n_1<+\infty$, is proportional to the integral of the two-form $\omega_F$ (\ref{omegaf}) over the relevant region of the $u^-_F,u^+_F$ space. In the considered case of a given number, $n_1$, of epochs, the relevant region is simply the box $F^{n_1}_{ba}$, as defined above, i.e. the domain $-2<u^-_F<-1$, $n_1-1<u^+_F<n_1$. To normalize this probability, we must then divide by the integral of $\omega_F$ over the full domain $F_{ba}$, as given in Eq. (\ref{intomegaf}), $2\ln2$. As a result, we obtain that the probability $P_{n_1}$ for an era to contain a number $n_1$ of epochs as
\be\label{BKLprob}
P_{n_1}=\frac{1}{\ln2}\int^{n_1}_{n_1-1}du^+_F\int^{-1}_{-2}du^-_F\frac{1}{(u^+_F-u^-_F)^2}=\frac{1}{\ln2}\ln\frac{(n_1+1)^2}{n_1(n_1+2)}.
\ee

This agrees with the result of BKL (which was obtained from the stationary probability distribution (\ref{2'}) of the Gauss iteration map).\\ 
For instance, the probabilities for the length of an era to take the values $n_1=1,2,3,4,5$ are
\begin{subequations}
\begin{align}
&P_1=0.4150\\
&P_2=0.1699\\
&P_3=0.0931\\
&P_4=0.0589\\
&P_5=0.0406,
\end{align}
\end{subequations}
whose sum is $\sum_{k=1}^5P_k=0.7775$. Hence, $77.75\% $ of the eras have lengths smaller or equal to $5$, and actually $58.49\% $ of the eras have lengths smaller or equal to $2$. This shows that most eras have rather small lengths. The asymptotic behaviour of $P_{n_1}$, as $n_1\rightarrow+\infty$ is
\be
P_{n_1}\simeq\frac{1}{cn_1^2},
\ee
with $c=\ln2\simeq0.69315$. Therefore, the probability to have $n_1\ge N_1$ is asymptotically given (for large $N_1$) by $P(n_1\ge N_1)\simeq\tfrac{1}{cN_1}$, which decreases rather slowly as $N_1$ increases. In other words, though most eras have rather small lengths, from time to time eras with unbounded lengths can arise. The rather slow decrease of $P_{n_1}$ as $n_1$ increases implies, in particular, that the mean value of $n_1$ is \textit{infinite} (being given by the logarithmically divergent series $\sum_{n_1\ge1}P_{n_1}n_1$). Note, however, that a finite result is obtained if one considers the expectation value of the \textit{geometric mean} of large sequences of independent era lengths, i.e. the exponential of the expectation value of $\ln n_1$ (known as the Khinchin number). This yields \cite{shanks1959}
\be
\exp\sum_{n_1\le1}P_{n_1}\ln n_1\simeq2.6854...
\ee
This result confirms that the ``typical'' length of an era is rather small.\\
\vspace{0.1cm}

Let us now consider the computation of a probability of a more specific event (not explicitly considered by BKL): namely the probability
 $P_{n_1,n_2}$ for an era of length $1<n_1<\infty$ to be followed by an era of length $1<n_2<\infty$. This probability is obtained by integrating the form  (\ref{5.19}) over the appropriate range of the variables $u^+_F, u^-_F$. This range is determined by Table (\ref{table4}) (as in the previous case), and by the properties of the CB-LKSKS map: combining the two information, the range $n_1-1+\tfrac{1}{n_2+1}<u^+_F<n_1-1+\tfrac{1}{n_2}$ is obtained. The range of $u^-_F$ is determined as in the previous case. As a result, we obtain
\be\label{pn1n2}
P_{n_1,n_2}=\tfrac{1}{\ln2}\int_{-2}^{-1}du^-_F\int^{n_1-1+\tfrac{1}{n_2}}_{n_1-1+\tfrac{1}{n_2+1}}\frac{du^+_F}{(u^+_F-u^-_F)^2}=  \tfrac{1}{\ln2}\ln\left(\tfrac{(n_1n_2+1)(n_1n_2+n_1+n_2+2)}{(n_1n_2+n_1+1)(n_1n_2+n_2+1)}\right).
\ee
Note that this probability is symmetric in $n_1$ and $n_2$. The previous probability $P_{n_1}$ is (as it should) recovered by summing $P_{n_1,n_2}$ over all the values of $n_2$.
\be\label{usualbklprob}
P_{n_1}=\tfrac{1}{\ln2}\sum_{n_2=1}^{n_2=\infty}P_{n_1,n_2}=\tfrac{1}{\ln2}\ln\left(\tfrac{(n_1+1)^2}{n_1(n_1+2)}\right).
\ee
Note that in both cases we were considering events that depend only on $u^+_F$ so that the result involved marginalizing the variable $u^-_F$, i.e. integrating the (normalized) $\omega$ form over the complete range of $u^-_F$. And, indeed, integrating the (normalized) two-form $\omega_F$ over the complete range of $u^-_F$ ($-2<u^-_F<-1$) yields the one-form
\be\label{150}
\frac{1}{\ln2}\int^{-1}_{-2}\frac{du^-_Fdu^+_F}{(u^+_F-u^-_F)^2}=\frac{1}{\ln2}\frac{du^+_F}{(u^+_F+1)(u^+_F+2)}
\ee
which yields the Gauss distribution (\ref{2'}), i.e. $w(x)dx=\tfrac{1}{\ln2}\tfrac{dx}{1+x}$ when parametrizing $u^+_F\in\rbrack0,+\infty\lbrack$ by $x\in\rbrack0,1\lbrack$, such that $u^+_F\equiv\tfrac{1}{x}-1$ (with $0<x<1$).\\
Note that the (integrable) invariant one-dimensional measure (\ref{150}) differs from the (non-integrable) invariant one-dimensional measure (\ref{omegaba}): while the former refers to the discrete dynamics of the $u^+$ value of the \textit{first} epoch era, the latter refers to the discrete dynamics of the $u^+$ values of \textit{all the epochs}. Note also that the probability $P_{n_1}$ to have an era of length $n_1$ is obtained (in view of the link $n_1=[u^+_F]+1$) by integrating the measure (\ref{150}) over the interval $n_1-1<u^+_F<n_1$.\\ 
It is interesting to note that the (pseudo-)random variables $n_1$ and $n_2$ (i.e. the lengths of two consecutive eras) are \textit{not} independent of each other, because $P_{n_1n_2}\neq P_{n_1}P_{n_2}$. However, the variables $n_1$ and $n_2$ are \textit{approximately} independent statistical variables. Indeed, using the explicit expression (\ref{BKLprob}) and (\ref{usualbklprob}) one finds that the ratio $R_{n_1n_2}\equiv P_{n_1n_2}/P_{n_1}P_{n_2}$, which would be, by definition, equal to one if $n_1$ and $n_2$ were independent random variables, takes values rather close to $1$. For instance, $R_{11}\simeq0.8826$, $R_{12}=R_{21}\simeq0.9985$, $R_{13}=R_{31}\simeq1.051$ and $R_{22}\simeq1.008$. Therefore the low values of $n_1$ and $n_2$ (which are of greatest importance for many issues) are approximately independent. As concerns the large values of $n_1$ and $n_2$, let us note that the asymptotic value of $P_{n_1n_2}$ is
\be
P_{n_1n_2}\simeq\frac{1}{cn_1^2n_2^2},
\ee
where $c=\ln2$ as above. This implies that the ratio $R_{n_1n_2}\equiv P_{n_1n_2}/P_{n_1}P_{n_2}$ is asymptotically constant, and equal to $c\approx0.69315$. Having seen that the lengths of \textit{consecutive} eras are approximately independent random variables, we expect that such an independence property will become more and more exact as one considers eras that are more and more separated. 
\subsection{Continued fractions}
Let us briefly recall (from \cite{BLK1971}) the usefulness of the continued-fraction representation of the variables $u^-,u^+$ in describing the effect of iterating the CB-LKSKS map.\\
Any number $0<y<+\infty$, can be uniquely decomposed as
\be\label{132}
y=[y]+\{y\},
\ee
where $[y]$ is its integer part, while $\{y\}$ is its fractional part. This decomposition can be iterated by considering the decomposition
(\ref{132}) of $1/\{y\}$. This leads to the unique continued-fraction decomposition of $y>0$
\be\label{cfd}
\{y\}= n_1+\frac{1}{n_2+\frac{1}{n_3+\frac{1}{...}}}\equiv[n_1;n_2, n_3, ...],
\ee
where $n_1, n_2, n_3, ...$ are natural integers. In Eq. (\ref{cfd}) we have introduced a notation for the continued-fraction expansion which distinguishes (by means of a semi-colon) the first integer $n_1\equiv[y]$. In the case where $n_1=0$ (i.e. in the case $0<y<1$, i.e. $y=\{y\}$), we shall also use the notation $\{y\}=[n_2, n_3, n_4, ...]$ (without semi-colon). The continued-fraction expansion contains a finite sequence of integers $n_1, n_2, ...$ if $y$ is rational, while it contains an infinite sequence of integers if $y$ is irrational.\\
\vspace{0.1cm}

With this notation, the continued-fraction expansions\footnote{Here, we adopt the definition of integer part of a negative number given in Footnote \ref{fn}, and we define the fractional part of a negative number accordingly, e.g. $\{-\pi\}=-0.14...$.}  of the first ($u^-,u^+$) values of a (quotiented) era starting (as above) in the box $F_{ba}$ can be written as
\begin{subequations}\label{contfrac}
\begin{align}
&u^+_F\equiv n_{ba}-1+ [ n_2,n_3,n_4,... ] \equiv [n_{ba}-1; n_2,n_3,n_4,...],\\
&u^-_F=-1-[m_1, m_2, m_3, ...].
\end{align}
\end{subequations}
Here, we have denoted the first integer of the decomposition of $u^+_F$ as $n_{ba}-1=[u^+_F]$, so that $n_{ba}=[u^+_F]+1$ denotes the length of the era starting with $u^+_F$. In terms of these decompositions pertaining to the first era, we can write the first $u^-,u^+$ values of the $N$-th era (with $N=2,3,...$), i.e. the ($N-1$)-th iteration of the $\mbox{\large{$\mathsf{T}$}}$ map
\begin{subequations}\label{137}
\begin{align}
&\mbox{\large{$\mathsf{T}$}}^{N-1}u^+_F=n_N-1+[n_{N+1}, n_{N+2}, n_{N+3}, ...],\\
&\mbox{\large{$\mathsf{T}$}}^{N-1}u^-_F=-1-[n_N, n_{N-1}, ..., n_2, n_{ba}, m_1, m_2, ...].
\end{align}
\end{subequations}
In other words, at each iteration of the CB-LKSKS map, the information about the length of the corresponding era is `transferred' from the $u^+$ variable to the $u^-$ variable. Note that the information contained in $u^+_F$ (i.e. the sequence of integers $m_1, m_2,...$) is progressively decaying in the ``tail'' of the iterates of $u^-_F$. By contrast, the iterates of $u^+_F$ progressively uncover the ``tail'' of $u^+_F$, thereby exhibiting the chaotic character of the $u^+_F$ dynamics which progressively amplifies smaller and smaller details of the continued-fraction expansion of the initial $u^+_F$.
\subsection{\label{recoveringinformation}Recovering information about the unquotiented dynamics}
Up to now, in this Section, we have discussed the quotiented dynamics (viewed within the representative box $B_{ba}$, or $F_{ba}$ when considering the first epoch of an era). This quotienting has ignored the fuller information contained in the original, unquotiented era hopscotch dynamics, namely the precise \textit{corner}, $\{x_N, y_N\}$ (with $x_N, y_N \in\{a, b, c\}$) and \textit{orientation}, ($x_N, y_N$), (i.e. $x_N\rightarrow y_N$), of the first epoch of the $N$-th hopscotch era. For each starting box, say $F_{xy}$, the specific Kasner transformation mapping the region $F_{xy}$ into the representative region $F_{ba}$ is given in Table \ref{table6}. Therefore, if we start an unquotiented era hopscotch dynamics in some specific region $F_{xy}$, we can first map it to the reference region $F_{ba}$ by some specific Kasner transformation $k_{xy}$ to get its $ba$-representative, $u^{+[ba]}_{F_{xy}}$, namely
\be
u^{+[ba]}_{F_{xy}}\equiv k_{xy}u^+_{F_{xy}}.
\ee
Then starting from $[u^+_{F_{xy}}]_{ba}$ we can define its continued-fraction decomposition, say
\be
u^{+[ba]}_{F_{xy}}=k_{xy}u^+_{F_{xy}}\equiv n_{xy}-1+[n_2, n_3, ...].
\ee
Here, the so-defined integers $n_{xy}, n_2, n_3, ...$ give us the values of the successive lengths of all the eras that will evolve from the initial value $u^+_{F_{xy}}$.\\
\vspace{0.1cm}

This reasoning shows how the knowledge of any era-starting values ($u^-_F, u^+_F$) in the ($u^-, u^+$) plane determine the subsequent era-length history. First, the location of $u_F$ in the plane determines the initial corner $xy$. Second, the knowledge of the initial corner uniquely determines $k_{xy}$ (mapping it to $F_{ba}$). And, third, the computation of $k_{xy}u^+_{F_{xy}}$ and of its continued-fraction expansion, determines all the era-lengths $n_{xy}, n_2, n_3, ...$. However, we need more information if we wish to recover the full hopscotch dynamics from the simpler quotiented hopscotch dynamics discussed above. Specifically, we need to recover information about: (i) what is the succession of \textit{corners} (among the three corners of the big billiard) that will be visited, and (ii) what is the succession of the \textit{directions} (clockwise or counter-clockwise) in which the oscillations within these corners will take place. Let us now show how one can recover this missing information from the knowledge of the era-starting values ($u^-_{F_{xy}}, u^+_{F_{xy}}$) in the ($u^-, u^+$) plane.\\
\vspace{0.1cm}

Before doing so, let us establish some notation. Looking at Fig. \ref{fig2}, we shall say that an era is \textit{clockwise} if its \textit{first epoch} connects two billiard walls in the clockwise sense with respect to the unit disk, i.e. if it is either $ba$, $ac$ or $cb$. In the other case (first epoch of the type $ab$, $bc$ or $ca$) we shall say that the era is \textit{counterclockwise}.\\
Now, we remark that the information about this direction of motion is contained in the determinant of the Kasner transformation $k_{xy}$ that maps each era-starting box $F_{xy}$ onto the reference box $F_{ba}$ (as given in Table \ref{table6}). More precisely, if the determinant of $k_{xy}$, say $D[k_{xy}]$ is equal to $+1$, the era $F_{xy}$ is clockwise, while if $D[k_{xy}]$ is equal to $-1$, the era is counter-clockwise. Given two era-starting regions $F_{xy}$ and $F_{x'y'}$, it will sometimes be convenient to say that they are ``parallel'' if they have the same direction of motion (clockwise or counter-clockwise), and ``antiparallel'' in the other case.\\
\vspace{0.1cm}

Le us now show how one can encode the information which is missing in the quotiented billiard in a pair $\rho, \eta$, where $\rho$ takes three different values, and $\eta$ two different ones. In more mathematical terms, $\rho\in Z_3$ and $\eta\in Z_2$, where $Z_n$ denotes the cyclic (multiplicative) group of order $n$. The values of $\rho$ can be $\{1, \exp(i2\pi/3), \exp(-i2\pi/3)\}$, and are encoding rotations in the disk model of the billiard by the angles $0$, $2\pi/3$ or $-2\pi/3$ respectively. The values of $\eta$ are $\{+1, -1\}$ and can encode the two possible ``directions of motion'' of an era (clockwise or counter-clockwise).\\
\vspace{0.1cm}

Our aim is, starting from some initial (era-starting) position ($u^-_{F_{xy}}, u^+_{F_{xy}}$)$\in F_{xy}$, to determine the \textit{ordered corner} $x_N, y_N$ within which the $N$-th unquotiented $\mathcal{T}$ iterate, $\mbox{\large{$\mathsf{T}$}}^N(u^-_F,u^+_F)$ of $(u^-_F,u^+_F)$ will oscillate (i.e. the first epoch of the $(N+1$-th era). We parametrize the ordered corner $x_N, y_N$ by the pair $\rho_N, \eta_N$ where $\rho_N$ is the rotation, in the disk model, mapping the initial (unordered) corner $\{x,y\}$ into $\{x_N,y_N\}$, and where $\eta_N$ gives us the \textit{relative} orientation between $x_N,y_N$ and $x,y$. (i.e $\eta_N=+1$ if they are parallel, and $\eta_N=-1$ if they are antiparallel).\\
One can iteratively build the values of $\rho_N$ and $\eta_N$ by using the following elementary facts:
\begin{itemize}
	\item an (intermediate) era $F_{x'y'}$ containing an odd (respectively, even) number of epochs is followed by an era $F_{x''y''}$ whose relative direction of motion is parallel (resp., antiparallel);
	\item the rotation (in the disk model) between some intermediate era $F_{x'y'}$ and the following $F_{x''y''}$ is equal to $e^{+i2\pi/3}$ (resp. $e^{-i2\pi/3}$) if $F_{x'y'}$ contains an even (resp., odd) number of epochs.
\end{itemize}
To exhibit the iterated effect of these elementary rules, it is convenient to define the following quantities (taking the values $\pm1$):
\be\label{epsilonj}
\epsilon_j\equiv(-)^{n_j+1},
\ee
where $n_j$ denotes the number of epochs contained in the $j$-th era defined by the initial value of $u^+_F$. With this notation, the $Z_3\times Z_2$ valued pair $(\rho_N, \eta_N)$ giving the rotation (with respect to the original era-starting domain $F_{xy}$) and the relative ``sense of motion'' (parallel or antiparallel to $xy$) corresponding to the ordered corner of $\mathcal{T}^N(u^-_F, u^+_F)$ is defined by (with $D[k_F]=\pm1$ denoting as above the determinant of the Kasner transformation $k_F$ mapping $(u^-_F, u^+_F)$ to $F_{ba}$)
\begin{subequations}\label{rhoeta}
\begin{align}
&\rho_N=e^{i\theta_N},\ \ {\rm with}\ \ \theta_N\equiv-D[k_F]\tfrac{2\pi}{3}\left(\epsilon_1+\epsilon_1\epsilon_2+...+\epsilon_1\epsilon_2...\epsilon_N\right),\\
&\eta_N=\epsilon_1\epsilon_2...\epsilon_N.
\end{align}
\end{subequations} 
Note that $\eta_N$ can be rewritten as $\eta_N=(-)^{Q_N}$, where
\be
Q_N=N+\sum_{k=1}^Nn_k.
\ee
Note also that the absolute sense of motion (clockwise or not) of the $N$-th era is given by
\be
D[k_F]\eta_N=D[k_F]\epsilon_1\epsilon_2...\epsilon_N.
\ee
\vspace{0.1cm}

Given this result, we can now write, for any starting point $(u^-_F, u^+_F)$, the explicit result of iterating $N$ times the unquotiented $\mathcal{T}$ map, i.e. 
\be
u^\pm_{F_{x^{N},y^{N}}}=\mathcal{T}^Nu^\pm_{F_{xy}}=\mathcal{T}\circ...\circ\mathcal{T}\circ\mathcal{T}u^\pm_{F_{xy}},
\ee
in terms of the simpler action of the quotiented $\mbox{\large{$\mathsf{T}$}}$ map, namely 
\be
u^\pm_{F_{x^{N},y^{N}}}=k_{x^{N},y^{N}}^{-1}\mbox{\large{$\mathsf{T}$}}^{N}k_{xy}u^\pm_{F_{xy}}.
\ee
[For brevity, we have denoted above the phase-space point ($u^-_F, u^+_F$) simply as $u^\pm_F$]. Here, $k_{xy}$ is, as above, the Kasner transformation mapping the initial era-starting box $F_{xy}$ to $F_{ba}$, and $k^{-1}_{x^{N},y^{N}}$  is the inverse of the Kasner transformation mapping the ordered corner of $\mathcal{T}^Nu^\pm_{F_{xy}}$ onto the ``standard'' $ba$ corner. The transformation is 
determined from the above-computed values of ($\rho_N, \eta_N$). More precisely, the procedure determining $k^{-1}_{x^{N},y^{N}}$ is : (i) starting from $k_{xy}$ and the continued-fraction decomposition of $k_{xy}u^+_F=n_1-1+[n_2, n_3, ...]$, one determines the $\epsilon_i$, Eq. (\ref{epsilonj}), and the $\rho_N$ and $\eta_N$; (ii) then $\rho_N$ determines the rotation between $xy$ and the final corner $x_N, y_N$, and $\eta_N$ determines whether this corner is ``parallel'' or ``antiparallel'' to the initial $xy$; (iii) finally, knowing the ordered corner $(x^N, y^N)$, Table \ref{table6} determines the transformation $k_{x^N, y^N}$ that maps it onto the $ba$ corner.\\
\vspace{0.1cm}

Let us note the dissymmetric roles of $u^+$ and $u^-$. In the above construction, it was the knowledge of the initial value of $u^+$ which allowed one to recover the full information about the future evolution of the unquotiented dynamics. The situation would be different if we wanted to describe the \textit{past} unquotiented dynamics. In that case, it would be the continued-fraction of the $ba$ transformation of the initial $u^-_F$ that would encode the needed information.\\
\vspace{0.1cm}

To make the above construction more concrete, let us end this subsection by working out an explicit example.\\
We consider (for simplicity) an initial era-starting box of the $ba$ type. For the convenience of the reader, we list in Table \ref{table7} the concrete meaning, for this case, of the six different values of ($\rho_N, \eta_N$) in determining the ordered corner of the $N$-th iterate of the initial point. Let us for instance consider $u^+_{F_{ba}}+1=[n_1; n_2, ...]$ of the form $u^+_{F_{ba}}+1=\sqrt{2}$, i.e. $u^+_{F_{ba}}+1=[1; 2, 2, ...]$, and consider the second iterate $\mbox{\large{$\mathsf{T}$}}^2u^+_{F_{ba}}$. In that case, $xy=ba$ and $k_{xy}=k_0$ (the identity), and therefore $D[k_{xy}]=+1$. As $n_1$ is odd and $n_2$ even, we easily find that $\rho_2=1$ and $\eta_2=-1$. This shows that the second iterate (i.e. the third era, if we count the initial one) is of the $ab$ type. The explicit expression of $\mathcal{T}^2u^\pm_{F_{ba}}$ is then
\be
\mathcal{T}^2u^\pm_{F_{ba}}=k_{ab}^{-1}\mbox{\large{$\mathsf{T}$}}^2u^\pm_{F_{ba}},
\ee 
where $k_{ab}=k_5$ in the list \ref{table6}. We have thereby reduced the computation of the iteration of $\mathcal{T}$ to the simpler computation of the iteration of its quotiented version $\mbox{\large{$\mathsf{T}$}}$.

\begin{table}
\begin{center}
    \begin{tabular}{ | l | l || l | }
    \hline
    $\rho_{M}$ & $\eta_{M}$ & $x^M y^M$\\ \hline
    $1$ & $1$ & $ba$\\ \hline
    $1$ & $-1$ & $ab$\\ \hline
    $e^{-i\frac{2}{3}\pi}$ & $1$ & $ac$\\ \hline
    $e^{-i\frac{2}{3}\pi}$ & $-1$ & $ca$\\ \hline
    $e^{-i\frac{4}{3}\pi}$ & $1$ & $cb$\\ \hline
    $e^{-i\frac{4}{3}\pi}$ & $-1$ & $bc$\\ \hline
    \end{tabular}
\end{center}
\caption{\label{table7} The exit possibilities for a sequence of eras starting with $F_{ba}$. The rotation $\rho_{M}$ is considered modulo $2\pi$.}
\end{table}
\subsection{\label{anisotropicbehaviour}On the anisotropic behavior of the unquotiented big billiard}
The aim of this subsection is to highlight one interesting feature of the unquotiented big billiard that is lost in its quotiented description: its \textit{anisotropy}, i.e. the fact that, after each given era (taking place in some corner, with some sense of motion for the first epoch) the next era has more probability to take place in a specific ordered corner, namely a corner obtained from the previous one by rotating it in the same direction as the first epoch in the disk model, and keeping the same sense of motion for the first epoch (by $\pm2\pi/3$). For instance, if the first era is, say, of the $ba$ type, the following era has more probability to be of the $ac$ type rather then the $bc$ one (which is the other possibility).\\
\vspace{0.1cm}

Indeed, the general formulas (\ref{rhoeta}) above show that, when $N=1$, i.e. after one iteration, the second era is obtained from the first by applying the rotation $\rho_1=e^{\theta_1}$, with $\theta_1=-\tfrac{2\pi}{3}D[K_F]\epsilon_1$, and that its \textit{relative} sense of motion is $\eta_1=\epsilon_1$. Here, $\epsilon_1\equiv(-)^{1+n_1}$ is determined by the parity of the length $n_1$ of the first era. On the other hand, the sign $\epsilon_1$, which determines both $\rho_1$ and $\eta_1$, is a statistical variable whose probability distribution is determined by that of $n_1$, i.e. by $P_{n_1}$, Eq. (\ref{BKLprob}). Among the two possible values of $\epsilon_1$, the most probable is $\epsilon_1=+1$, corresponding to $n_1$ being odd. Indeed, this probability is obtained by summing (\ref{BKLprob}) over $n_1=2k+1$, $k\in\mathbb{N}$, and reads
\be
P(\epsilon_1=+1)=p_{\rm odd}=\sum_{k=0}^{\infty}\frac{1}{\ln 2}\ln  \left(\frac{(2k+2)^2}{(2k+1)(2k+3)}\right)=\frac{\ln\pi-\ln2}{\ln2}\simeq0.6515
\ee 
The complementary probability that $\epsilon_1=-1$ i.e that $n_1$ be even, is
\be
P(\epsilon_1=-1)=p_{\rm even}=\sum_{k=1}^{\infty}\frac{1}{\ln 2}\ln  \left(\frac{(2k+1)^2}{(2k)(2k+2)}\right)=\frac{2\ln2-\ln\pi}{\ln2}\simeq0.3485
\ee
In other words, we have a strongly anisotropic behavior after one era: if the first era is, say, of the $ba$ type, the following era will be of the $ac$ type in $61.15\% $ of cases, and of the $bc$ one in only $34.85\% $ of cases.\\
\vspace{0.1cm}

Let us now see what happens after two iterations. Fixing for simplicity the initial ordered corner to be $ba$, the ordered corner after two iterations can be of four different types:
\begin{itemize}
	\item if $(\epsilon_1, \epsilon_2)=(+1,+1)$ (corresponding to $\theta_2=-4\pi/3$, $\eta_2=+1$ ), it will be $cb$;
	\item if $(\epsilon_1, \epsilon_2)=(+1,-1)$ (corresponding to $\theta_2=0$, $\eta_2=-1$ ), it will be $ab$;
	\item if $(\epsilon_1, \epsilon_2)=(-1,+1)$ (corresponding to $\theta_2=4\pi/3$, $\eta_2=-1$ ), it will be $ca$;
	\item if $(\epsilon_1, \epsilon_2)=(-1,-1)$ (corresponding to $\theta_2=0$, $\eta_2=+1$ ), it will be $ba$.
\end{itemize}
The probabilities corresponding to each one of these cases is easily computed from the probability distribution (\ref{pn1n2}), for $n_1, n_2$ (remembering that $\epsilon_1=(-)^{n_1+1}$, $\epsilon_2=(-)^{n_2+1}$). For instance, $P(\epsilon_1=+1,\epsilon_2=+1)$ $=\sum_{k_2\ge0,k_1\ge0}P_{2k_1+1,2k_2+1}\equiv$ $ P_{\rm odd, odd}$ is found to be
\be
P(\epsilon_1=+1,\epsilon_2=+1)=P_{\rm odd, odd}\simeq0.4199.
\ee
Similarly,
\begin{subequations}
\begin{align}
&P(\epsilon_1=+1,\epsilon_2=-1)=P_{\rm odd, even}=\sum_{k_2\ge1,k_1\ge0}P_{2k_1+1,2k_2}\simeq0.2316,\\
&P(\epsilon_1=-1,\epsilon_2=+1)=P_{\rm even, odd}=\sum_{k_2\ge0,k_1\ge1}P_{2k_1,2k_2+1}=P_{\rm odd, even}\simeq0.2316,\\
&P(\epsilon_1=-1,\epsilon_2=-1)=P_{\rm even, even}=\sum_{k_2\ge1,k_1\ge1}P_{2k_1,2k_2}\simeq0.1169
\end{align}
\end{subequations}
Again we see a strong `anisotropy' among the various possibilities. In particular, the most probable case is again that corresponding to applying the two rotations (by $\pm2\pi/3$) in the direction indicated by the first epoch, keeping the same sense of motion. The calculation gets more involved for higher iterations. One expects that, after many iterations, the memory of the initial ordered corner will get lost, and that one will end up with asymptotically equal probabilities in any one of the six possible ordered corners. Note, however, that the anisotropic behavior we are discussing here will continue to be present \textit{locally}: after each era (whether or not it is much ``later'' than the initial era), the next era will take place in a preferred corner with respect to the previous one.
\section{Periodic orbits}
Let us briefly discuss periodic orbits in the big billiard with a focus on the differences between the notion of periodic orbit in the  unquotiented billiard, and the corresponding notions either in the quotiented billiard or in the BKL map (acting solely on $u^+$).\\
\vspace{0.1cm}

Given any discrete map, say ${\mathcal T}$, acting on some space $X$, a periodic orbit is a set of successive ${\mathcal T}$ images of a point $x \in X$, say $\{ x , {\mathcal T} x , \ldots , {\mathcal T}^{m-1} x \}$, such that ${\mathcal T}^m x = x$. The (minimal possible) integer $m$ is called the period of the discrete map ${\mathcal T}$. As the CB-LKSKS $\mbox{\large{$\mathsf{T}$}}$ map is a quotiented version of the full hopscotch map ${\mathcal T}$, it is easily seen that any $n$-periodic orbit of ${\mathcal T}$ will automatically ``descend'' to a corresponding periodic orbit of $\mbox{\large{$\mathsf{T}$}}$. However, the period of the corresponding $\mbox{\large{$\mathsf{T}$}}$ orbit might be a {\it divisor} of $m$. On the other hand, it is a priori possible that periodic orbits of $\mbox{\large{$\mathsf{T}$}}$ could not be ``lifted'' to periodic orbits of ${\mathcal T}$. To study these two issues (the change in period from ${\mathcal T}$ to $\mbox{\large{$\mathsf{T}$}}$, and the possibility of lifting periodic orbits from $\mbox{\large{$\mathsf{T}$}}$ to ${\mathcal T}$) let us start from some given $n$-periodic orbit of $\mbox{\large{$\mathsf{T}$}}$.\\
\vspace{0.1cm}

A first issue that should be discussed is the relation between periodic orbits of the BKL map $T_{\rm BKL}$ (i.e. the restriction of the two-dimensional map $\mbox{\large{$\mathsf{T}$}}$ to the one-dimensional map $u'_+ = T_{\rm BKL} (u_+)$), and periodic orbits of $\mbox{\large{$\mathsf{T}$}}$ in the $(u^- , u^+)$ plane. The periodicity condition of $T_{\rm BKL}$ involves only one condition, namely $u_+ = T_{\rm BKL}^m (u_+) \equiv \mbox{\large{$\mathsf{T}$}}^m (u_+)$, while the periodicity condition of $\mbox{\large{$\mathsf{T}$}}$ looks much more restrictive as it involves two separate conditions, namely $u_+ = \mbox{\large{$\mathsf{T}$}}^m (u_+)$ and $u_- = \mbox{\large{$\mathsf{T}$}}^m (u_- , [u_+])$ (we recall that the action of $T$ on $u_+$ only depends on $u_+$, while its action on $u_-$ depends, in addition, on the integer part of $u_+$). However, we have seen above that the $\mbox{\large{$\mathsf{T}$}}$ map was always {\it contracting} in the $u^-$ direction (see Eq. (\ref{139})). Therefore, we expect that the iterated effect on {\it any} starting value of $u^-$ of the finite collection of maps indexed by the various values of $[u_+]$ in the periodic orbit $\mbox{\large{$\mathsf{T}$}} (u^- , [u^+])$ will converge to some corresponding fixed orbit of $u^-$ values.\\
\vspace{0.1cm}

Let us then start by an arbitrary $m$-periodic orbit of the one-dimensional BKL map: $T_{\rm BKL}^m (u^+) = u^+$. It is well-known, \cite{BLK1971}, \cite{Cornish:1996yg}, \cite{Cornish:1996hx}, and evident from the explicit form of the action of $\mbox{\large{$\mathsf{T}$}}$ on the continued-fraction expansion of $u^+$, that any such $m$-periodic orbit is parametrized by the special values of $u^+$ that admit a (regular) periodic continued-fraction expansion (cfe) of the type
\begin{equation}
\label{eq:mcfe}
u^+ + 1 = [n_1 ; n_2 , n_3 , \ldots, n_m , n_1 , n_2 , n_3 , \ldots] \, .
\end{equation}
By well-known theorems going back to Euler and Lagrange all such values of $u^+ + 1$ are quadratic irrational numbers, i.e. irrational real roots of quadratic equations of the form $ax^2 + bx + c = 0$ with integer coefficients and a positive discriminant $b^2 - 4ac$ (that is not a perfect square). The simplest example of such periodic-cfe numbers is the (large) golden ratio $u^+ + 1 = \Phi = [1;1,1,1,1,\ldots] = (\sqrt 5 + 1) / 2 \simeq 1.618$ (so that $u^+ = \Phi - 1 \equiv \phi \equiv (\sqrt 5 -1)/2$ is equal to the small golden ratio $\phi = \Phi - 1 \simeq 0.618$). The second simplest examples are of the type $u^+ + 1 = [n;n,n,n,\ldots] = (\sqrt{n^2 + 4} + n)/2$, with some integer $n \geq 1$ and are sometimes called ``silver ratios''. Note that in the cases of the golden ratio, or of the silver ratios, the period $m$ of the BKL orbit is $m=1$.\\
\vspace{0.1cm}

Let us first show that any $m$-periodic orbit of the one-dimensional BKL map $T_{\rm BKL}$ gives rise to a unique corresponding $m$-periodic orbit of the two-dimensional $\mbox{\large{$\mathsf{T}$}}$ map. This follows from the explicit expression (\ref{137}) of the iterated action of $\mbox{\large{$\mathsf{T}$}}$ on an arbitrary starting value of $u^-$, written as Eq. (\ref{2.28}). By repeatedly iterating the $(u^+$-dependent) action of $T$ on $u^-$ one sees that the information contained in the initial value of $u^-$ is lost in the receding tail of the cfe of $\mbox{\large{$\mathsf{T}$}}^N u^-$ so that the sequence of values of $u^-$ tends to a fixed orbit that is entirely defined by the periodic cfe of $u^+$. More precisely, the limiting value of $u^-$, which pairs with the given $u^+$ to define a two-dimensional $m$-periodic orbit of $\mbox{\large{$\mathsf{T}$}}$ is given by
\be
-(1+u^-) = [n_m , n_{m-1} , \ldots , n_2 , n_1 , n_m , n_{m-1} , \ldots] \, .
\ee
Having shown that any $m$-periodic orbit of the one-dimensional BKL map $T$, $u^+ \to T_{\rm BKL} (u^+)$, i.e. any (regular) periodic cfe of the type, Eq. (\ref{eq:mcfe}), uniquely determines a corresponding $m$-periodic orbit of the two-dimensional $T$ map, $(u^+ , u^-) \to T(u^+ , u^-)$, we now discuss the issue whether any $m$-periodic orbit of $T$ can be lifted to some periodic orbit of the full, unquotiented billiard. This question can be answered {\it positively} by studying the $m$-iterated action of the unquotiented map ${\mathcal T}$ on the initial point $(u^- , u^+) \in B_{ba}$ of a $m$-periodic orbit of the quotiented map $\mbox{\large{$\mathsf{T}$}}$. The fact that $\mbox{\large{$\mathsf{T}$}}^m (u^- , u^+) = (u^- , u^+)$ means that the $S_3$-symmetry orbit of ${\mathcal T}^m (u^- , u^+)$ coincides with that of $(u^- , u^+)$. This means that there exists a particular Kasner transformation $k_*$ (which depends on $m$ and on the considered periodic orbit of $\mbox{\large{$\mathsf{T}$}}$) such that
\be\label{171}
{\mathcal T}^m (u^- , u^+) = k_* (u^- , u^+) \, .
\ee
The set of six Kasner transformations is a realization of the $S_3$ permutation group (of order $3!=6$). In fact, this permutation group consists of the identity, $3$ transpositions [$(12), (23)$ and $(31)$], and $2$ cyclic transformations [$(213)$ and $(321)$].We recall that the \textit{order} of a particular group element, such as $k_*$, is the smallest integer $p$ such that $k_*^p=k_0$. As a transposition is of order $2$, and a cyclic permutation, ($123$) or ($321$), of order $3$, we see that the order $p$ of $k_*$ must be equal to $p=1, 2$ or $3$. Therefore, by iterating (\ref{171}), we get
\be
{\mathcal T}^{mp} (u^- , u^+)=k_*^p (u^- , u^+)= (u^- , u^+) \, ,
\ee
and $mp$ will be the smallest such integer. In other words, $(u^- , u^+)$ is the initial point of a periodic orbit under the unquotiented billiard map ${\mathcal T}$, with period $pm$, where $p=1,2,3$ is the order of $k_*$.\\
Actually, the specific value of $p$ can be algorithmically derived from the value of the $Z_2\times Z_3$ pair $(\eta_m, \rho_m)$
corresponding to $\mathcal{T}^m$. More precisely: (i) if $\eta_m=-1$, then $p=2$, independently of the value of $\rho_m$; (ii)
if $\eta_m=1$ and $\rho_m=1$, then $p=1$; while (iii) if $\eta_m=1$ and $\rho_m\neq1$, then $p=3$. (Note that, when $m=1\Rightarrow\rho_m\neq1$.)\\
\vspace{0.1cm}

We have therefore proven that any periodic orbit of the quotiented map $\mbox{\large{$\mathsf{T}$}}$ (or, even, any periodic orbit of the one-dimensional BKL map $T_{\rm BKL}$) can be lifted to a periodic orbit of the unquotiented big billiard map ${\mathcal T}$. Note that this property extends to the corresponding continuous billiard motion in the unquotiented big billiard (simply by considering the geodesic segments corresponding to all the $(u^- , u^+)$'s belonging the periodic orbit). To make this general result more concrete, let us consider a particular example. The simplest periodic orbit of the quotiented billiard is that given by the golden ratio, namely
\begin{subequations}
\begin{align}
&u^+ + 1 = [1;1,1,1,\ldots] = \Phi = \frac{\sqrt 5 + 1}{2}\\
&-(u^- + 1) = [1,1,1,\ldots] = \phi = \frac{\sqrt 5 -1}{2} \, .
\end{align}
\end{subequations}
``Downstairs'' its period is $m=1$. However, the ${\mathcal T}$ transform of the above ``golden-ratio'' point $(u^- , u^+) = (-1-\phi , \phi)$ is given by reflection in the $a$ wall, i.e. by the matrix $A$ of Eq. (\ref{5.14}), so that
\begin{subequations}
\begin{align}
&{\mathcal T} u^+ = - u^+ = -\phi,\\
&{\mathcal T} u^- = -u^- = 1+\phi \, .
\end{align}
\end{subequations}

The transformation $u' = -u$ is {\it not} one of the Kasner transformation. However, in keeping with the general reasoning above one can use the fixed-point property of this golden-ratio periodic orbit (namely $\phi (1+\phi) = 1$) to rewrite the r.h.s.'s of the above equations as
\begin{subequations}\label{goldenratio}
\begin{align}
&{\mathcal T} u^+ = - \frac{1}{1+\phi} = - \frac{1}{u^+ + 1} \equiv k_4 (u^+),\\
&{\mathcal T} u^- = \frac{1}{\phi} = - \frac{1}{u^- + 1} \equiv k_4 (u^-)
\end{align}
\end{subequations}
where $k_4 (u) \equiv -1 / (u+1)$ is one of the Kasner transformations of Table \ref{table1}. Therefore, the specific Kasner transformation $k_*$ corresponding to the particular ``golden-ratio'' periodic orbit of $\mathcal{T}$ is $k_* = k_4$. The latter Kasner transformation correspond to the element of $S_3$ realizing the cyclic permutation $(p_1 , p_2 , p_3) \to (p_2 , p_3 , p_1)$ of Kasner exponents. The order of such a cyclic transformation is $p=3 : k_4 \circ k_4 \circ k_4 =$ identity. This shows that the golden-ratio initial conditions above define a periodic orbit of the unquotiented billiard of order $pm = 3 \times 1 = 3$. We recover the periodic orbit of Eq. (\ref{5.7}) made of three successive one-epoch eras between the middle of the three successive gravitational walls $a$, $c$ and $b$.\\
\vspace{0.1cm}

Note that, in the general case of a starting value for $u^+$ of the type (\ref{eq:mcfe}), an $m$-periodic orbit of the quotiented billiard will contain $m$ eras containing, successively, $n_1 , n_2 , \ldots , n_m$ epochs (so that it contains $n_1 + n_2 + \ldots + n_m$ epochs in all), while the lift of this periodic orbit onto the unquotiented big billiard will contain $pm$ eras, containing $p(n_1 + n_2 +  \ldots + n_m)$ epochs in all.\\
\vspace{0.1cm}

Finally, let us note that the billiard periodic orbits discussed here are the projection down to hyperbolic space ${\mathcal H}_2$ of Lorentzian-billiard motions in $\beta$-space which are {\it not} periodic there. Indeed, the spatial metric $g_{ij} (T , \bf{x})$ corresponding to these dynamics is expressed in terms of the $\beta$'s, rather than the projected $\gamma$'s, say (for the diagonal Bianchi IX case of relevance to the big billiard)
\be
g_{ij} (T , {\bf x}) = \sum_a e^{-2\beta^a(T)} \, e_i^a ({\bf x}) \, e_j^a ({\bf x}) \, ,
\ee
with
\be
\beta^a (T) = \rho (T) \, \gamma^a (T) \, ,
\ee
and 
\be
\rho (T) \simeq \exp (cT) \, , \qquad (\mbox{with} \ c > 0)
\ee
where we used the result \cite{Damour:2000hv}, \cite{Damour:2002et} that $\lambda = \ln \rho$ is (asymptotically) a linear function of the coordinate time $T$ defined in Eq. (\ref{2.21}). A periodic orbit of the big billiard is such that $\gamma^a (T + n {\mathcal P}) = \gamma^a (T)$ for $n \in {\mathbb N}$ and some period ${\mathcal P}$ in $T$-time. This periodicity ``downstairs'' in $T$-time does not correspond to a periodicity of the metric coefficients $g_{ij} (T , {\bf x})$. It does not even correspond, as one might have thought, to a discrete self-similar symmetry of the metric (i.e. $g_{ij} (T + n {\mathcal P} , {\bf x}) = \lambda^n g_{ij} (T,{\bf x})$) but to a rather different discrete transformation under which the ``scale factors'' $a_a (T) \equiv e^{-\beta^a (T)}$ (i.e. the BKL variables $a,b,c$) change as
\begin{equation}
\label{eq:aevolution}
a_a (T + n {\mathcal P}) = [a_a (T)]^{\lambda^n}
\end{equation}
with
\be
\lambda = e^{c{\mathcal P}} > 1 \, .
\ee
As the $\gamma^a$'s are confined by the big billiard walls to remain (non strictly) positive, the (periodic) scale factors $a_a (T)$ stay $\leq 1$, and we see on Eq.~(\ref{eq:aevolution}) that (apart when they collide on a gravitational wall, where the corresponding scale factor becomes equal to $1$) the scale factors tend to zero {\it super exponentially} with the number $n$ of periods.

\section{Small billiard}
Up to now we have been discussing the \textit{big billiard}, with three walls $a, b, c$ making up an ideal triangle in hyperbolic space (see Fig. \ref{fig2}  or Fig. \ref{fig3}.). This billiard corresponds to the dynamics of the diagonal Bianchi IX model, i.e. the $a, b, c$ system of BKL. However, as recalled above, in the most general (non-diagonal, inhomogeneous) case, the use of an Iwasawa decomposition of the spatial metric, as in Eq. (\ref{2.4}), leads to a closely related but slightly different billiard, namely the \textit{small billiard} made of one gravitational wall, and two symmetry walls. As in Fig. \ref{fig2} or Fig. \ref{fig3}, we shall denote the (partial) gravitational wall as $G$ (for gravity or green), and its symmetry walls as $B$ (for blue) and $R$ (for red). More precisely, in the notation of the Poincar\'e model of Fig. \ref{fig3},
\begin{itemize}
	\item the $G$ wall is the portion $u=0, v>1$ of the gravitational wall $a=0$;
	\item the $B$ wall is the portion $u=-1/2, v>\sqrt{3}/2$ of the symmetry wall $a=b$;
	\item the $R$ wall is the portion $u^2+v^2=1, -1/2<u<0$ of the symmetry wall $a=c$.
\end{itemize}
Our aim in this section is to relate the dynamics within the small billiard to the dynamics within the big billiard studied above. Somewhat surprisingly, though the two billiard tables are closely related, the two corresponding dynamics cannot be straightforwardly mapped among themselves.

\subsection{ Dynamics of the unquotiented small billiard}

We start by considering the dynamics within the three walls $G,B,R$ of the small billiard, without introducing any extra quotienting. Indeed, the small billiard table is already a fundamental domain of the six-fold symmetry group $S_3$ of the big billiard acting on ${\mathcal H}_2$, therefore one could a priori expect that the small-billiard dynamics be equivalent to the quotient of the big-billiard one by $S_3$ (as studied above). Actually, this is not the case. We shall find that the small billiard dynamics is {\it not} equivalent to the quotiented big billiard dynamics. The basic reason for this non equivalence is that the small billiard table is obtained by quotienting only the  {\it configuration space} ($q$ space) of the big billiard dynamics by $S_3$, while the $S_3$-quotienting we considered above was done in {\it phase-space} ($q,p$ space). When considering on its own the small billiard dynamics it is natural to introduce analogs of the notions introduced (by BKL) in the big billiard context. First, we shall define an {\it epoch} of the small billiard as a geodesic segment (i.e. a Kasner motion) connecting two successive walls. For example, a $B \to G$ epoch (or, for short, a $BG$ epoch) is an epoch starting from the blue $(B)$ wall and ending on the green or gravitational $(G)$ wall, etc.\\
\vspace{0.1cm}

The dynamics of successive epochs of the small billiard is similar to that of the big billiard. For instance, the following sequence of epochs
\begin{equation}
\label{eq:169}
R \to G \to B \to G \to R
\end{equation}
corresponds to a succession of ``collisions'' on the $G$, $B$ and $G$ walls for a dynamics which started on the red wall and returned on it. As in the big billiard case, it is convenient to parametrize each epoch by a point in the $(u^- , u^+)$ plane, with $u^+$, resp. $u^-$, parametrizing the end, resp. beginning, of the extended geodesic corresponding to the considered epoch. A first difference with the big billiard case is that the regions of the $u^- u^+$ plane which describe the small billiard dynamics are quite dissimilar to the corresponding big-billiard regions drawn in Fig.~6. 
\begin{figure}[htbp]
\begin{center}
\includegraphics[width=17cm]{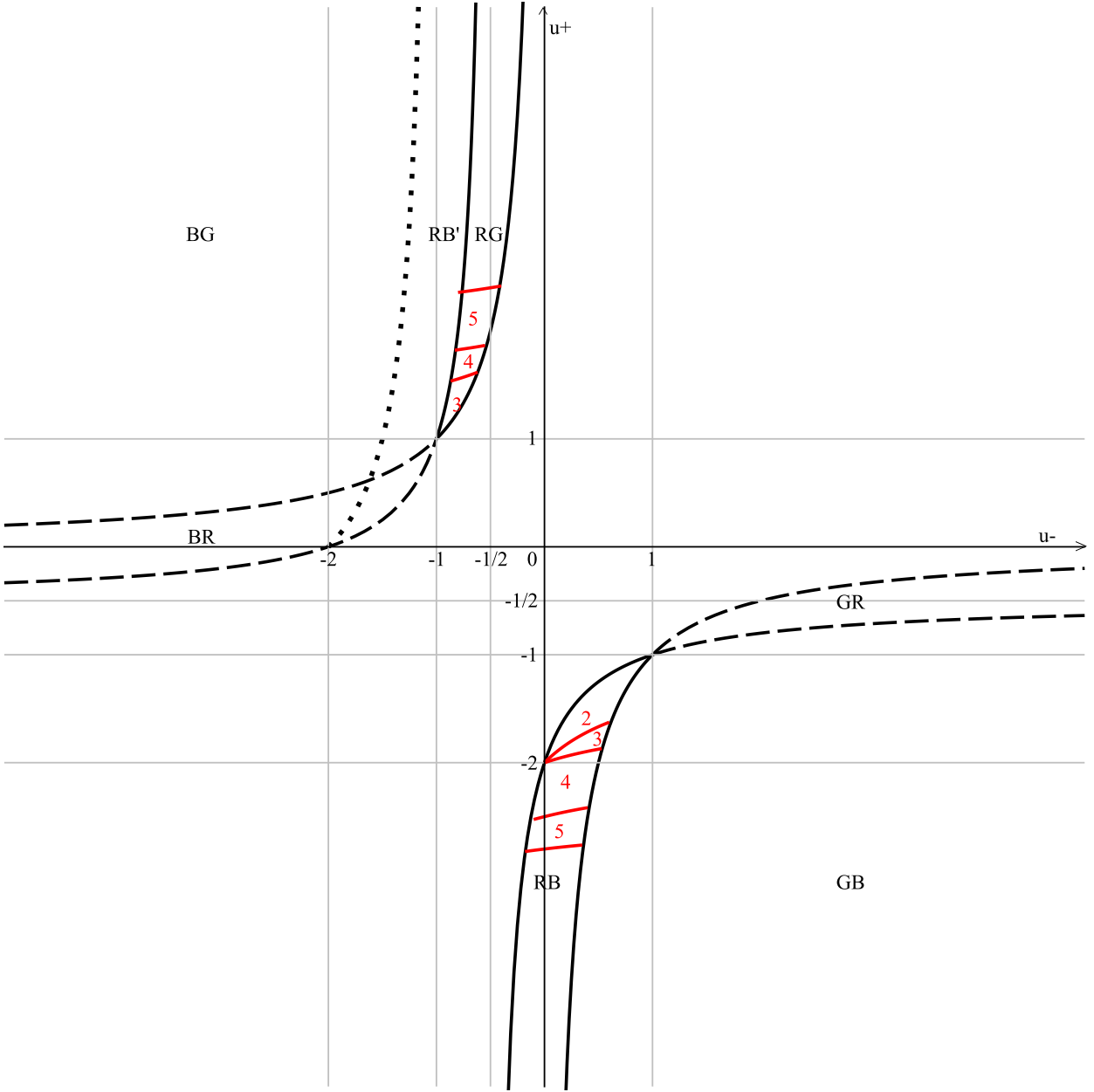}
\caption{\label{smallbilliard}The phase space of the small billiard in the $u^-,u^+$ parametrization. The regions $RG$ and $RB$ are delimited by solid thick lines, the regions $BR$ and $GR$ are delimited by dashed thick lines. The Kasner intervals are indicated on both axes by thin gray lines. The subdomains corresponding to the first few eras of $RG$ and $RB$ are also sketched.} 
\end{center}
\end{figure}

\begin{table}
\begin{center}
    \begin{tabular}{ | l | l | }
    \hline
    $ BG $ & $ u^-<-1,\ \ u^+>u_\alpha$\\ 
           & $ -1<u^-<-1/2,\ \ u^+>u_\beta$\\ \hline
    $ BR $ & $ u^-<-1,\ \ u_\beta<u^+<u_\alpha$\\  \hline
    $ RG $ & $ -1<u^-<-1/2,\ \ u_\alpha<u^+<u_\beta$\\ 
           & $ -1/2<u^-<0,\ \ u^+>u_\alpha$\\ \hline
    $ RB $ & $ -1/2<u^-<0,\ \ u^+<u_\beta$\\ 
           & $ <u^-<1,\ \ u_\alpha<u^+<u_\beta$\\ \hline
    $ GR $ & $ u^->1,\ \ u_\alpha<u^+<u_\beta$\\  \hline
    $ GB $ & $ 0<u^-<1,\ \ u^+<u_\alpha$\\ 
           & $ u^->1,\ \ u^+<u_\beta$\\ \hline
    \end{tabular}
\end{center}
\caption{\label{table11} The regions of the $u^+u^-$ plane, where the dynamics of the small billiard takes place.}
\end{table}

They are drawn in Fig. \ref{smallbilliard}: they comprise six allowed regions, labelled as $BG$, $BR$, $RG$, $RB$, $GR$ and $GB$, and a large, connected forbidden (``vacuum'') region which occupies the central part of Fig.~6 (between the upper hyperbola-like curves and the lower ones). The precise definition of the allowed regions is given in Table \ref{table11} where we used a short-hand notation for the following functions of $u^-$:
\begin{subequations}\label{182}
\begin{align}
&u_{\alpha} \equiv - \frac{1}{u^-}\\
&u_{\beta} \equiv - \frac{u^- + 2}{2u^- + 1}.
\end{align}
\end{subequations}
As we see the boundaries of the various allowed (and forbidden) regions are made of segments of hyperbolas: $u^+ = (au^- + b) / (cu^- + d)$. This is different from the big billiard case, where all the boundaries corresponded to horizontal or vertical lines (see Fig. \ref{biliardino11}). In addition, the forbidden region of the big billiard (white domains in Fig. \ref{biliardino11}) was made of three disconnected pieces. If we compare Fig. \ref{biliardino11} and Fig. \ref{smallbilliard} we can roughly consider that the small-billiard $u^- u^+$ picture, Fig. \ref{smallbilliard}, is obtained by ``morphing'' the big-billiard one, Fig.~6, via a deformation where the $B_{ba}$ box becomes the $BG$ one, $B_{ca} \to RG$, $B_{bc} \to BR$, $B_{ac} \to GR$, $B_{cb} \to RB$ and $B_{ab} \to GB$, while the three disconnected forbidden regions of Fig.~6 ``percolate'' among themselves into the connected central forbidden domain of Fig. \ref{smallbilliard}. This correspondence between the two pictures exists because, for instance, the extension to the big billiard of a small-billiard $BG$ epoch corresponds to a $ba$ epoch, etc. However, contrary to what one might have naively expected, it is not possible to find a globally-defined transformation $u^{+'} = U (u^- , u^+)$, $u^{-'} = V(u^- , u^+)$ (leaving invariant the $2$-form $\omega$, Eq. (\ref{5.3})) that maps Fig. \ref{biliardino11} into Fig. \ref{smallbilliard}. The main obstacle to the existence of such a transformation is the fact that the forbidden regions of the small billiard comprise (say in the Poincar\'e model) not only all the forbidden regions of the big billiard (geodesics in the Poincar\'e half-plane that do not intersect the $abc$ triangle), but, in addition, new forbidden regions (geodesics that intersect the $abc$ triangle but not its $GBR$ subtriangle) that were allowed before.\\
\vspace{0.1cm}

Let us now briefly describe the dynamics of the small billiard, as seen in the $u^- u^+$ plane, Fig. \ref{smallbilliard}. It is similar to the ``hopscotch game'' associated to Fig. \ref{biliardino11}. Namely, each initial point $(u^- , u^+)$ in the occupied regions of Fig. \ref{smallbilliard} will ``jump'' to another position according to the following rules:
\begin{enumerate}
\item[$\bullet$] if the point $(u^- , u^+)$ belongs either to the $BG$ region or the RG one (so that it corresponds to an epoch starting either on $B$ or on $R$ and ending by a collision on $G$) it will jump by a $G$-collision, i.e. by the transformation:
\be\label{green}
G \ \mbox{wall} \ (u=0) : u^{\pm} \to G(u^{\pm}) \equiv -u^{\pm}
\ee
\item[$\bullet$] if $(u^- , u^+)$ belongs either to the $RB$ or $GB$ regions, it jumps by the transformation:
\be\label{blue}
B \ \mbox{wall} \ (u = -\frac{1}{2}) : u^{\pm} \to B(u^{\pm}) \equiv -u^{\pm} - 1
\ee
\item[$\bullet$] if $(u^- , u^+)$ belongs either to the $BR$ or $GR$ regions, it jumps by the transformation:
\be\label{red}
R \ \mbox{wall} \ (u^2 + v^2 = 1) : u^{\pm} \to R(u^{\pm}) \equiv 1/u^{\pm} \, .
\ee
\end{enumerate}

Note that a point in some $XY$ region will jump either to the $YX$ or the $YZ$ region (with $\{ X,Y,Z \} = \{ G,B,R \}$). Note also that the ``jumping rules'' above are similar to, but different from, the corresponding $A,B,C$ jumping rules of the big billiard, see Eqs. (\ref{5.12}).\\
\vspace{0.1cm}

The small-billiard hopscotch game defined above leaves invariant the 2-form $\omega$, Eq.~(\ref{4.29}). However, as in the big billiard case the integral of $\omega$ over the allowed regions of the small billiard is logarithmically infinite. As in the big billiard case, this logarithmic divergence comes (when writing $\omega$ in terms of Birkhoff coordinates along the sides of the billiard table) from the infinite hyperbolic length of the $G$ and $B$ walls that meet on the absolute. This suggests a natural way of by-passing this divergence problem: to consider the (Poincar\'e) {\it return map of the small billiard on its red $(R)$ wall}, which is the only wall having a finite length. In other words, it is very natural, within the small billiard, to collect together all the epochs corresponding to bounces between the $G$ and $B$ walls into {\it small-billiard eras}, and to focus on the {\it small-billiard era hopscotch} dynamics which maps the beginning of such an era to the beginning of the next one. For instance, a small-billiard era comprising four epochs was indicated in Eq. (\ref{eq:169}). For definiteness, we shall define the {\it first} epoch of a small-billiard era as the epoch which {\it starts} on $R$ (for instance the leftmost $R \to G$ epoch in Eq. (\ref{eq:169})). Clearly, the era-hopscotch transformation $u_F^{\pm'} = f(u_F^{\pm})$ mapping the $u^{\pm}$ coordinates of the first epoch in an era to the coordinates of the first epoch in the next era, being obtained by composing the individual $G,B,R$ rules above, will be given by some {\it diagonal fractional-linear} transformation:
\begin{equation}
\label{eq:716}
u'^+_F = T_{\!SB} \, u_F^+ \, , \quad u'^-_F = T_{SB} \, u'_F \, , \quad \mbox{with} \quad T_{SB} u = \frac{au + b}{cu + d} \, ,
\end{equation}
and will leave invariant the restriction to the $R$ wall of the usual two-form $\omega$, Eq.~(\ref{4.29}). More precisely, the analog of what was before the ``first-epoch'' domains (such as the $F_{ba}$ subregion of $B_{ba}$ represented in Fig.~7) become the ``$R$-leaving'' domains. Contrary to what happened in the big-billiard case, we do not need here to delineate these regions as subregions of the full small-billiard hopscotch court of Fig. \ref{smallbilliard}. Indeed, by definition the era-starting region of the small billiard consists of the union of the $RG$ and $RB$ regions in Fig. \ref{smallbilliard}. The $RG$ region represents the first-epochs of the small-billiard eras that leave $R$ towards the $G$ wall, while the $RB$ region represents the first-epochs of the small-billiard eras that leave $R$ towards the $B$ wall. They are indicated by solid thick lines in Fig. \ref{smallbilliard}. The era-hopscotch rule $T_{\!SB}$ (\ref{eq:716}) will map $RG \cup RB$ onto itself (mapping sometimes, say, $RG$ to itself or to $RB$, etc.). Now, the small-billiard era-map $T_{\!SB}$ leaves invariant $\omega$ on a space $(RG \cup RB)$ on which $\omega$ has a finite integral. Therefore we are now in the good conditions for applying ergodic theory, and considering that $\omega$ defines a probability measure on $RG \cup RB$. More precisely, one finds that
\be
\int_{RG} \frac{1}{2} \omega = \frac{1}{2}\ln\left(\frac{3}{2}\right) \, , \qquad \int_{RB} \frac{1}{2} \omega =\frac{1}{2}\ln2 
\ee
so that the normalized probability measure on $RG \cup RB$ is
\be
c \, \frac{du^+ \wedge du^-}{(u^+ - u^-)^2} \qquad \mbox{with} \quad \frac{1}{c} = \frac{1}{2}\ln\left(\frac{3}{2}\right) + \frac{1}{2}\ln2=\frac{1}{2}\ln 3 \, .
\ee

It would seem that, at this stage, we have obtained a small-billiard era-dynamics that is very similar to the quotiented big-billiard era dynamics, i.e. the CB-LKSKS map $\mbox{\large{$\mathsf{T}$}}$ between the big-billiard first-epoch region $F_{ba}$ to itself. An apparent qualitative difference is the fact that the small-billiard era-starting region $RG \cup RB$ is made of two disconnected pieces. However, it is easily found that if we transform the $RB$ region by the diagonal transformation
\be
u'_{\pm} = -u_{\pm}-1
\ee
(which leaves invariant $\omega$), the $RB$ region will be mapped onto a new region say $RB'$, of the $u^- u^+$ plane which is contiguous to $RG$ (along its left boundary). Therefore, we can simply replace the $RG \cup RB$ domain by a thicker, connected domain $RG \cup RB'$ (delimited in Fig. \ref{smallbilliard} by a dotted line), and consider the dynamics of the (suitably transformed) era-hopscotch map $\tilde T_{SB}$ from $RG \cup RB'$ onto itself. At this stage, we have an era-hopscotch map which looks quite similar to the quotiented map $\mbox{\large{$\mathsf{T}$}}$ from $F_{ba}$ onto itself. Moreover, if we look at the asymptotic region of long eras, i.e. the $u^+ \gg 1$ region of $RG \cup RB'$ (where the $RB'$ part came from the $u^+ \ll -1$ region of $RB$) we see that the shape of our new small-billiard first-epoch domain is simply
\be\label{rb'}
 -1 < u^- < 0 \, , \qquad u^+ \gg 1 \,
\ee
In other words, it is asymptotically {\it rectangular} and, modulo a simple shift of $u^-$ by one unit, seems to coincide with the (exactly rectangular) quotiented big billiard domain $F_{ba}$.\\
\vspace{0.1cm}

This asymptotic coincidence (modulo some suitable identifications) between the small-billiard red-return map and the quotiented big-billiard era-map $\mbox{\large{$\mathsf{T}$}}$ was physically expected because long eras of the big billiard (with $u^+ \gg 1$) correspond to many bounces between the $b$ and $a$ walls which clearly (see Fig.~5 together with Fig.~3) will correspond to roughly twice as many bounces between the $B$ and $G$ walls of the small billiard. Here, we have in mind folding some, say, $a \to b \to a$ oscillation of the big billiard back onto the small billiard by introducing a $B$ ``mirror'' in the middle of the $ba$ corner, so that the big billiard bounces $a \to (B) \to b \to (B) \to a$ (which do not ``see'' the $B$ wall) become transformed in small-billiard oscillations of the type $G \to B \to G \to B \to G$, where we used the fact that, deep into the corner, $a$ and $G$ coincide.
However, this asymptotic (large $u^+$) coincidence does not extend to the small-$u^+$ (short-era) region. In other words, there does not exist a $\omega$-preserving transformation of the $u^- u^+$ plane that would map the $RG \cup RB' \overset{T_{SB}}{-\!\!-\!\!\!\longrightarrow} RG \cup RB'$ dynamics onto the $F_{ba} \overset{\mbox{\large{$\mathsf{T}$}}}{-\!\!-\!\!\!\longrightarrow} F_{ba}$ one. Indeed, if such a transformation existed the $\omega$-area of $FR \cup RB$ (or equivalently $RG \cup RB'$) would be equal to that $\omega$-area of $F_{ba}$. However, we have
\be
\int_{RG} \frac{1}{2} \omega + \int_{RB} \frac{1}{2} \omega = \frac{1}{2}\ln\frac{3}{2}+\frac{1}{2}\ln2 = \frac{1}{2}\ln3 \ne \ln 2 = \int_{F_{ba}} \frac{1}{2} \omega \, .
\ee

However, this non equivalence does not mean that the small-billiard is unrelated to the big-billiard. [Actually, we shall discuss in the following subsection a well-defined, accurate relation between the two billiards.] It mainly means that the natural definition of eras in the small-billiard, as $R$-return maps, cannot be identified with the natural definition of eras in the quotiented big billiard. It is true that, roughly speaking, the small-era definition (which means, when viewed in the big billiard, that an era ends when one crosses one of the three symmetry segments connecting the center of the disk to the middles of the three gravitational walls in Fig. \ref{fig2}) signals the passage from an oscillation in one of the corners of the $abc$ billiard to an oscillation in another corner. However, the problem is that the precise definition used in the big billiard of the transition between an $ab$-type oscillation to, say, a $bc$-one cannot be equivalently characterized as the crossing of one of the three symmetry segments (similar to the $R$ segment in Fig.~2). This non equivalence essentially stems from the fact that a big-billiard epoch (going to a gravitational wall to another gravitational wall) can, on its way, cross either $1$, $2$, or even $3$ symmetry walls. And even if one considers the three symmetry segments of the type of the $R$ one in Fig.~2, some big-billiard epochs can cross either $1$ or $2$ $R$-type symmetry segments. This non constancy in the number of crossing of symmetry walls or segments prevents one from being able to define, in a uniform manner, a transformation between big-billiard eras, and small-billiard ones, which respects, say, the number of epochs during an era.\\
\vspace{0.1cm}

In addition to this non equivalence in the definition of eras, a technical inconvenience of the unquotiented small-billiard era-hopscotch map defined above is that it is more difficult to find the generic, explicit expression of the small-billiard era map $T_{SB}$, Eq.~(\ref{eq:716}). Indeed, while it was easy to define the small-billiard {\it epoch} hopscotch map, see Table \ref{tablelast} and Eqs. (\ref{green}), (\ref{blue}), and (\ref{red}) for any starting point $u^- u^+$, the corresponding definition of the small-billiard {\it era} map $T_{SB}$ (as the return map on the $R$ wall) has remained rather implicit. Given some $(u^- , u^+) \in RG \cup RB$ one needs to iterate the epoch map a certain number of times (which depends on the starting point) to find the explicit expression of the red-return map $T_{SB}$. Actually, with some more effort it is possible to define $T_{SB}$ nearly explicitly. Let us indicate how. First, we can delineate the small-billiard analogs of the $F_{xy}^k$ boxes of the big billiard, i.e. the subregions of the $RG$ and $RB$ regions that will lead to (small-billiard) eras having some given number of epochs. Let us call $RG_n$ (resp. $RB_n$) the sub-region of $RG$ (resp. $RB$) which provides a starting point for an $n$-epoch era. Here, as it is easily seen, $n \geq 3$ for $RG$ and $n \geq 2$ for $RB$. [For instance, any $RG$-starting era must go through $R \to G \to B \to R$ before returning to $R$.] These regions are defined by the domains described in Table \ref{tablelast}. In this table the last column lists the equations of the four curves that delimit, for each $n$, the corresponding era-starting subdomains. These curves are given by equations of the type $u^+=f(u^-)$, where the functions $f(u^-)$ are either the functions $u_\alpha(u^-)$, $u_\beta(u^-)$ defined in Eqs. (\ref{182}), or the following functions 
\begin{itemize}
	\item $u_b^n=\tfrac{1}{2}\frac{-2nu^-+2u^-+n^2-2n+5}{-2u^-+n-1}$;
	\item $u^a_n=\tfrac{1}{2}\frac{-4n+7-2nu^-+4u^-+n^2}{-2-2u^-+n}$;
	\item $U^b_n=-\tfrac{1}{2}\frac{2nu^--2u^-+n^2-2n+5}{2u^-+n-1}$;
	\item $U_a^n=-\tfrac{1}{2}\frac{3+2nu^-+2n^2}{2u^-+n}$.
\end{itemize}


Some of the corresponding regions, for $n \leq 5$, are represented in Fig. \ref{smallbilliard}. As we see, contrary to the big-billiard case where the boundaries between the $F_{xy}^k$ boxes where always horizontal (i.e. of the type $u^+ = {\rm const}$), here the boundaries between $RG_n$ or $RB_n$ boxes are curved. This contributes to the difficulty of writing an explicit expression for the red-return map $T_{SB}$. Anyway, given these results, we can semi-explicitly define $T_{SB}$ by the following algorithm. Given some red-starting point $(u^- u^+)$, one must first find to which box, $RG^n$ or $RB^n$ it belongs by using Table \ref{tablelast}. Then, knowing the type ($RG$ or $RB$), and the length $(n)$ of the era starting from $(u^- u^+)$, one can write $T_{SB}$ by composing $n-1$ transformations of the $G$ or $B$ type, and one final $R$ transformation. Let us give an explicit example. If $(u^- u^+) \in RG_3$, we will have
\be
T_{SB} (u^- , u^+) = R \circ B \circ G (u^- , u^+) \, ,
\ee
where the explicit expressions of the $R$, $B$ and $G$ transformations have been given in Eqs. (\ref{green}), (\ref{blue}), (\ref{red}) above.\\
\begin{table}
\begin{center}
    \begin{tabular}{ | l | l |}
    \hline
    $RG^n,\ \ n\ \ \rm{odd}$ & $u^+>1,\ \ u^-<-1/2,\ \ (u_\alpha, u_\beta, u_b^n, u_a^n)$ \\ \hline
    $RG^n,\ \ n\ \ \rm{even}$ & $u^+>1,\ \ u^-<-1/2,\ \ (u_\alpha, u_\beta, _b^{n-1}, u_a^{n+1})$ \\ \hline
    $RB^n,\ \ n\ \ \rm{odd}$ & $u^+<0,\ \ u^->-1/2, \ \ (u_\alpha, u_\beta, U_b^n, U_a^n)$\\ \hline 
    $RB^n,\ \ n\ \ \rm{even}$ & $u^+<0,\ \ u^->-1/2, \ \ (u_\alpha, u_\beta, U_b^{n-1}, U_a^{n+1})$\\ \hline
    \end{tabular}
\end{center}
\caption{\label{tablelast} Small billiard hopscotch}
\end{table}

Collecting all these ingredients together, it is easy to evaluate the probability $P_{RX^n}$ for a small-billiard era $RX$ to contain a certain number $n$ of epochs. These probabilities are obtained by integrating the $\omega$ form over the suitable domain, as explained in Table \ref{tablelast}. Because the domains are not defined by straight lines, but by hyperbolas, the explicit expression for such probabilities would be somewhat awkward to write down. Nevertheless it is nevertheless always possible and straightforward to calculate it.\\
As the most direct example, we evaluate the probability $P_{RB^2}$, i.e. the probability for a $RB$ era to consist of $2$ epochs. As we see from Fig. \ref{smallbilliard} and Table \ref{tablelast}, the subregion of $RB$ corresponding to this case is the simplified domain
\begin{subequations}
\begin{align}
&0<u^-<\phi,\ \ u_a^3<u^+<u_\beta,\\
&\phi<u^-<1,\ \ u_\alpha<u^+<u_\beta,
\end{align}
\end{subequations}
where $u_a^3=-(u^-+2)/(u^-+1)$, and $u_\alpha=u_a^3$ at $u^-=\phi$. The probability then reads
\be
P_{RB^2}=c\left[\int_0^\phi du^-\int^{u_\beta}_{u_a^3}\tfrac{du^+}{(u^+-u^-)^2}+\int_\phi^1 du^-\int^{u^\beta}_{u_\alpha}\tfrac{du^+}{(u^+-u^-)^2}\right]\simeq0.1240
\ee
We remark that, while, in the BKL case, the probability for an era to consist of $k$ epochs was a monotonically decreasing function of $k$, in the small billiard, this probability is a non-monotonic function of $k$: it increases for the lowest values of $k$ and then starts to decrease. Furthermore, the probabilities $P_{RB'G^n}$ for an era to contain $n$ epochs in the $RG\cup RB'$ domain can also be investigated. The most interesting information is obtained in the asymptotic region for large $u^+$, as described in the above. 
For the $RG$ region, in this asymptotic regime, the $u^-$ boundaries can be considered as vertical straight lines, i.e. $-1/2<u^-<0$, and the $u^+$ boundaries can be considered as horizontal straight lines, as given by the asymptotic behavior of the functions $u^n_b\rightarrow(n-1)/2$, $u_a^n\rightarrow(n-2)/2$. On the other hand, the probabilities for the region $RB'$ are obtained applying the transformation (\ref{rb'}) to the appropriate functions in Table \ref{tablelast}. For large values of $u^+$, the region $RB'$ is delimited by vertical straight lines $u^-=-1$ and $u^-=-1/2$, and the horizontal lines delimiting the $n$-epoch era starting region are $u^+=(n-2)/2$ and $u^+=(n-3)/2$.\\
This way, the probability, say $P_{RB'G^n}$, to obtain a $n$-epoch era starting from the $RG\cup RB'$ extended region is given by the integral of the $\omega$ form over the appropriate domain, i.e.:
\be
P_{RB'G^n}\simeq\frac{1}{\tfrac{1}{2}\ln3}\left[\int^{\tfrac{n-1}{2}}_{\tfrac{n-2}{2}}du^+\int^0_{-1/2}du^-\frac{1}{(u^+-u^-)^2}+\int^{\tfrac{n-2}{2}}_{\tfrac{n-3}{2}}du^+\int^{-1/2}_{-1}du^-\frac{1}{(u^+-u^-)^2}\right]=\frac{1}{\tfrac{1}{2}\ln3}2\ln\frac{(n-1)^2}{n(n+2)}.
\ee
For large values of $n$ one finds
\be\label{prgbn}
P_{RB'G^n}\simeq\frac{4}{\ln3}\frac{1}{n^2}.
\ee
Note that the result (\ref{prgbn}) \textit{differs} from the corresponding quotiented big-billiard (or BKL) result which is $P_{n_1}\simeq\tfrac{1}{\ln2}\tfrac{1}{n_1^2}$. This difference shows that the natural notion of small-billiard era cannot be precisely mapped on the usual notion of BKL eras.

\subsection{ Quotiented small billiard}

The direct, seemingly natural red-return map approach to the small billiard discussed in the previous subsection leads to a rather complex description of its dynamics. In particular, the fact that the $n$-epoch boxes $RG_n$ and $RB_n$ defined in the previous subsection have curved boundaries would make it difficult to define a small-billiard analog of the nice continued-fraction BKL description of the (big billiard) era dynamics. Here, we wish to show how, starting from the definition of the small-billiard, one can recover its deep relation with the big billiard, and uncover the technically simple BKL-like description of its dynamics. To start with, let us show several ways of relating the small-billiard dynamics to the big-billiard one. First, we can see this relation by using a graphical representation of the big-billiard dynamics which has been introduced by BKL. Namely, we mean the plot of the three logarithmic scale factors $\alpha \equiv \beta^1 \equiv -\ln a$, $\beta \equiv \beta^2 \equiv - \ln b$, $\gamma \equiv \beta^3 \equiv - \ln c$ as functions of $\tau$. Modulo a conventional change of sign in the definition of $\alpha , \beta , \gamma$ (introduced here to ensure that $\alpha , \beta , \gamma$ are always $\geq 0$), this representation was, e.g., used in Fig. $2$ of the review \cite{BLK1971}. In terms of this graphical representation, the dynamics of the variables $\alpha' , \beta' , \gamma'$ of the small billiard is simply related to that of the variables $\alpha , \beta , \gamma$ of the big billiard in the following way: starting from a BKL graph of $\alpha (\tau)$, $\beta (\tau)$, $\gamma (\tau)$ [in which the three lines $\alpha (\tau)$, $\beta (\tau)$, $\gamma (\tau)$ can cross and keep (nearly) constant slopes except when one of them touches, from above, the horizontal axis (gravitation wall)] one can define the corresponding graph of $\alpha' (\tau)$, $\beta' (\tau)$, $\gamma' (\tau)$ simply by defining $\alpha' (\tau)$, for each $\tau$, as $\alpha' (\tau) \equiv \inf [\alpha (\tau) , \beta (\tau) , \gamma (\tau)]$,  $\gamma' (\tau)$ as $\gamma' (\tau) \equiv \sup [\alpha (\tau) , \beta (\tau) , \gamma (\tau)]$ and $\beta' (\tau)$ as the remaining middle curve among the three curves $\alpha (\tau)$, $\beta (\tau)$, $\gamma (\tau)$. In other words, $\alpha' (\tau)$ is defined as the lower envelope of the three original $\alpha , \beta , \gamma$ curves, $\gamma' (\tau)$ the upper envelope, and $\beta' (\tau)$ the intermediate curve. The three curves $\alpha' (\tau)$, $\beta' (\tau)$, $\gamma'(\tau)$ satisfy $\alpha' (\tau) \leq \beta' (\tau) \leq \gamma' (\tau)$ and keep constant slopes expect when either two of them ``collide'' (symmetry wall), or when the lower one, i.e. $\alpha' (\tau)$ ``collides'' with the $\alpha' = 0$ axis ($\alpha'$-gravitational wall). One easily sees that $\alpha' , \beta' , \gamma'$ are the logarithmic scale factors of a small billiard (with gravitational wall $\alpha ' = 0$ and symmetry walls $\alpha' = \beta'$ and $\beta' = \gamma'$). In other words, a suitable (time-dependent) re-ordering of the three big-billiard variables $\alpha , \beta , \gamma$ transforms the big-billiard dynamics in the small-billiard one. Reciprocally, starting from the graph giving the three curves $\alpha' (\tau)$, $\beta' (\tau)$, $\gamma' (\tau)$ of a small-billiard dynamics one can, starting from an arbitrary identification of $\alpha , \beta , \gamma$ with $\alpha' , \beta' , \gamma'$ at some initial time, extend the definition of the $\alpha (\tau)$, $\beta (\tau)$, $\gamma (\tau)$ curves by the condition that the only changes of slopes of these curves happen when one of them touches the zero axis. This graphical reasoning shows that there is an essential equivalence between the two billiards. Note in passing that this graphical approach can also clarify why the definition used in the previous subsection of the red-return eras introduces an artificial difference between the two billiards. Indeed, the red-return eras are defined by the ``collisions'' $\beta' (\tau) = \gamma' (\tau)$ between the two upper curves. By contrast, the usual big-billiard eras are defined by a different condition consisting in finding when the ``oscillations'' between the crossing  curves $\alpha (\tau)$ and $\beta (\tau)$ cease to give rise to oscillations between either $\alpha$ and $\gamma$ or $\beta$ and $\gamma$. The difference between these two definitions of ``eras'' give rise to the technical differences found in the previous subsection.\\
\vspace{0.1cm}

A second way of relating the two dynamics consists of introducing some further quotienting of the small billiard. Explicitly, if we replace the small billiard by its kaleidoscopic version (as explained above in the big billiard case), i.e. by replacing the single moving ball of the small billiard by its six images with respect to the symmetry walls, we end up with a quotiented small billiard where the symmetry walls have no effect (because of the equivalence of the $S_3$ orbits) and where the only effective collisions take place on the gravitational walls. Finally, we conclude that the symmetry-wall-quotiented small-billiard coincides with the symmetry-wall-quotiented big-billiard. This shows again that, modulo some discrete relabellings, the two billiards are essentially identical.

\section{Brief concluding remarks}
In this work, we have analyzed several aspects of ``chaotic'' cosmological billiards, in ($3+1$)-dimensional gravity, and of their statistical behavior as one approaches the cosmological singularity. We have reviewed how the dynamics of the diagonal degrees of freedom (logarithmic scale factors $\beta^a$) of the spatial metric near the singularity is conveniently described by Lorentzian, or (after projection) hyperbolic-space billiards. We emphasized that the hyperbolic-space billiard table for the usual, Bianchi IX $a b c$ system is an ideal triangle (with three vanishing angles), which contains six copies of the ``small billiard'' that is naturally ``related'' to the Weyl chamber of an hyperbolic Kac-Moody algebra, see Fig. \ref{fig2}. We reviewed several useful facts about integral invariants in Hamiltonian systems, and showed how their application to cosmological billiards allowed one to derive several forms and measures that are invariant under both the continuous and the discrete billiard dynamics. Contrary to previous treatments of cosmological billiards (starting with the classic work of Belinski, Khalatnikov and Lifshitz, BKL), we did not use the six-fold symmetry group ($S_3$) of the Bianchi IX $a$, $b$, $c$ system to symmetry-quotient its dynamics. This led us to defining a richer ``hopscotch dynamics'' between several sub-regions of the two-dimensional phase-space ($u^-, u^+$) parametrizing successive Kasner epochs. Several aspects of this hopscotch dynamics have been discussed in detail: (i) the existence of a non-normalizable measure on the two-dimensional ($u^-, u^+$) hopscotch court; (ii) the existence of a non-normalizable measure on the single variable $u^+$, i.e. on the Kasner circle parametrizing the exponents of successive Kasner epochs; (iii) the existence of a normalizable measure on the subset of the hopscotch court describing the first epochs of successive eras; (iv) the link between the unquotiented hopscotch dynamics, and its quotiented version, equivalent to the usual BKL dynamics.\\
Several statistical features of the hopscotch dynamics have been discussed, e.g. ($1$) the joint probability $P_{n_1, n_2}$ for two successive eras to have specified lengths $n_1$ and $n_2$, and the fact that the random variables $n_1$ and $n_2$ are not statistically independent; ($2$) the ``anisotropic'' behavior of the hopscotch dynamics, i.e. the fact that the successive corners, between which the billiard ball representing the metric bounces, are statistically correlated. We briefly discussed the link between periodic orbits in the unquotiented hopscotch court, and the usually discussed periodic orbits in the quotiented, BKL description. Finally, we discussed the relation between the billiard dynamics within the full ideal triangle associated with the (diagonal) Bianchi IX model, and the dynamics between the ``small billiard'' that naturally arises in the treatment of the gravitational dynamics that uses an Iwasawa decomposition of the spatial metric.
\appendix*\section{\label{integralinvariants}Integral invariants in Hamiltonian systems}
As a complement to Section \ref{reducedforms}, we recall here some (not always well known) facts about integral invariants in Hamiltonian systems. When considering a general (possible time-dependent) Hamiltonian dynamics with (Hamiltonian) action (\ref{4.1new}), one defines the Poincar\'e-Cartan one-form
\be\label{4.2a}
\sigma^{(1)}_{PC}:=p_idq^i-H(q,p,t)dt,
\ee
defined in the ($2n+1$)-dimensional \textit{extended} phase space $X^{(2n+1)}=\{(q^i, p_i, t)\}$. The Poincar\'e-Cartan one-form (\ref{4.2a}) then defines a \textit{relative} integral invariant of the \textit{unparametrized} Hamiltonian flow. This means that the integral
\be\label{4.3a}
I(C)=\oint_C\sigma^{(1)}_{PC}
\ee
of $\sigma^{(1)}$ over any \textit{closed} curve $C$ in extended phase space remains the same if one displaces $C$ in an arbitrary manner along the \textit{unparametrized} flow lines of the Hamiltonian dynamics. [By \textit{unparametrized} flow lines we mean here the unparametrized lines in $X^{(2n+1)}$ whose tangents are everywhere parallel to the Hamiltonian flow, i.e. proportional (without being necessarily equal) to $(\partial H/\partial p_i, -\partial H/\partial q^i, 1)$.] An equivalent formulation of this property is to say that the Poincar\'e-Cartan two-form
\be\label{4.4a}
\omega^{(2)}_{PC}:=-d\sigma^{(1)}_{PC}=dq^i\wedge dp_i-dt\wedge dH(q,p,t)
\ee
is an \textit{absolute} integral invariant of the unparametrized Hamiltonian flow, i.e. that the integral of $\omega^{(2)}$ over an arbitrary two-surface $\Sigma$ (with boundary) is invariant as $\Sigma$ is moved, in an arbitrary manner, along the unparametrized flow lines. Note that these invariance properties are stronger than those that are usually stated, which only refer to the invariance of the simpler `Liouville' forms in the \textit{unextended} phase space $X^{(2n)}=\{ (q^i, p_i)\}$,
\begin{subequations}
\begin{align}
&\sigma^{(1)}_L:=p_idq^i,\label{4.5a}\\
&\omega^{(2)}_L:=-d\sigma^{(1)}_L=dq^i\wedge dp_i,\label{4.6a}
\end{align}
\end{subequations}
under the time-parametrized Hamiltonian flow $(\partial H/\partial p, -\partial H/\partial q, 1)$.\\
Moreover, the absolute invariance of the two-form $\omega^{(2)}_{PC}$, Eq. (\ref{4.4a}), allows one to construct an invariant measure on any ($2n$)-dimensional transverse section moving along the unparametrized Hamiltonian flow, namely
\be\label{4.7a}
\Omega^{(2n)}_{PC}:=c(n)\left(\omega^{(2)}_{PC}\right)^{\wedge n}\equiv c(n)\omega^{(2)}_{PC}\wedge...\wedge\omega^{(2)}_{PC}\ \  ({\rm with}\ \ n \ \  {\rm factors}).
\ee
Here $c(n)$ is a numerical factor which can be taken to be $(-)^{n(n-1)/2}(n!)^{-1}$ if one wishes to recover the usual normalization. In the case when one restricts oneself to displacements along the time-parametrized Hamiltonian flow one can replace the extended phase space form $\Omega^{(2n)}_{PC}$ by the Liouville measure 
\be\label{4.8a}
\Omega^{(2n)}_L:=c(n)\left(\omega^{(2)}_{L}\right)^{\wedge n}=dq^1\wedge dq^2\wedge...\wedge dq^n\wedge dp_1\wedge dp_2\wedge...\wedge dp_n
\ee
on the \textit{unextended} phase space $X^{(2n)}=\left\{(q^i,p_i)\right\}$.\\
\vspace{0.1cm}

In addition, if one is considering a time-independent Hamiltonian $H(p,q)$, and if one wishes to restrict oneself to the dynamics on a specific $(2n-1)$-dimensional energy hypersurface, say $\mathcal{E}^{(2n-1)}_E$, satisfying $H(p,q)=E$ in unextended phase space $X^{(2n)}$, the above results simplify in that one can drop the $H$-dependent contribution in (\ref{4.2a}) and (\ref{4.4a}) (because $dH(q,p)=0$ on the energy shell) and conclude that the simpler Liouville-type two-form $\omega^{(2)}_L$, Eq. (\ref{4.6a}), is invariant not only under the usual time-parametrized Hamiltonian flows $\dot{q}=\partial_pH$, $\dot{p}=-\partial_qH$ in $\mathcal{E}^{(2n-1)}$, but also under more general `many fingered time flows' $\dot{q}=F\partial_pH$, $\dot{p}=-F\partial_qH$, involving an arbitrarily varying time-rescaling function $F(q,p,t)$.\\
This leads to introducing two possible measures associated to the dynamics on the energy-surface $\mathcal{E}^{(2n-1)}_E$. A first measure is the standard energy-shell reduced Liouville measure
\be\label{4.9a}
\Omega^{(2n-1)}_{L,E}=\Omega^{(2n)}_{L}\delta(H(q,p)-E),
\ee 
which is a ($2n-1$)-form, and yields a smooth measure on the ($2n-1$)-dimensional energy shell $\mathcal{E}^{(2n-1)}$.\\
\vspace{0.1cm}

A second possible construction (which is linked to the general theory of the reduction of phase spaces with symmetry) is to use the invariance of the two-form $\omega^{(2)}_{PC}$, Eq. (\ref{4.4a}), or simply $\omega^{(2)}_{L}$, Eq. (\ref{4.6a}), under arbitrary `glidings' along the Hamiltonian flow (which takes place within $\mathcal{E}^{(2n-1)}$) to define both a reduced (symplectic) $2$-form
\be\label{4.10a}
\omega^{(2)}_{\rm red}:=\left[\omega^{(2)}_L\right]_{Q^{(2n-2)}_E}
\ee
and a corresponding  measure
\be\label{4.11a}
\Omega^{(2n-2)}_{\rm red}:=c(n-1)\left(\omega^{(2)}_{\rm red}\right)^{\wedge (n-1)}
\ee 
on the ($2n-2$)-dimensional \textit{quotient space} $Q^{(2n-2)}_E=\mathcal{E}^{(2n-1)}_E/\mathcal{F}_H$ (where $\mathcal{F}_H$ denotes the unparametrized Hamiltonian flow on $\mathcal{E}^{(2n-2)}_E$). In other words, $\mathcal{E}^{(2n-1)}_E/\mathcal{F}_H$ is the space of unparametrized Hamiltonian motions on $\mathcal{E}^{(2n-1)}_E$. A concrete representation of this quotient space can be obtained by considering any transverse section of $\mathcal{F}_H$ on $\mathcal{E}^{(2n-1)}$, i.e. any `initial conditions' for $\mathcal{F}_H$. Note that this transverse section does not need to be taken at some fixed time $t$, but can have an arbitrary `slope' in extended phase space. The invariance of $\omega^{(2)}_{PC}=\omega^{(2)}_{L}$ (on $\mathcal{E}^{(2n-1)}$) under $\mathcal{F}_H$ then guarantees that the ($2n-2$)-form (\ref{4.10a}) lifts to the quotient space $Q^{(2n-2)}_E=\mathcal{E}^{(2n-1)}_E/\mathcal{F}_H$, i.e. defines a measure on the ($2n-1$)-dimensional space of (unparametrized) motions with energy $E$.\\
\vspace{0.1cm}

Let us mention that there is a simple link between the reduced ($2n-2$)-form $\Omega^{(2n-2)}_{\rm red}$, Eq. (\ref{4.10a}) and the usually considered energy-shell reduced Liouville measure $\Omega^{(2n-2)}_{L,E}$, Eq. (\ref{4.9a}). First, note that these forms define measures on different spaces: $\Omega^{(2n-1)}_{L,E}$ `lives' on the ($2n-1$)-dimensional energy surface $\mathcal{E}^{(2n-1)}_E$, while $\Omega^{(2n-2)}_{\rm red}$ `lives' on the ($2n-2$)-dimensional quotient $Q^{(2n-1)}_E$ of   $\mathcal{E}^{(2n-1)}_E$ by the Hamiltonian flow $\mathcal{F}_H$. To see the link between these constructs, we can introduce (at least locally) in the full, ambient ($2n$)-dimensional phase space $X^{(2n)}=\{(q^i,p_i)\}$ a new canonical coordinate system where the $n$-th momentum coordinate $p_n^{\rm new}$ is equal to the Hamiltonian: $H(p^{\rm old},q_{\rm old})=p_n^{\rm new}$. In this new canonical coordinate system, the $n$-th conjugate position coordinate $q^n_{\rm new}$ is such that
\be\label{4.14a}
\frac{dq^n_{\rm new}}{dt}=\frac{\partial H}{\partial p_n^{\rm new}}=1,
\ee   
while the remaining coordinates ($q^{\bar{i}},p_{\bar{i}}$), with $\bar{i}=1,...,n-1$ all satisfy $\dot{q}^{\bar{i}}_{\rm new}=0=\dot{p}^{\rm new}_{\bar{i}}$, i.e. are invariant under the Hamiltonian flow $\mathcal{F}_H$ (considered in unparametrized form, i.e. with many fingered time displacements: $\Delta t=F(p,q)$).\\
\vspace{0.1cm}

If we denote the conjugate pair ($q^n_{\rm new},p_n^{\rm new}$) simply by ($s,H$), and the ($n-1$)-other pairs by ($\bar{q}^{\bar{i}},\bar{p}_{\bar{i}}$), we see that the symplectic form $\omega^{(2)}_L$, Eq. (\ref{4.6a}), in the ambient phase space $X^{(2n)}$ reads
\be\label{4.15a}
\omega^{(2)}_L(q,p)=\omega^{(2)}_{\rm red}(\bar{q},\bar{p})+ds\wedge dH,
\ee
where 
\be\label{4.16a}
\omega^{(2)}_{\rm red}(\bar{q},\bar{p})=\sum_{\bar{i}=1}^{n-1}d\bar{q}^{\bar{i}}\wedge d\bar{p}_{\bar{i}}
\ee
is clearly equal to the reduced symplectic form (\ref{4.10a}) on the quotient space $Q^{(2n-2)}_E=\mathcal{E}^{(2n-1)}_E /\mathcal{F}_H$ (independently of the values of $E$ and $s$).\\
\vspace{0.1cm}

If we now insert (\ref{4.15a}) in the general definition (\ref{4.9a}) of the energy-shell-reduced Liouville measure, one easily sees, using $\delta(H-E)dH=1$, that
\be\label{4.17a}
\Omega^{(2n-1)}_{L,E}=\Omega^{(2n-2)}_{\rm red}\wedge ds,
\ee
where, as explained above, $s$ is a phase-space coordinate which is canonically conjugate to the Hamiltonian (which implies that $ds/dt=1$ along the Hamiltonian flow).\\
\vspace{0.1cm}

So far, we have been considering any autonomous Hamiltonian system ($\partial H/\partial t=0$). We can, in particular, apply the above results to general \textit{billiard systems}, i.e. to a Hamiltonian of the form
\be\label{4.18a}
H(p,q)=\sum_{i,j=1}^n\frac{1}{2}g^{ij}(q)p_ip_j+V_{\infty}(q),
\ee 
where $g^{ij}(q)$ is the matrix inverse of some (pseudo-)Riemannian metric $g_{ij}(q)dq^idq^j$, and where the (formal) potential function $V_\infty(q)$ is equal to zero within some domain, say $\mathcal{B}$ (the ``billiard table''), of the $q$ variables, and equal to $+\infty$ outside of this domain. In that case, the general invariance of the reduced symplectic form $\omega^{(2)}_{\rm red}$ under arbitrary, many fingered time motions on the energy hypersurface (see Eq. (\ref{4.15a})) can be concretely used to show that the restriction of the ambient phase space symplectic form $\omega^{(2)}_L(q,p)$ on the boundary $\partial\mathcal{B}$ (with the constraint $H(q,p)=E$) of the billiard , say $\omega^{(2)}_{\rm restr}(q^{\rm restr},p_{\rm restr})$, is invariant both under each collision on $\partial\mathcal{B}$ and under each `free flight' between two successive collisions. In other words, we are here considering transverse (Poincar\'e-type) cross sections of the energy-reduced Hamiltonian flow on $\mathcal{E}^{(2n-1)}_E$, defined, in a `stroboscopic' manner, by the successive collisions. This allows one to extract from the continuous Hamiltonian flow $\phi_t$ [$x(t)=\phi_t(x(0))$ with $x\equiv(q,p)$], the discrete `billiard map' say $\mathcal{T}$, such that $x_{N+1}=\mathcal{T}(x_N)$ where $x_N=(q_N,p_N)$ is the phase-space position just after the $N$-th collision on $\partial\mathcal{B}$ and $\mathcal{T}(x)=\phi_{\tau(x)+0}(x)$ the stroboscopic Hamiltonian evolution\footnote{Note that the time $\tau(x)$ between two successive collisions generally depends on the starting position $x$, so that we need to use here the invariance of the (Poincar\'e-Cartan (\ref{4.4a}) or Liouville (\ref{4.6a})) two-form under the \textit{unparametrized} many-fingered Hamiltonian flow.} between two successive collisions (including the `reflection' effect of the second collision). In other words, the restrictions $\omega^{(2)}_{\rm restr}(q_{\rm restr},p_{\rm restr})$ of $\omega^{(2)}_L$ on the ($2n-2$)-dimensional phase space of $\left[\partial\mathcal{B}\right]_{H(p,q)=E}$ after each collision gives us an infinite collection of concrete realizations of the reduced 2-form $\omega^{(2)}_{\rm red}$ on the abstract quotient space $Q^{(2n)}=\mathcal{E}^{(2n+1)}/\mathcal{F}_H$. It also yields several absolute integral invariants of the billiard map $\mathcal{T}$, namely $\omega^{(2)}_{\rm restr}(q_{\rm restr},p_{\rm restr})$ itself and all its exterior powers, and notably the reduced measure $\Omega^{(2n-2)}_{\rm red}$, Eq. (\ref{4.11a}). Note that the link (\ref{4.17a}) between the energy-shell Liouville measure and the reduced measure $\Omega^{(2n-2)}_{\rm red}$ (invariant under the billiard map $\mathcal{T}$) is well-known in the mathematical literature on billiards (see, e.g., \cite{corn1982} ).\\
\vspace{0.1cm}

Let us now discuss the various ways in which the above results can be applied to cosmological billiards. We start by recalling that the dynamics of the `diagonal' degrees of freedom (i.e. the logarithmic scale factors $\beta^a$ entering the Iwasawa decomposition (\ref{2.4})) is described, near a cosmological singularity, by an action of the general form \cite{Damour:2002et}
\be\label{4.1}
S_\beta=\int dx^0\left[\frac{1}{2\tilde{N}}G_{ab}\dot{\beta}^a\dot{\beta}^b-\tilde{N}V(\beta)\right],
\ee
where $\dot{\beta}^a=\frac{d\beta^a}{dx^0}$, and where $V(\beta)$ is a sum of `exponential walls':
\be\label{4.2}
V(\beta)=\sum_Ac_A\exp\left(-2w^A(\beta)\right),
\ee
with linear forms $w^A(\beta)$, see Eqs. (\ref{2.11}, \ref{2.12}, \ref{2.13}). The spatial gradients of the metric and of the other fields enter only in the coefficient $c_A$ of the exponential walls. In the near-spacelike-singularity limit (`BKL limit') the time- and space- dependent coefficients $c_A(x^0, \mathbf{x})$ tend to some finite limits so that one can describe the asymptotic dynamics of the $\beta^a(x^0,\mathbf{x})$ at each point of space by means of the Toda-like billiard (\ref {4.1}, \ref{4.2}) (with $c_A$ replaced by their limits). A further approximation (which also holds in the BKL limit) consists in replacing the exponential walls (\ref{4.2}) by their formal `sharp wall limit', namely
\be\label{4.3}
V_{\infty}(\beta)=\sum_A\Theta_{\infty}(-2w_A(\beta))
\ee
where the formal sharp-wall $\Theta_\infty$-function is defined as: $\Theta_\infty(x):=0$ if $x<0$ and $\Theta_\infty(x):=+\infty$ if $x>0$.\\
 The action (\ref{4.1}) with $V(\beta)\rightarrow V_\infty(\beta)$ given by (\ref{4.3}) defines a Lorentzian billiard dynamics in the $\beta$-space. This dynamics can be equivalently described by the Hamiltonian action
 \be\label{4.4}
 S_\beta=\int dx^0\left[\pi_a\dot{\beta}^a-H_\beta(\beta^a,\pi_a)\right],
 \ee
 \be\label{4.5}
 H_\beta(\beta,\pi)=\tilde{N}\left[\frac{1}{2}G^{ab}\pi_a\pi_b+V_\infty(\beta)\right];
 \ee
which is of the general type (\ref{4.18a}) (with a flat Lorentzian-signature metric $G_{ab}$). Here, $\pi_a$ denotes the conjugate momentum of $\beta^a$, i.e.
\be\label{4.6}
\pi_a=\frac{1}{\tilde{N}}G_{ab}\frac{d\beta^b}{dx^0}.
\ee 
 Note that $\pi_a$ is invariant under the redefinitions of the time coordinate $x^0$ (which affect both $dx^0$ and $\tilde{N}\equiv N/\sqrt{g}$ but leave invariant the product $\tilde{N}dx^0\propto Ndx^0$).\\
\vspace{0.1cm}

 As the (rescaled) lapse $\tilde{N}$ is a Lagrange multiplier in the action $S_\beta$, Eq. (\ref{4.1}), we have the well-known Hamiltonian constraint stating that $H_\beta$ must vanish, i.e. that we must constrain ourselves to the specific energy hypersurface
 \be\label{4.7}
 H_{\beta}(\beta,\pi)=E_\beta=0.
 \ee
 We are therefore in the condition where we can apply the general results recalled above. [For simplicity, we can assume that we are working in any gauge where $\tilde{N}$ is given as some autonomous function of $\beta$ and $\pi$. This is the case both of the $\tau$-time gauge $\tilde{N}=1$, Eq. (\ref{2.3}), and of the $T$-time one $\tilde{N}=\rho^2=-G_{ab}\beta^a\beta^b$, Eq. (\ref{2.21}). ] In particular, we see that 
 \be\label{4.8}
 \omega^{(2)}_\beta=-d[\pi_ad\beta^a]=d\beta^a\wedge d\pi_a
 \ee 
is an absolute integral invariant of the Hamiltonian flow, as well as the corresponding energy-shell measure
\be\label{4.9}
\Omega^{(2d-1)}_{E_\beta=0}=c(d)\left(\omega^{(2)}_\beta\right)^{\wedge d}\delta(H_\beta).
\ee 
In addition, we can consider the (double) reduction of the $2$-form $\omega^{(2)}_\beta$ on the quotient space $Q^{(2d-2)}_{\beta}=\mathcal{E}^{(2d-1)}_{E_\beta=0}/\mathcal{F}_{H_\beta}$,
\be\label{4.10}
\omega^{(2)}_{\beta {\rm red}}=\left[d\beta^a\wedge d\pi_a\right]_{Q^{(2d-2)}_{\beta}},
\ee
and the corresponding measure on $Q^{(2d-2)}_{\beta}$
\be\label{4.11}
\Omega^{(2d-2)}_{\beta {\rm red}}=c(d-1)\left(\omega^{(2)}_{\beta {\rm red}}\right)^{\wedge(d-1)}.
\ee
 The latter reduced measure is related to the Liouville-type measure (\ref{4.9}) via the general result (\ref{4.17a}) where $s$ is a $\beta$-phase-space coordinate which is canonically conjugate to $H_\beta$, Eq. (\ref{4.5}). These results are particularly simple if one uses the $\tau$-time gauge where $\tilde{N}=1$, so that the Hamiltonian (\ref{4.5}) is the sum of a constant-Lorentzian-metric kinetic term $\frac{1}{2}G^{ab}\pi_a\pi_b$ and of a sharp-wall billiard potential.\\
 The integral invariants we have just discussed concern the dynamics of the Lorentzian billiard in $\beta$-space. They would be useful if one were studying the full $\beta$-space billiard dynamics. However, in this paper, we are interested in discussing the projection of the $\beta$-billiard on the hyperbolic space $\mathcal{H}_{d-1}$, e.g. described by the unit hyperboloid (\ref{2.22}) in $\beta$-space. This projection is obtained by separating out the motion along the `radial direction' $\rho$, Eq. (\ref{2.19}). Indeed, setting $\bar{N}\equiv\tilde{N}/\rho^2$ and $\lambda\equiv\ln\rho$, the billiard action (\ref{4.1}), (\ref{4.3}) can be rewritten as (see, e.g. \cite{Misner:1974qy} for the $d=3$ case and \cite{Damour:2000hv} for the general case)
 \be\label{4.12}
 S=\int dx^0 \left[\frac{1}{2\bar{N}}\left(-\dot{\lambda}^2+G_{ab}\dot{\gamma}^a\dot{\gamma}^b\right)-\bar{N}V_\infty(\gamma)\right],
 \ee
where $V_\infty(\gamma)=\sum_A\Theta_\infty(-2w_A(\gamma))$. [As usual, we are using here the simplifying fact that, in the limit $\rho\rightarrow+\infty$, the potential term becomes independent of $\rho$, i.e. becomes $T$-time independent.]\\
In the $T$ gauge (\ref{2.21}), i.e. $\bar{N}\equiv\tilde{N}/\rho^2=1$, the radial kinetic energy term $-\frac{1}{2}\dot{\lambda}^2$ decouples from the angular motion terms and leads to a uniform radial motion: $d\lambda/dT={\rm const}$. in that gauge, one can simply work with the ``angular action''
\be\label{4.13}
S_\gamma=\int dT\left[\frac{1}{2}G_{ab}\frac{d\gamma^a}{dT}\frac{d\gamma^b}{dT}-V_\infty(\gamma)\right],
\ee
submitted to the constraint that the (constant) angular-motion energy
\be\label{4.14}
E_\gamma=\frac{1}{2}G_{ab}\frac{d\gamma^a}{dT}\frac{d\gamma^b}{dT}+V_\infty(\gamma)
\ee
be equal to $(d\lambda/dT)^2={\rm const}$.\\
\vspace{0.1cm}

Note that while the $\beta$-space action $S_\beta$, Eq. (\ref{4.4}), corresponded to a phase-space ($\beta^a, \pi_a$) with $2d$ dimensions, the reduced action $S_\gamma$, Eq. (\ref{4.13}), corresponds to a phase space with $2(d-1)$ dimensions. In order to explicitly describe the reduced dynamics, one needs to choose some parametrization of the $(d-1)$ dimensional hyperbolic space, say $q^i$, where the index $i$ takes only $d-1$ values. The hyperbolic metric on $\mathcal{H}_{d-1}$ will have some expression, say ($i,j=1, ..., d-1$)
\be\label{4.15}
G_{ab}d\gamma^a d\gamma^b=ds^2=g_{ij}(q)dq^idq^j,
\ee
and the angular action (\ref{4.11}) will read
\be\label{4.16}
S_\gamma=\int dT\left[\frac{1}{2}g_{ij}(q)\frac{dq^i}{dT}\frac{dq^j}{dT}-V_\infty(\gamma(q))\right].
\ee
The conjugate momenta to the $q^i$'s read
\be\label{4.17}
p_i=g_{ij}(q)\frac{dq^j}{dT},
\ee
while the angular-motion Hamiltonian will read
\be\label{4.18}
H_\gamma(q^i,p_i)=\frac{1}{2}g^{ij}(q)p_ip_j+V_\infty(\gamma(q)),
\ee
where $g^{ij}(q)$ denotes the inverse of the (covariant) metric $g_{ij}(q)$.\\
\vspace{0.1cm}

Similarly to the discussion above, we can now introduce the (reduced) Poincar\'e-Cartan one-form
\be\label{4.19}
\sigma^{(1)}_\gamma:=p_idq^i-H_\gamma(q,p)dT,
\ee 
and the corresponding two-form
\be\label{4.20}
\omega^{(2)}_\gamma:=-d\sigma^{(1)}_\gamma=dq^i\wedge dp_i-dT\wedge dH_\gamma(q,p).
\ee
As before $\sigma^{(1)}_\gamma$ (respectively $\omega^{(2)}_\gamma$) defines a \textit{relative} (resp. \textit{absolute}) integral invariant of the unparametrized Hamiltonian flow in extended phase-space ($q,p,T$). And, as before, we can use the absolute invariance of the two-form $\omega^{(2)}_\gamma$ to construct an invariant measure. As we are again in a situation where we can work on a fixed-energy hypersurface (here $H_\gamma=E_\gamma={\rm const}$, after eliminating the uniform radial motion $\lambda(T)$) we can drop the last, $H_\gamma$-dependent term in $\omega^{(2)}_\gamma$, Eq. (\ref{4.20}), and work with the usual ($\gamma$-space) symplectic form $\omega^{(2)}_\gamma=dq^i\wedge dp_i$. As before we end up with having a whole set of integral invariants of the billiard dynamics in $\gamma$-space (i.e. on the hyperboloid $\mathcal{H}_{d-1}$): the two-form $\omega^{(2)}_\gamma$ itself, and its various exterior powers, and its energy-shell restricted measure
\be\label{4.21}
\Omega^{(2d-3)}_{E_\gamma}=c(d-1)\left(\omega^{(2)}_\gamma\right)^{\wedge(d-1)}\delta\left(H_\gamma(p,q)-E_\gamma\right).
\ee
Moreover, we can also use the invariance of the reduction of the symplectic form on the quotient space $Q^{(2d-4)}_\gamma=\mathcal{E}^{(2d-3)}_{E_\gamma}/\mathcal{F}_{H_{\gamma}}$,
\be\label{4.22}
\omega^{(2)}_{\gamma \rm red}=\left[dq^i\wedge dp_i\right]_{Q^{(2d-4)}_\gamma}
\ee
and its maximal exterior power 
\be\label{4.23}
\Omega^{(2d-4)}_{\gamma {\rm red}}=c(d-2)\left(\omega^{(2)}_{\gamma {\rm red}}\right)^{\wedge(d-2)}.
\ee
In addition, we still have a link of the type (\ref{4.17a}) (where $s$ is a $\gamma$-phase-space coordinate such that $ds/dT=1$ along the Hamiltonian flow), and we also know that the abstract quotient-space reduced symplectic form $\omega^{(2)}_{\gamma {\rm red}}$ can be concretely computed by restricting $dq^i\wedge dp_i$ by two conditions: $H_\gamma(q,p)=E_\gamma$ and any cross-section condition transverse to the Hamiltonian flow. In particular, we can use the events of collisions on successive walls of the billiard as cross-sections, and thereby prove that $\omega^{(2)}_{\gamma {\rm collision}}$ and $\Omega^{(2d-4)}_{\gamma \rm{collision}}$ are invariants of the discrete $\gamma$-billiard map $\mathcal{T}$ which connects a collision to the next.\\
\vspace{0.1cm}

In Section (\ref{reducedforms}), we apply these general results to the case of the BKL cosmological billiards, in $d=3$ spatial dimensions, working within the radially-projected picture on the $\gamma$-space, i.e. on the hyperbolic plane $\mathcal{H}_{2}$.
\section*{Acknowledgments} We thank Volodia Belinski and David Ruelle for informative discussions. One of us (O.M.L.) gratefully acknowledges the support of an Angelo Della Riccia grant for the early stages of this work.

\end{document}